\documentclass[aps,prd,reprint,superscriptaddress,nofootinbib,floatfix,10pt]{revtex4-2}
\usepackage{amsmath,amssymb,bm,graphicx,diagbox}
\usepackage{slashed}
\usepackage[colorlinks=true,linkcolor=blue,citecolor=blue,urlcolor=blue]{hyperref}

\begin{document}

\title{Self-interacting dark matter and neutrino masses \\from MeV-scale lepton-number breaking}

\author{Manoranjan Dutta}
\email{md@nlu.ac.in}%
\affiliation{Department of Physics, North Lakhimpur University, Khelmati, North Lakhimpur, Lakhimpur, Assam 787031, India}


\begin{abstract}
We show that a scalar whose vacuum expectation value (vev) breaks lepton number at the MeV scale, generating the light-neutrino masses through the inverse seesaw, can also serve as the light mediator of self-interacting dark matter. Dark matter is a pseudo-Dirac fermion whose mass splitting arises from the same vev and is predicted to be proportional to the self-interaction coupling times the vev. A strong central interaction between the dark states requires this splitting to be sizable. The CP-even mediator decays into neutrinos at a rate fixed by the neutrino masses, independent of its mixing with the Higgs boson. Fixing the coupling by the relic abundance and evaluating exact partial-wave cross sections, the model reproduces the velocity dependence of the self-interaction cross section inferred from dwarf galaxies to clusters along a narrow band, $m_s\simeq1.9~{\rm MeV}\times(m_\chi/{\rm GeV})$, extending up to $m_\chi\simeq12$~GeV with a best fit at $m_\chi\simeq7.6$~GeV and $m_s\simeq15$~MeV, and on a sequence of resonant islands of comparable fit quality at higher masses. The cluster data bound the dark matter mass from above, so that the allowed masses extend from about $3$ to $70$~GeV ($95\%$ confidence level), with correlated mediator masses of $6$-$28$~MeV. Combining constraints from the effective number of neutrino species, SN~1987A and direct detection, we find the lepton-number-breaking scale confined to a window of a few hundred MeV and a strong preference for normal neutrino mass ordering. The allowed parameter space is testable by future measurements of $N_{\rm eff}$, low-threshold direct detection and the neutrino mass ordering.
\end{abstract}

\maketitle

\section{Introduction}
\label{sec:introduction}

The existence of dark matter (DM) and the nonzero masses of neutrinos are two of the most robust indications of physics beyond the Standard Model (SM). On large scales, the observed structure of the Universe is well described by cold, collisionless DM. On galactic and sub-galactic scales, however, the inferred inner density profiles of dwarf and low-surface-brightness galaxies and the diversity of galactic rotation curves~\cite{Oman:2015xda} have motivated the possibility that DM scatters off itself with an appreciable cross section~\cite{Carlson:1992fn,deLaix:1995vi,Spergel:1999mh,Ackerman:2008kmp,Vogelsberger:2012ku,Rocha:2012jg,Peter:2012jh,Zavala:2012us,Kamada:2016euw,Tulin:2017ara,Bullock:2017xww,Dutta:2024ydz}. A combined analysis of systems ranging from dwarf galaxies to galaxy clusters prefers a velocity-dependent cross section per unit mass, of order $\sigma/m_\chi\sim1$-$10~{\rm cm^2/g}$ at the characteristic velocities of dwarf galaxies and falling to $\sigma/m_\chi\sim0.1~{\rm cm^2/g}$ in clusters~\cite{Kaplinghat:2015aga}. Such a velocity dependence arises naturally when DM interacts through a light MeV-scale mediator~\cite{Feng:2009mn,Feng:2009hw,Buckley:2009in,Loeb:2010gj,Tulin:2012wi,Tulin:2013teo,Kamada:2020buc}.

A light mediator must, however, be embedded in a consistent cosmology. It has to decay before Big Bang nucleosynthesis (BBN), and the energy it injects must not distort the cosmic microwave background (CMB). If the mediator is a scalar that couples to the SM only through mixing with the Higgs boson, these requirements are in strong tension with direct detection: for MeV-scale masses it can decay only into electron-positron pairs, at a rate suppressed by the electron Yukawa coupling, so that a lifetime below one second requires a mixing angle $\theta_{hs}\gtrsim10^{-5}$, whereas the absence of a signal in direct search experiments requires $\theta_{hs}\lesssim10^{-8}$~\cite{Kaplinghat:2013yxa,Kouvaris:2014uoa,Kainulainen:2015sva,Bringmann:2016din}. A mediator that decays dominantly into neutrinos avoids this tension, and such scenarios have been considered with the neutrino coupling introduced as an additional ingredient~\cite{vandenAarssen:2012vpm,Ma:2018dilepton,Hambye:2019tjt,PandaX:2025scalar,Wang:2026cyl}.

Neutrino masses provide an independent motivation for new physics at low scales. In the inverse seesaw~\cite{Mohapatra:1986bd,Bernabeu:1987gr,Deppisch:2015qwa}, the light-neutrino masses are proportional to a small lepton-number-violating parameter $\mu_S$, so that the heavy neutral fermions can lie near the electroweak or TeV scale with sizable Yukawa couplings. If $\mu_S$ is generated dynamically by the vacuum expectation value (vev) $v_\phi$ of a scalar singlet $\phi$ that breaks lepton number~\cite{Chikashige:1980ui,GonzalezGarcia:1988rw}, then $v_\phi$ becomes a physical scale of the theory, and the excitations of $\phi$ couple to the light neutrinos with a strength fixed by the neutrino masses. Connections between self-interacting dark matter (SIDM) and neutrino masses or leptogenesis have been explored in several models~\cite{Borah:2021pet,Borah:2021rbx,Dutta:2022knf,Mahapatra:2023oyh,Borah:2024wos,Dutta:2025xlv}. It is therefore natural to ask whether the same scalar sector can also provide the light mediator required by SIDM. In this work we show that it can, and that the connection is predictive. 

We consider a Dirac fermion $\chi$ that carries lepton number and is stabilized by a $\mathbb Z_2$ symmetry. After lepton-number breaking, $\phi$ generates small Majorana masses for both chiral components of $\chi$, which becomes a pseudo-Dirac pair of Majorana states $\chi_{1,2}$ with a mass splitting $\Delta m_\chi$. The same vev generates $\mu_S$, so that the dark-matter spectrum and the light-neutrino masses originate from a common scale. A central part of our analysis concerns the structure of the couplings of the CP-even singlet $s$ and of the pseudo-Majoron $a$ to the pseudo-Dirac pair. Because the two Majorana mass eigenstates carry opposite intrinsic CP parity, the CP-even scalar couples to $\chi_1\chi_1$ and $\chi_2\chi_2$ through scalar interactions, but connects $\chi_1$ with $\chi_2$ only through a pseudoscalar interaction, while the pseudo-Majoron has the opposite pattern.  We show that every coupling capable of producing a central, spin-independent force is fixed by the mass splitting, $|g^{\rm central}|\simeq\Delta m_\chi/(2v_\phi)$, for arbitrary values of the Yukawa couplings and of their CP-violating phases, consistent with the Goldstone theorem~\cite{Goldstone:Majorana:2002}. A strong central coupling therefore requires a splitting too large for inelastic scattering in halos~\cite{TuckerSmith:2001hy,Schutz:2014nka,Dutta:2021wbn,Borah:2021yek}, so a scalar whose vev generates the pseudo-Dirac splitting cannot mediate a strong, purely inelastic self-interaction. In particular, a scalar transition coupling of unsuppressed strength, as assumed in recent constructions~\cite{Das:2026modular}, is not compatible with the pseudo-Dirac structure. 

The viable regime is instead the one in which the Majorana Yukawa couplings of the two chiral components are approximately equal. The CP-even singlet then mediates an attractive, elastic Yukawa interaction between $\chi_1$ particles, and the model predicts \begin{equation} \Delta m_\chi\simeq2\,g_{11}\,v_\phi , \label{eq:intro_prediction} \end{equation} where $g_{11}$ is the self-interaction coupling. For MeV-scale $v_\phi$, the splitting exceeds the kinetic energies available in halos and in terrestrial detectors, so both self-scattering and nuclear scattering are purely elastic. For $\Delta m_\chi>m_s$ the heavier state decays promptly, and DM today consists of $\chi_1$ alone.

The neutrino sector also fixes the fate of the mediator. Because $m_\nu\propto v_\phi$, the CP-even singlet couples to the light neutrinos with strength $m_\nu/v_\phi$ and decays into neutrinos within tens of milliseconds for $v_\phi$ of a few hundred MeV. The tension between BBN and direct detection described above is thereby removed without any additional coupling, and the Higgs mixing that controls the direct-detection rate is automatically small because it is proportional to $v_\phi$. Annihilation of $\chi_{1,2}$ into pairs of mediators is $p$-wave, and all its products end up as neutrinos, so that CMB limits on energy injection do not apply. 

We fix the self-scattering coupling by the relic abundance, allowing for a separate temperature of the dark sector, and compute the self-scattering cross section by solving the Schr\"odinger equation in partial waves for identical Majorana fermions. The model then reproduces the velocity dependence of the self-interaction cross section inferred from dwarf galaxies to galaxy clusters with only two free parameters, the masses of DM and of the mediator. Earlier analyses of these data used a fixed coupling~\cite{Kaplinghat:2015aga}, the Born approximation~\cite{Girmohanta:2022dog}, a Breit-Wigner resonance~\cite{Chu:2018fzy} or benchmark comparisons~\cite{Aboubrahim:2020lnr}.  To our knowledge, the present analysis is the first in which the coupling is fixed by the relic abundance at every point, including the thermal history of the dark sector, and the cross section is computed exactly throughout the parameter space, including the resonant regime with several bound states. The data prefer a narrow band of DM masses around $8$~GeV with mediator masses near $15$~MeV, together with resonant islands of comparable fit quality at higher masses, and the cluster data bound the DM mass from above, $m_{\chi_1}\lesssim70$~GeV at $95\%$ CL.

The couplings of the mediator to neutrinos, which make it decay before BBN, also make it a target for SN~1987A. Because they are dominated by the heaviest neutrino, the supernova bound is weakened for normal ordering of the neutrino masses, while for inverted ordering it conflicts with the $N_{\rm eff}$ bound, so that the model strongly favors normal ordering. Together with the lifetime of the mediator, it confines $v_\phi$ to a window of a few hundred MeV. A thermal origin of the dark sector through the Higgs portal is compatible with current direct-detection limits, which next-generation experiments will probe from above.  

The remainder of the paper is structured as follows. In Sec.~\ref{sec:model} we introduce the model and analyze the structure of the singlet couplings. Sections~\ref{sec:relic} and~\ref{sec:selfint} derive the coupling required by the relic abundance and compare the resulting self-interaction cross section with astrophysical data. The neutrino sector is discussed in Sec.~\ref{sec:neutrino_pheno}. Section~\ref{sec:cosmo} derives the early-Universe and supernova constraints and the resulting window of the lepton-number-breaking scale, Sec.~\ref{sec:dd} treats direct detection, and Sec.~\ref{sec:other} collects the remaining constraints. All requirements are combined in Sec.~\ref{sec:window}, and we conclude in Sec.~\ref{sec:conclusions}. Technical details are given in the appendices. 
\section{The model} 
\label{sec:model} 
\subsection{Field content and symmetries} \label{sec:fields} 
We extend the SM by three generations of gauge-singlet fermions $N_R$ and $S_L$, a gauge-singlet Dirac fermion $\chi$, and a complex scalar singlet $\phi$. Lepton number $U(1)_L$ is a global symmetry that is broken spontaneously by $\langle\phi\rangle$ and explicitly only by soft terms in the scalar potential. The stability of $\chi$ is ensured by an exact $\mathbb Z_2$ symmetry under which it is the only odd field. The charge assignments are listed in Table~\ref{tab:particles}. 
\begin{table}[!htbp]
\centering
\renewcommand{\arraystretch}{1.2}
\begin{tabular}{|c|c|c|c|c|c|}
\hline
Field & $SU(3)_C$ & $SU(2)_L$ & $U(1)_Y$ & $U(1)_L$ & $\mathbb{Z}_2$ \\
\hline
$L$    & $\mathbf{1}$ & $\mathbf{2}$ & $-\frac{1}{2}$ & $+1$ & $+$ \\
$e_R$  & $\mathbf{1}$ & $\mathbf{1}$ & $-1$           & $+1$ & $+$ \\
$N_R$  & $\mathbf{1}$ & $\mathbf{1}$ & $0$            & $+1$ & $+$ \\
$S_L$  & $\mathbf{1}$ & $\mathbf{1}$ & $0$            & $+1$ & $+$ \\
$H$    & $\mathbf{1}$ & $\mathbf{2}$ & $+\frac{1}{2}$ & $0$  & $+$ \\
$\phi$ & $\mathbf{1}$ & $\mathbf{1}$ & $0$            & $-2$ & $+$ \\
$\chi$ & $\mathbf{1}$ & $\mathbf{1}$ & $0$            & $+1$ & $-$ \\
\hline
\end{tabular}
\caption{Field content and charges under the gauge and global symmetries.}
\label{tab:particles}
\end{table}
The Lagrangian of the model is
\begin{align}
\mathcal L={}&\mathcal L_{\rm SM}+\overline{N_R}\,i\slashed\partial N_R+\overline{S_L}\,i\slashed\partial S_L
+\overline\chi\,i\slashed\partial\chi+(\partial_\mu\phi)^\dagger(\partial^\mu\phi)
\nonumber\\
&+\mathcal L_\nu+\mathcal L_\chi-V(H,\phi),
\label{eq:Lagrangian}
\end{align}
where the scalar interactions \(V(H,\phi)\), the neutrino interactions $\mathcal L_\nu$ and
the dark-sector interactions $\mathcal L_\chi$ are given in Secs.~\ref{sec:scalar}, \ref{sec:neutrino}
and~\ref{sec:darkmass}, respectively.

\subsection{Scalar sector}
\label{sec:scalar}

The scalar potential is
\begin{align}
V(H,\phi)={}&
-\mu_H^2H^\dagger H+\lambda_H(H^\dagger H)^2-\mu_\phi^2|\phi|^2
\nonumber\\
&+\lambda_\phi|\phi|^4+\lambda_{H\phi}(H^\dagger H)|\phi|^2
\nonumber\\
&-\frac{\mu'^2}{2}\left(\phi^2+\phi^{*2}\right)
-\frac{\kappa^3}{\sqrt2}\left(\phi+\phi^*\right).
\label{eq:potential}
\end{align}
The first five terms are invariant under $U(1)_L$. The last two terms break lepton number softly, by four and two units respectively. The
$\Delta L=4$ term gives mass to the pseudo-Goldstone boson of lepton-number breaking, and the small $\Delta L=2$ tadpole removes the residual $\mathbb Z_4$ subgroup of $U(1)_L$ that would otherwise lead to stable domain walls (Sec.~\ref{sec:walls}). For $\kappa^3\ll\lambda_\phi v_\phi^3$ its effect on the spectrum is negligible.

We expand around the vacuum as
\begin{equation}
\begin{gathered}
H=\frac{1}{\sqrt2}\begin{pmatrix}0\\ v_h+h\end{pmatrix}, \\
\phi=\frac{1}{\sqrt2}\left(v_\phi+s+ia\right),
\end{gathered}
\label{eq:vevs}
\end{equation}
with $v_h=246$~GeV. The minimization conditions are
\begin{equation}
\begin{gathered}
\mu_H^2=\lambda_Hv_h^2+\tfrac12\lambda_{H\phi}v_\phi^2, \\
\mu_\phi^2=\lambda_\phi v_\phi^2+\tfrac12\lambda_{H\phi}v_h^2-\mu'^2-\frac{\kappa^3}{v_\phi}.
\end{gathered}
\label{eq:minimization}
\end{equation}
The CP-even states $(h,s)$ have the mass matrix
\begin{equation}
\mathcal M^2_{\rm even}=
\begin{pmatrix}
2\lambda_Hv_h^2 & \lambda_{H\phi}v_hv_\phi\\
\lambda_{H\phi}v_hv_\phi & 2\lambda_\phi v_\phi^2+\kappa^3/v_\phi
\end{pmatrix},
\label{eq:Meven}
\end{equation}
and the mass of the pseudo-Majoron $a$ is given by
\begin{equation}
m_a^2=2\mu'^2+\frac{\kappa^3}{v_\phi}.
\label{eq:ma}
\end{equation}
For $\kappa^3\ll\lambda_\phi v_\phi^3$ and small mixing, the singlet mass is $m_s^2\simeq2\lambda_\phi v_\phi^2$ and the $h$-$s$ mixing angle is
\begin{equation}
\theta_{hs}\simeq\frac{\lambda_{H\phi}v_hv_\phi}{m_h^2-m_s^2}
\simeq\frac{\lambda_{H\phi}v_hv_\phi}{m_h^2}.
\label{eq:mixing}
\end{equation}
As CP is conserved in the scalar sector, $a$ does not mix with $h$.

The singlet $s$ will be the mediator of DM self-interactions, which requires $m_s$ in the range of a few to tens of MeV. Perturbativity of $\lambda_\phi$ then implies $v_\phi\gtrsim m_s/\sqrt{8\pi}$, so that the lepton-number-breaking scale is itself low, and Eq.~\eqref{eq:mixing} shows that the Higgs mixing is suppressed by $v_\phi/v_h$ for any given portal coupling. We take the pseudo-Majoron to be heavier than the pair of dark states, $m_a>m_{\chi_1}+m_{\chi_2}$, so that all annihilation channels with $a$ in the final state are closed (Sec.~\ref{sec:relic}) and $a$ decays promptly into $\chi_1\chi_2$ (Sec.~\ref{sec:other}).

\subsection{Neutrino masses}
\label{sec:neutrino}

\begin{figure}[!htbp]
\centering
\includegraphics[width=\columnwidth]{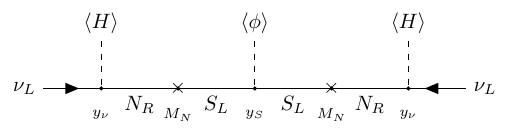}
\caption{Light-neutrino mass in the inverse seesaw. Lepton number is broken by $\langle\phi\rangle$, which generates $\mu_S=y_Sv_\phi/\sqrt2$.}
\label{fig:diag_numass}
\end{figure}

The most general renormalizable neutrino-sector Lagrangian consistent with
the symmetries is
\begin{align}
\mathcal L_\nu={}&-y_\nu\,\overline L\,\widetilde H\,N_R-M_N\,\overline{N_R}\,S_L
\nonumber\\
&-\frac{y_S}{2}\,\phi\,\overline{S_L^c}\,S_L-\frac{y_N}{2}\,\phi\,\overline{N_R^c}\,N_R+{\rm h.c.},
\label{eq:Lnu}
\end{align}
where $\widetilde H=i\sigma_2H^*$ and flavor indices are suppressed. 
After symmetry breaking,
\begin{equation}
\begin{gathered}
m_D=\frac{y_\nu v_h}{\sqrt2}, \quad
\mu_S=\frac{y_Sv_\phi}{\sqrt2}, \quad
\mu_R=\frac{y_Nv_\phi}{\sqrt2},
\end{gathered}
\label{eq:mD_muS}
\end{equation}
and in the basis $n_L=(\nu_L,N_R^c,S_L)^T$, the neutral-fermion mass matrix is
\begin{equation}
\mathcal M_\nu=
\begin{pmatrix}
0&m_D&0\\
m_D^T&\mu_R&M_N\\
0&M_N^T&\mu_S
\end{pmatrix}.
\label{eq:Mnu}
\end{equation}
For $\|\mu_{S,R}\|\ll\|m_D\|\ll\|M_N\|$, the light-neutrino mass matrix is (Fig.~\ref{fig:diag_numass})
\begin{equation}
m_\nu\simeq m_D\,M_N^{-1}\,\mu_S\,(M_N^{-1})^T\,m_D^T ,
\label{eq:inverse_seesaw}
\end{equation}
while $\mu_R$ contributes only at one loop. The heavy states form pseudo-Dirac pairs with masses $M_{N_i}\pm(\mu_{S,i}+\mu_{R,i})/2$.
The phenomenology of the neutrino sector, including the couplings of $s$ and $a$ to the light neutrinos that follow from Eq.~\eqref{eq:inverse_seesaw}, is discussed in Sec.~\ref{sec:neutrino_pheno}.

\subsection{Pseudo-Dirac dark matter}
\label{sec:darkmass}

The dark-sector Lagrangian is
\begin{equation}
\mathcal L_\chi=
-M_\chi\,\overline\chi\chi
-\frac{y_L}{2}\,\phi\,\overline{\chi_L^c}\chi_L
-\frac{y_R}{2}\,\phi\,\overline{\chi_R^c}\chi_R
+{\rm h.c.}
\label{eq:Lchi}
\end{equation}
It is convenient to work with the left-handed Weyl spinors $\xi_1\equiv\chi_L$ and $\xi_2\equiv\chi_R^c$, which carry lepton number
$+1$ and $-1$, respectively. Since $\overline\chi\chi=\xi_1\xi_2+{\rm h.c.}$, $\overline{\chi_L^c}\chi_L=\xi_1\xi_1$ and
$\overline{\chi_R^c}\chi_R=\bar\xi_2\bar\xi_2$, lepton-number invariance pairs $\xi_1\xi_1$ with $\phi$ and $\xi_2\xi_2$ with $\phi^*$:
\begin{equation}
\mathcal L_\chi=-\frac12\Big[2M_\chi\,\xi_1\xi_2+y_L\,\phi\,\xi_1\xi_1
+y_R^*\,\phi^*\,\xi_2\xi_2\Big]+{\rm h.c.}
\label{eq:Lchi_weyl}
\end{equation}
We take the Yukawa couplings to be real; the conclusions of this section hold
for arbitrary phases, as shown in Appendix~\ref{app:takagi}. Inserting
Eq.~\eqref{eq:vevs},
\begin{equation}
\mathcal L_\chi=-\frac12\,\xi^T\Big[\mathcal M_\chi+s\,\mathcal S+a\,\mathcal A\Big]\xi+{\rm h.c.},
\label{eq:Lchi_matrix}
\end{equation}
with $\xi=(\xi_1,\xi_2)^T$ and
\begin{equation}
\begin{gathered}
\mathcal M_\chi=\begin{pmatrix}m_L&M_\chi\\ M_\chi&m_R\end{pmatrix}, \\
\mathcal S=\frac{1}{\sqrt2}\begin{pmatrix}y_L&0\\0&y_R\end{pmatrix}, \\
\mathcal A=\frac{i}{\sqrt2}\begin{pmatrix}y_L&0\\0&-y_R\end{pmatrix},
\end{gathered}
\label{eq:MSA}
\end{equation}
where $m_{L,R}=y_{L,R}v_\phi/\sqrt2$. The relative sign in $\mathcal A$ arises because $\xi_2\xi_2$ couples to $\phi^*$.

Defining
\begin{equation}
\begin{gathered}
\bar m=\frac{m_L+m_R}{2}, \\
\bar\Delta=\frac{m_L-m_R}{2}, \\
R=\sqrt{M_\chi^2+\bar\Delta^2},
\end{gathered}
\end{equation}
the eigenvalues of $\mathcal M_\chi$ are $\bar m\pm R$. In the pseudo-Dirac regime $|m_{L,R}|\ll M_\chi$ one of them is negative, and the physical masses
are
\begin{equation}
\begin{gathered}
m_{\chi_{1,2}}=R\mp|\bar m|, \\
\Delta m_\chi\equiv m_{\chi_2}-m_{\chi_1}=|m_L+m_R| ,
\end{gathered}
\label{eq:masses}
\end{equation}
where the expression for the splitting is exact. The mass eigenstates are obtained as $\xi=U\eta$ with
\begin{equation}
\begin{gathered}
U=\big(\,i\,v_-,\;v_+\big), \\
v_+=\begin{pmatrix}\cos\vartheta\\ \sin\vartheta\end{pmatrix},\quad
v_-=\begin{pmatrix}-\sin\vartheta\\ \cos\vartheta\end{pmatrix},\quad
\tan2\vartheta=\frac{M_\chi}{\bar\Delta},
\end{gathered}
\label{eq:Takagi}
\end{equation}
for $\bar m>0$. The factor $i$ is required to render both masses real and
positive,
\begin{equation}
U^T\mathcal M_\chi U={\rm diag}(m_{\chi_1},m_{\chi_2}),
\end{equation}
and it cannot be removed by a field redefinition. The Majorana fields $\chi_i=(\eta_i,\bar\eta_i)^T$ are the physical dark states.

\subsection{Couplings of the singlets to the dark states}
\label{sec:couplings}

We define
\begin{equation}
\begin{gathered}
\kappa_\chi\equiv\frac{y_L+y_R}{2\sqrt2}=\pm\frac{\Delta m_\chi}{2v_\phi}, \\
\eta_\chi\equiv\frac{y_L-y_R}{2\sqrt2}=\frac{\bar\Delta}{v_\phi}, \\
c\equiv\cos2\vartheta=\frac{\bar\Delta}{R}\simeq\frac{\bar\Delta}{M_\chi}.
\end{gathered}
\label{eq:kappa_eta}
\end{equation}
Rotating Eq.~\eqref{eq:MSA} to the mass basis gives
\begin{equation}
\begin{gathered}
U^T\mathcal S\,U=\begin{pmatrix}-\kappa_\chi+\eta_\chi c&-i\eta_\chi\sin2\vartheta\\
-i\eta_\chi\sin2\vartheta&\kappa_\chi+\eta_\chi c\end{pmatrix}, \\
U^T\mathcal A\,U=\begin{pmatrix}-i(\eta_\chi-\kappa_\chi c)&\kappa_\chi\sin2\vartheta\\
\kappa_\chi\sin2\vartheta&i(\eta_\chi+\kappa_\chi c)\end{pmatrix}.
\end{gathered}
\end{equation}
A real (imaginary) coefficient of $\eta_i\eta_j+{\rm h.c.}$ corresponds to a
scalar (pseudoscalar) bilinear of the Majorana fields. With
$\sin2\vartheta\simeq1$, the interactions are
\begin{align}
\mathcal L\supset{}&
-\frac{s}{2}\Big[g_{11}\,\overline{\chi_1}\chi_1+g_{22}\,\overline{\chi_2}\chi_2\Big]
\nonumber\\
&-\eta_\chi\,s\,\overline{\chi_1}\,i\gamma^5\chi_2-\kappa_\chi\,a\,\overline{\chi_1}\chi_2
\nonumber\\
&-\frac{a}{2}\Big[(\eta_\chi-\kappa_\chi c)\,\overline{\chi_1}\,i\gamma^5\chi_1
\nonumber\\
&\qquad-(\eta_\chi+\kappa_\chi c)\,\overline{\chi_2}\,i\gamma^5\chi_2\Big],
\label{eq:couplings}
\end{align}
up to convention-dependent signs, with
\begin{equation}
\begin{gathered}
g_{11}=-\kappa_\chi+\eta_\chi c, \\
g_{22}=\kappa_\chi+\eta_\chi c .
\end{gathered}
\label{eq:g11}
\end{equation}
The structure is summarized in Table~\ref{tab:couplings}.

\begin{table}[!htbp]
\centering
\setlength{\tabcolsep}{4pt}
\renewcommand{\arraystretch}{1.25}
\begin{tabular}{|c|c|c|}
\hline
\diagbox[height=2.7\baselineskip]{field}{coupling}
 & \begin{tabular}{@{}c@{}}scalar\\(central)\end{tabular}
 & \begin{tabular}{@{}c@{}}pseudoscalar\\(spin-dependent)\end{tabular} \\
\hline
\begin{tabular}{@{}c@{}}$s$\\(CP-even)\end{tabular}
 & \begin{tabular}{@{}c@{}}$\chi_1\chi_1,\ \chi_2\chi_2$:\\$\mp\kappa_\chi+\eta_\chi c$\end{tabular}
 & $\chi_1\chi_2$: $\eta_\chi$ \\
\hline
\begin{tabular}{@{}c@{}}$a$\\(CP-odd)\end{tabular}
 & $\chi_1\chi_2$: $\kappa_\chi$
 & \begin{tabular}{@{}c@{}}$\chi_1\chi_1,\ \chi_2\chi_2$:\\$\simeq\eta_\chi$\end{tabular} \\
\hline
\end{tabular}
\caption{Lorentz structure and strength of the singlet couplings to the
pseudo-Dirac pair.}
\label{tab:couplings}
\end{table}

Two independent arguments confirm this pattern. First, the diagonal couplings of the radial mode satisfy $g_{ii}=\partial m_{\chi_i}/\partial v_\phi$, as follows from Eq.~\eqref{eq:Lchi_matrix} (see Appendix~\ref{app:radial} for arbitrary phases). Second, in the limit $\mu',\kappa\to0$ the pseudo-Majoron is an exact Goldstone boson, whose non-derivative scalar coupling between two mass eigenstates must be proportional to their mass difference~\cite{Goldstone:Majorana:2002}; indeed $\kappa_\chi=\Delta m_\chi/(2v_\phi)$, and Appendix~\ref{app:majoron} shows that this relation holds for arbitrary complex Yukawa couplings and any soft breaking in the scalar potential. Physically, $\chi_1$ and $\chi_2$ carry opposite intrinsic CP parity, so the CP-even $s$ connects them only through $i\gamma^5$ and the CP-odd $a$ only through a scalar bilinear. 

The central couplings in Table~\ref{tab:couplings} are thus all of order $\kappa_\chi$, up to corrections of relative size $\eta_\chi c\sim y^2v_\phi/M_\chi$. In Appendix~\ref{app:takagi} we show that this holds for arbitrary complex Yukawa couplings:
\begin{equation}
|g^{\rm central}|=\frac{\Delta m_\chi}{2v_\phi}
+\mathcal O\!\left(\frac{y^2v_\phi}{M_\chi}\right).
\label{eq:central_bound}
\end{equation}
Consequently, a scalar whose vev generates the pseudo-Dirac splitting cannot mediate a strong central interaction between the dark states unless the splitting is itself large. Quantitatively, the self-interaction and the relic abundance require $|g^{\rm central}|\sim0.1$ (Secs.~\ref{sec:relic} and~\ref{sec:selfint}), and hence $\Delta m_\chi\gtrsim0.2\,v_\phi$. Since perturbativity of $\lambda_\phi$ requires $v_\phi\gtrsim m_s/\sqrt{8\pi}$ (Sec.~\ref{sec:scalar}), the splitting exceeds $0.04\,m_s\gtrsim0.2$~MeV for any viable mediator mass, and for $v_\phi$ in the window of a few hundred MeV found in Sec.~\ref{sec:vphi_window} it is tens of MeV, far above the kinetic energies of dark-matter pairs in halos, which remain below an MeV even at cluster velocities. The inelastic channel is then closed, and a strong self-interaction mediated by $s$ or $a$ is necessarily elastic. In particular, the antisymmetric regime $y_R\simeq-y_L$, in which $\Delta m_\chi$ is small, yields only pseudoscalar couplings of unsuppressed strength, and no strong inelastic self-interaction can arise from either $s$ or $a$.


We therefore work in the symmetric regime
\begin{equation}
\begin{gathered}
y_R=y_L+\delta y, \quad |\delta y|\ll|y_L|,
\end{gathered}
\label{eq:symmetric_regime}
\end{equation}
in which $\eta_\chi\simeq0$ and the CP-even singlet couples to the lighter
dark state through an unsuppressed scalar interaction,
\begin{equation}
\begin{gathered}
g_{11}\simeq-g_{22}\simeq-\frac{y_L}{\sqrt2}, \\
\alpha_{11}\equiv\frac{g_{11}^2}{4\pi}.
\end{gathered}
\label{eq:alpha11}
\end{equation}
The splitting is then fixed by the self-coupling and the lepton-number-breaking
scale. From Eq.~\eqref{eq:masses}, 
\[\Delta m_\chi=|y_L+y_R|\,v_\phi/\sqrt2
\simeq\sqrt2\,|y_L|\,v_\phi. \]
Using Eq.~\eqref{eq:alpha11} and $\mu_S=y_Sv_\phi/\sqrt2$
(Eq.~\eqref{eq:mD_muS}), we obtain
\begin{equation}
\begin{gathered}
\Delta m_\chi\simeq2\,|g_{11}|\,v_\phi\,, \\
\frac{\Delta m_\chi}{\mu_S}\simeq\frac{2\sqrt2\,|g_{11}|}{|y_S|}.
\end{gathered}
\label{eq:prediction}
\end{equation}
The pseudo-Dirac condition $|m_{L,R}|\ll M_\chi$ is automatically satisfied for MeV-scale $v_\phi$ and GeV-scale DM. The small pseudoscalar coupling $\eta_\chi\propto\delta y$ allows the heavier state to decay via $\chi_2\to\chi_1 s$ whenever this channel is open. Since
$m_s\simeq\sqrt{2\lambda_\phi}\,v_\phi$, the condition $\Delta m_\chi>m_s$ reads
\begin{equation}
\lambda_\phi<2g_{11}^2 .
\label{eq:chi2_decay}
\end{equation}

The dark-matter phenomenology is therefore controlled by four parameters: $m_{\chi_1}$, $m_s$, $\alpha_{11}$ and $v_\phi$. The first three determine the self-interaction cross section and the relic abundance. The vev $v_\phi$ sets the mass splitting $\Delta m_\chi$ and the lifetime of $s$ and, for a given portal coupling $\lambda_{H\phi}$, it also fixes the Higgs mixing through Eq.~\eqref{eq:mixing}. The remaining parameters, $m_a$ and the neutrino-sector couplings, are constrained by the light-neutrino masses and by the requirement that the pseudo-Majoron be heavy. 

\section{Relic abundance}
\label{sec:relic}

\subsection{Annihilation channels}
\label{sec:channels}

\begin{figure}[!htbp]
\centering
\includegraphics[width=0.85\columnwidth]{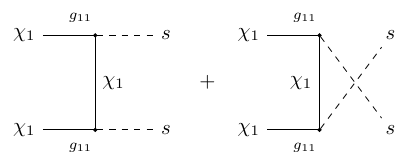}
\caption{Annihilation $\chi_1\chi_1\to ss$ through $t$- and $u$-channel exchange of $\chi_1$.}
\label{fig:diag_ann}
\end{figure}

At freeze-out, the mass splitting is far smaller than the temperature, $\Delta m_\chi\ll T_f\simeq m_{\chi_1}/20$, so $\chi_1$ and $\chi_2$ are
both present in comparable numbers and annihilate together. The allowed channels follow from the coupling structure of Table~\ref{tab:couplings}. In the symmetric regime, the CP-even singlet couples only diagonally and the pseudo-Majoron only off-diagonally, while the pseudo-Majoron is heavier than the pair, $m_a>m_{\chi_1}+m_{\chi_2}$ (Sec.~\ref{sec:scalar}). As a result:
\begin{itemize}
\item $\chi_1\chi_1\to ss$ and $\chi_2\chi_2\to ss$ proceed through
$t$- and $u$-channel exchange of $\chi_1$ and $\chi_2$, respectively
(Fig.~\ref{fig:diag_ann}), with the
scalar couplings $g_{11}$ and $g_{22}=-g_{11}$;
\item $\chi_1\chi_2\to ss$ requires the pseudoscalar transition coupling
$\eta_\chi\propto\delta y$ and is negligible;
\item all channels with a pseudo-Majoron in the final state are kinematically
closed.
\end{itemize}
Since an $s$-wave pair of identical Majorana fermions ($^1S_0$) has $CP=-1$, while two CP-even scalars have $CP=+1$, annihilation into $ss$ is $p$-wave.
The $t$- and $u$-channel exchange of $\chi_i$ gives, after averaging over initial spins and including the factor $1/2$ for identical scalars in the
final state,
\begin{equation}
\sigma(\chi_i\chi_i\to ss)=\frac{g_{11}^4}{32\pi s\,(s-4m^2)}
\left[\frac{N\,L}{s-2m_s^2}-K\,R\right],
\label{eq:sigma_exact}
\end{equation}
where $s$ here denotes the squared center-of-mass energy, $m\equiv m_{\chi_1}$ (the splitting is negligible), and
\begin{equation}
\begin{gathered}
N=s^2+16m^2s-32m^4-4m_s^2(s+4m^2)+6m_s^4,\\
K=2+\frac{(4m^2-m_s^2)^2}{m^2(s-4m_s^2)+m_s^4},\\
R=\sqrt{(s-4m^2)(s-4m_s^2)},\qquad
L=\ln\frac{s-2m_s^2+R}{s-2m_s^2-R}.
\end{gathered}
\label{eq:sigma_exact_aux}
\end{equation}
In the non-relativistic limit Eq.~\eqref{eq:sigma_exact} reduces to
\begin{equation}
\sigma v_{\rm rel}(\chi_i\chi_i\to ss)\simeq\frac{3\pi\alpha_{11}^2}{8m_{\chi_1}^2}\,v_{\rm rel}^2,
\label{eq:sigmav_pwave}
\end{equation}
where $v_{\rm rel}$ is the relative velocity in the center-of-mass frame.

Relativistic corrections reduce Eq.~\eqref{eq:sigma_exact} below Eq.~\eqref{eq:sigmav_pwave} by $3\%$ at $v_{\rm rel}=0.3$ and $13\%$ at $v_{\rm rel}=0.6$; hence we use Eq.~\eqref{eq:sigma_exact} throughout. The $s$-channel diagram with an intermediate $s$ and the cubic scalar coupling is suppressed by $m_s^2/(v_\phi m_{\chi_1})$ relative to the $t$- and $u$-channel contributions and is negligible. The attractive Yukawa potential between the annihilating particles, Eq.~\eqref{eq:yukawa}, enhances the annihilation (Sommerfeld effect)~\cite{Hisano:2003ec,ArkaniHamed:2008qn}. Since the range of the potential is much larger than the de~Broglie wavelength during freeze-out, $m_s\ll m_{\chi_1}v_{\rm rel}$, we use the Coulomb limit of the $p$-wave enhancement factor~\cite{Iengo:2009ni,Cassel:2009wt},
\begin{equation}
S_1=\big(1+\eta^2\big)\,\frac{2\pi\eta}{1-e^{-2\pi\eta}},\qquad
\eta=\frac{\alpha_{11}}{v_{\rm rel}},
\label{eq:sommerfeld}
\end{equation}
by which we multiply Eq.~\eqref{eq:sigma_exact}. For $\alpha_{11}\sim10^{-3}$-$10^{-2}$ and $v_{\rm rel}\simeq0.3$-$0.6$ at freeze-out the enhancement is modest: it lowers the required coupling by $0.7\%$ at the best-fit point BP-A (Sec.~\ref{sec:selfint_results}), by $2.5\%$ at $m_{\chi_1}=30$~GeV and by $7\%$ at $100$~GeV.


\subsection{Boltzmann equation ($T_{D} = T$)}
\label{sec:boltzmann}

We follow the evolution of the total comoving abundance $Y=(n_{\chi_1}+n_{\chi_2})/s_{\rm tot}$, where $s_{\rm tot}$ is the total entropy density. For coannihilating species with rapid conversions between them, $Y$ obeys~\cite{Griest:1990kh,Edsjo:1997bg}
\begin{equation}
\begin{gathered}
\frac{dY}{dx}=-\frac{s_{\rm tot}\,\langle\sigma_{\rm eff}v\rangle}{x\,H}
\left(1+\frac13\frac{d\ln g_{*s}}{d\ln T}\right)
\left(Y^2-Y_{\rm eq}^2\right), \\
x=\frac{m_{\chi_1}}{T},
\end{gathered}
\label{eq:boltzmann}
\end{equation}
with the effective cross section
\begin{equation}
\begin{gathered}
\langle\sigma_{\rm eff}v\rangle=\sum_{i,j=1,2}\langle\sigma_{ij}v\rangle\,r_ir_j, \\
r_i=\frac{g_i(1+\Delta_i)^{3/2}e^{-x\Delta_i}}{\sum_kg_k(1+\Delta_k)^{3/2}e^{-x\Delta_k}},
\end{gathered}
\label{eq:sigma_eff}
\end{equation}
where $g_i=2$ and $\Delta_i=(m_{\chi_i}-m_{\chi_1})/m_{\chi_1}$. For each
channel the thermal average is computed with the relativistic expression of
Gondolo and Gelmini~\cite{Gondolo:1990dk},
\begin{align}
\langle\sigma v\rangle={}&\frac{1}{8m_{\chi_1}^4\,T\,K_2^2(x)}
\int_{4m_{\chi_1}^2}^\infty\sigma(s)\,\big(s-4m_{\chi_1}^2\big)
\nonumber\\
&\times\sqrt s\,K_1\!\left(\frac{\sqrt s}{T}\right)ds ,
\label{eq:gondolo}
\end{align}
with $\sigma(s)$ from Eq.~\eqref{eq:sigma_exact} multiplied by Eq.~\eqref{eq:sommerfeld}. 
Without the Sommerfeld factor, the non-relativistic estimate obtained from Eq.~\eqref{eq:sigmav_pwave} exceeds Eq.~\eqref{eq:gondolo} by $32\%$ at the typical freeze-out temperature $x\simeq20$, and the two agree to within $3\%$ for $x>200$. The SM contributions to the energy and entropy densities are taken from lattice-based determinations that include the QCD crossover~\cite{Saikawa:2018rcs}. We solve Eq.~\eqref{eq:boltzmann} from $x=3$, where both states are in equilibrium, to $x=2000$. The asymptotic yield $Y_\infty$ gives the relic density
\begin{equation}
\Omega_\chi h^2=2.744\times10^8\,(m_{\chi_1}/{\rm GeV})\,Y_\infty .
\label{eq:omega}
\end{equation}

\subsection{Thermal history of the dark sector ($T_{D} \neq T$)}
\label{sec:darktemp}

Equation~\eqref{eq:boltzmann} assumes a common temperature for the visible and dark sectors. The dark sector at freeze-out consists of $\chi_{1,2}$ and $s$; the pseudo-Majoron is heavier and plays no role at freeze-out. The dark sector is brought into equilibrium with the SM by the Higgs portal, $\lambda_{H\phi}(H^\dagger H)|\phi|^2$. Dark-sector particles are produced mainly in the decays $h\to ss$ and, for $m_a<m_h/2$, $h\to aa$, with $\Gamma(h\to ss)=\lambda_{H\phi}^2v_h^2/(32\pi m_h)$. Relative to the Hubble rate, this production is most efficient at $T\simeq m_h/3$, before the Higgs population becomes Boltzmann suppressed. Requiring the production rate to exceed the Hubble rate times the equilibrium density of the dark sector, whose effective number of degrees of freedom is five ($s$, $a$ and the two Majorana fermions), gives
\begin{equation}
\lambda_{H\phi}\gtrsim\lambda_{\rm th}\simeq6\times10^{-8},
\label{eq:lambda_th}
\end{equation}
which rises to $8\times10^{-8}$ if $h\to aa$ is closed (Appendix~\ref{app:thermal}). Scattering processes $HH^\dagger\to\phi\phi^*$ above the electroweak scale lower the threshold somewhat, so that Eq.~\eqref{eq:lambda_th} is conservative. Once the Higgs bosons disappear from the plasma, the portal can no longer maintain equilibrium: for $\lambda_{H\phi}\lesssim10^{-6}$ the dark sector decouples at temperatures of about $10$-$30$~GeV, and the decays $s\to\nu\nu$ are far slower than the expansion rate at GeV temperatures. Freeze-out therefore generically takes place in a dark sector with its own temperature $T_D$~\cite{Feng:2008mu,Bernal:2015ova}. After decoupling at $T_{\rm dec}$, the two sectors conserve their entropies separately, and
\begin{equation}
\frac{g_{D,s}(T_D)\,T_D^3}{g_{*s}(T)\,T^3}=\frac{g_{D,s}(T_{\rm dec})}{g_{*s}(T_{\rm dec})},
\label{eq:xi}
\end{equation}
where $g_{D,s}$ counts the entropy degrees of freedom of $\chi_{1,2}$ and $s$ at temperature $T_D$. With a separate dark-sector temperature, Eq.~\eqref{eq:boltzmann} is modified as
\begin{equation}
\frac{dY}{dx}=-\frac{s_{\rm tot}\,\langle\sigma_{\rm eff}v\rangle(T_D)}{x\,H}
\left(1+\frac13\frac{d\ln g_{*s}}{d\ln T}\right)\left(Y^2-Y_{\rm eq}^2\right),
\label{eq:boltzmann_TD}
\end{equation}
with $x=m_{\chi_1}/T$, where $T$ is the SM temperature, and $Y=(n_{\chi_1}+n_{\chi_2})/s_{\rm tot}$. The equilibrium density and the effective cross section are evaluated at $T_D$, the Hubble rate is determined by the energy density of both sectors, and $s_{\rm tot}$ is the sum of the
separately conserved entropies:
\begin{equation}
\begin{gathered}
s_{\rm tot}=\frac{2\pi^2}{45}\,g_{*s}(T)\,T^3\left[1+\frac{g_{D,s}(T_{\rm dec})}{g_{*s}(T_{\rm dec})}\right],\\
H=\left[\frac{8\pi}{3M_{\rm Pl}^2}\,\frac{\pi^2}{30}\Big(g_*(T)\,T^4+g_{D,\rho}(T_D)\,T_D^4\Big)\right]^{1/2},\\
Y_{\rm eq}=\frac{n_{\rm eq}(T_D)}{s_{\rm tot}}, {\rm with}\\
n_{\rm eq}=\frac{m_{\chi_1}^2T_D}{\pi^2}\sum_{i=1,2}(1+\Delta_i)^2K_2\big((1+\Delta_i)x_D\big),\\
\langle\sigma_{\rm eff}v\rangle(T_D)=\langle\sigma v\rangle_{11}(T_D)\,
\frac{w_1^2+w_2^2}{(w_1+w_2)^2}, {\rm with}\\
w_i=(1+\Delta_i)^{3/2}e^{-x_D\Delta_i},
\end{gathered}
\end{equation}
where $x_D=m_{\chi_1}/T_D$, $\Delta_1=0$, $\Delta_2=\Delta m_\chi/m_{\chi_1}$, and
$T_D(T)$ follows from Eq.~\eqref{eq:xi}. We adopt the decoupled dark sector as our default, since the portal cannot maintain equilibrium once the Higgs bosons have disappeared from the plasma. Unless stated otherwise, all results in this work are obtained from Eq.~\eqref{eq:boltzmann_TD} with $T_{\rm dec}=20$~GeV; the common-temperature Eq.~\eqref{eq:boltzmann} is used only for comparison, to assess the dependence on the thermal history of the dark sector.

As $\chi_{1,2}$ become non-relativistic, their entropy is transferred to $s$ and the dark sector becomes hotter than the SM. For $m_{\chi_1}\gtrsim5$~GeV we find $T_D/T\simeq1.3$-$1.5$ at freeze-out, and correspondingly a larger coupling is needed to obtain the observed abundance. Since the dark sector decouples at about $10$-$30$~GeV for portal couplings between $\lambda_{\rm th}$ of Eq.~\eqref{eq:lambda_th} and $10^{-6}$ (Appendix~\ref{app:thermal}), we take $T_{\rm dec}=20$~GeV as representative; the larger couplings still allowed by direct detection at low DM masses give somewhat lower values. Varying $T_{\rm dec}$ between $10$ and $40$~GeV changes the required $\alpha_{11}$ by at most $3.5\%$ for $m_{\chi_1}\lesssim20$~GeV. At $m_{\chi_1}=60$~GeV, where DM is already partly non-relativistic at decoupling, the change grows to $-13\%$ for $T_{\rm dec}=10$~GeV and $+7\%$ for $T_{\rm dec}=40$~GeV.

\subsection{Self-coupling from the relic density}
\label{sec:relic_results}

\begin{figure}[h]
\centering
\includegraphics[width=\columnwidth]{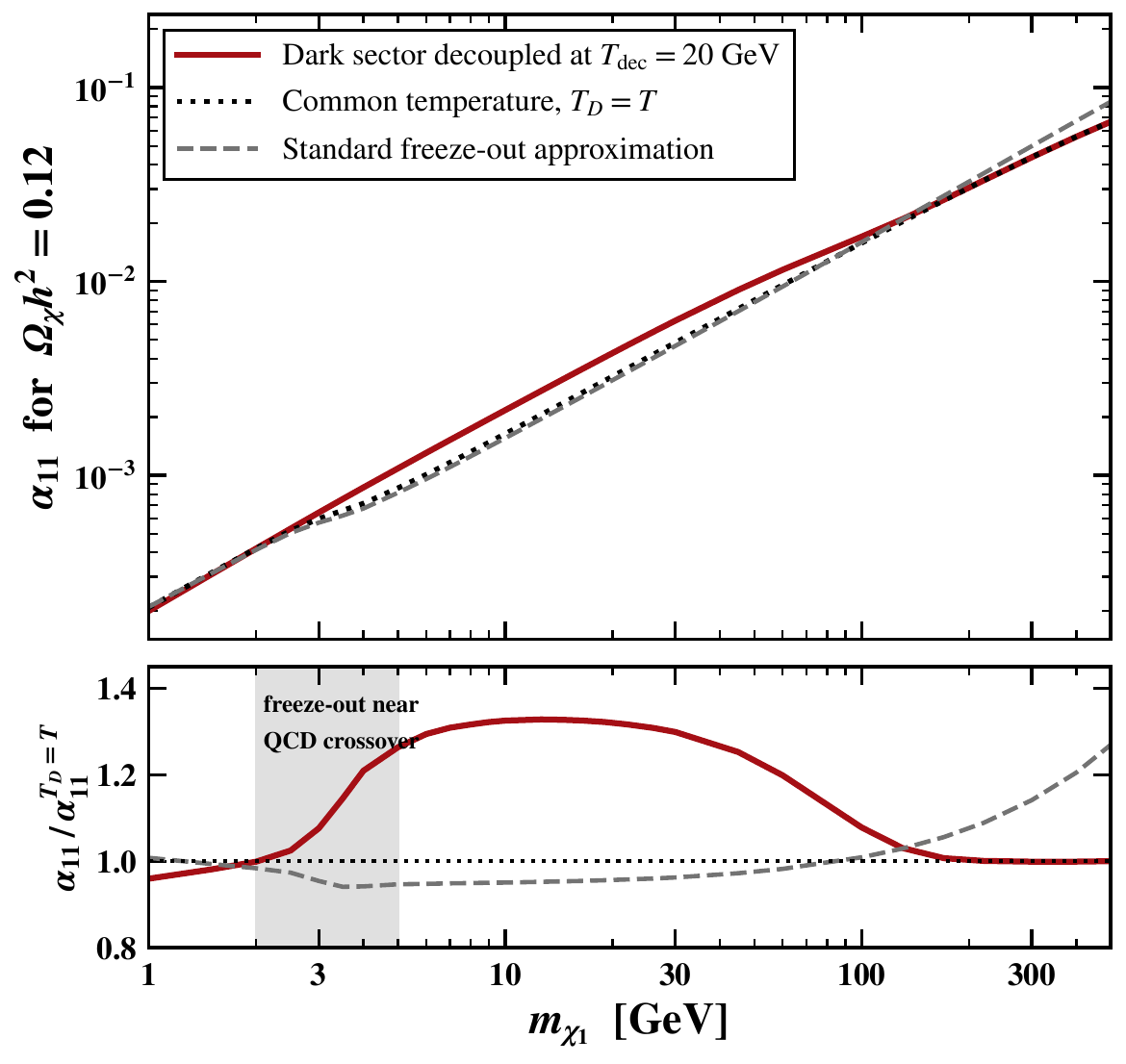}
\caption{Upper panel: Coupling $\alpha_{11}$ required for $\Omega_\chi h^2=0.12$: decoupled dark
sector (solid), and common temperature $T_D=T$ from the full Boltzmann
equation (dotted) and from the standard freeze-out
approximation~\cite{Kolb:1990vq} (dashed). Lower panel: ratio to the dotted curve. Shaded:
freeze-out during the QCD crossover.}
\label{fig:alpha_relic}
\end{figure}

Figure~\ref{fig:alpha_relic} shows the coupling $\alpha_{11}$ required for $\Omega_\chi h^2=0.12$~\cite{Planck:2018vyg} as a function of the DM mass, for a common temperature and for a dark sector that decouples at $T_{\rm dec}=20$~GeV. The required coupling is independent of the mediator mass to better than $0.1\%$ for $m_s=5$-$30$~MeV. Through the splitting $\Delta m_\chi=2g_{11}v_\phi$ it depends weakly on $v_\phi$; we evaluate it at $v_\phi=200$~MeV, inside the windows derived in Sec.~\ref{sec:vphi_window}. Across the windows of the benchmark points, $v_\phi\simeq130$-$390$~MeV, the required coupling changes by $+1\%$ to $-3\%$ at BP-A and by at most $+2\%$ to $-5\%$ at the lower end of the allowed mass range, $m_{\chi_1}=3$~GeV, where the effect is largest. The relic abundance thus defines an essentially universal relation $\alpha_{11}(m_{\chi_1})$ that we impose in the self-interaction analysis. Representative values are listed in Table~\ref{tab:alpha_relic}. Compared with a common temperature, a decoupled dark sector requires a coupling larger by $20$-$33\%$ for $m_{\chi_1}=5$-$60$~GeV; for $m_{\chi_1}\gtrsim200$~GeV freeze-out precedes the decoupling, and the two results coincide. The standard approximation of instantaneous freeze-out~\cite{Kolb:1990vq}, evaluated for a common temperature, underestimates the coupling by about $5\%$ for $m_{\chi_1}\lesssim30$~GeV; at higher masses the Sommerfeld enhancement, which it does not include, reverses the sign of the difference. The feature at $m_{\chi_1}\simeq2$-$5$~GeV arises because freeze-out coincides with the QCD crossover, where $g_*$ changes rapidly.

\begin{table}[h]
\centering
\renewcommand{\arraystretch}{1.2}
\begin{tabular}{|c|c|c|c|}
\hline
$m_{\chi_1}$ [GeV] & $T_D=T$ & $T_{\rm dec}=20$~GeV & difference \\
\hline
1 & $2.08\times10^{-4}$ & $2.00\times10^{-4}$ & $-4\%$ \\
3 & $5.98\times10^{-4}$ & $6.43\times10^{-4}$ & $+8\%$ \\
10 & $1.64\times10^{-3}$ & $2.17\times10^{-3}$ & $+33\%$ \\
16.5 & $2.68\times10^{-3}$ & $3.55\times10^{-3}$ & $+33\%$ \\
30 & $4.83\times10^{-3}$ & $6.28\times10^{-3}$ & $+30\%$ \\
100 & $1.58\times10^{-2}$ & $1.70\times10^{-2}$ & $+8\%$ \\
\hline
\end{tabular}
\caption{Coupling $\alpha_{11}$ required for $\Omega_\chi h^2=0.12$.}
\label{tab:alpha_relic}
\end{table}

\begin{figure}[h]
\centering
\includegraphics[width=\columnwidth]{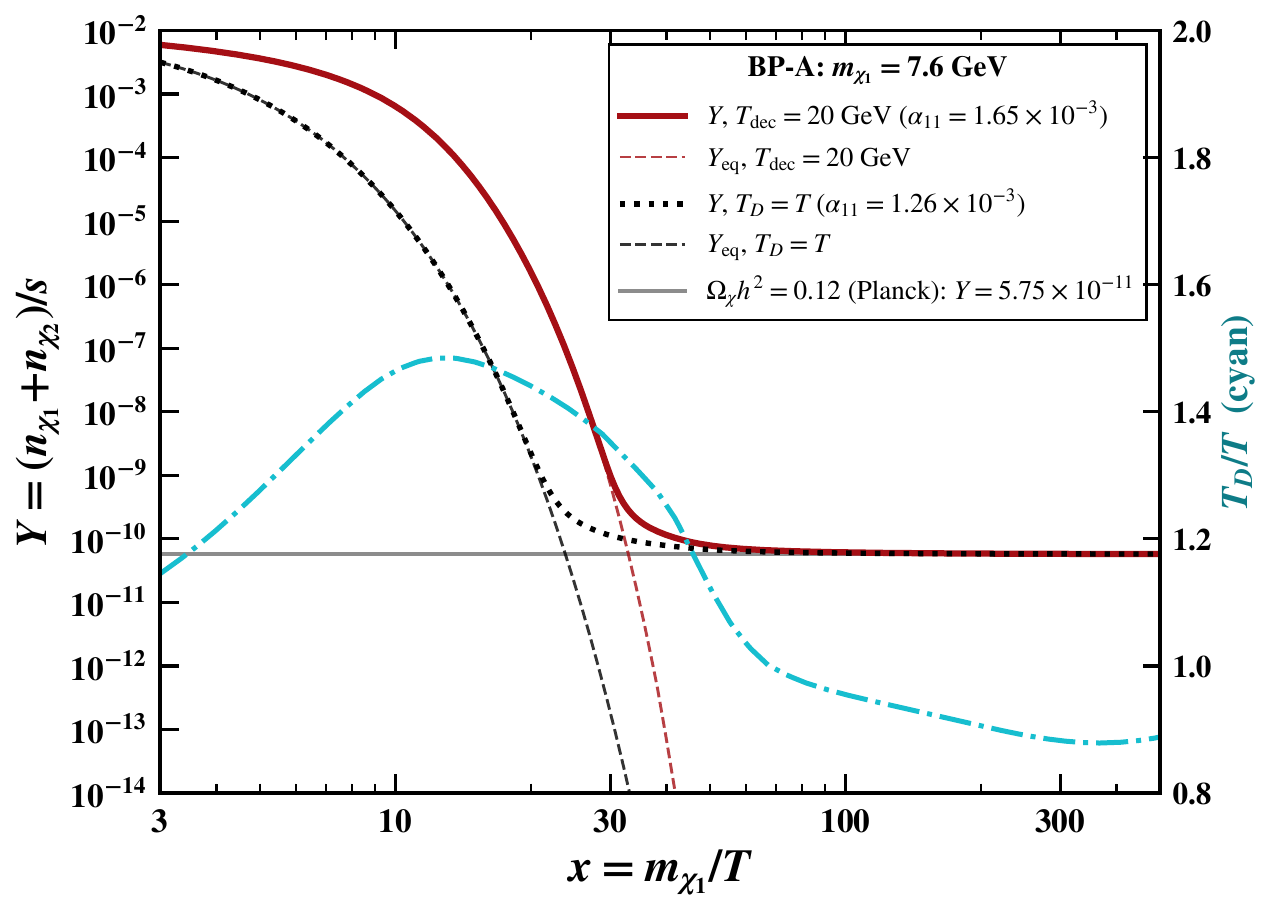}
\caption{Comoving abundance $Y$ (thick) and its equilibrium value (thin dashed) for BP-A, $m_{\chi_1}=7.6$~GeV, for a decoupled dark sector (solid) and a common temperature (dotted). Gray line: $Y$ giving $\Omega_\chi h^2=0.12$. Dash-dotted: $T_D/T$ (right axis).}
\label{fig:Y_evolution}
\end{figure}

As an example, Fig.~\ref{fig:Y_evolution} shows the evolution of the abundance and of $T_D/T$ for the best-fit point BP-A (Sec.~\ref{sec:selfint_results}), $m_{\chi_1}=7.6$~GeV. With the coupling fixed by $\Omega_\chi h^2=0.12$, both thermal histories reach the same final abundance. The ratio $T_D/T$ rises during freeze-out, as $\chi$ annihilations heat the decoupled dark sector, and falls at the QCD crossover, when the SM bath is heated in turn.

\section{Dark matter self-interactions}
\label{sec:selfint}

\subsection{Scattering problem}
\label{sec:scattering}

\begin{figure}[h]
\centering
\includegraphics[width=0.75\columnwidth]{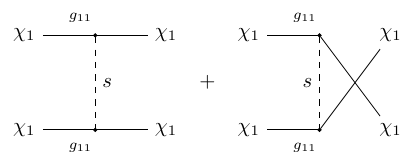}
\caption{Self-scattering of $\chi_1$ through $s$ exchange; the crossed diagram arises from the identical Majorana fermions.}
\label{fig:diag_scatter}
\end{figure}

In the symmetric regime, two $\chi_1$ particles interact through the exchange of the CP-even singlet with the scalar coupling $g_{11}$, which produces the attractive Yukawa potential (Fig.~\ref{fig:diag_scatter})
\begin{equation}
V(r)=-\alpha_{11}\,\frac{e^{-m_sr}}{r},
\label{eq:yukawa}
\end{equation}
with $\alpha_{11}$ of Eq.~\eqref{eq:alpha11}.
Exchange of the pseudo-Majoron connects $\chi_1\chi_1$ with $\chi_2\chi_2$, which is kinematically closed in halos for the MeV-scale splittings of Eq.~\eqref{eq:prediction}; its second-order effect on elastic scattering is suppressed by $\alpha_{11}m_s/\Delta m_\chi\lesssim10^{-2}$. Pseudoscalar couplings produce only spin-dependent interactions suppressed by $(q/m_{\chi_1})^2$~\cite{Kahlhoefer:2017umn}. Self-scattering is therefore governed by Eq.~\eqref{eq:yukawa} alone, and it depends on three parameters, $m_{\chi_1}$, $m_s$ and $\alpha_{11}$.

Defining $x=m_sr$, the radial Schr\"odinger equation for the partial wave
$\ell$ reads
\begin{equation}
\begin{gathered}
u_\ell''(x)+\left[\hat k^2-\frac{\ell(\ell+1)}{x^2}+b\,\frac{e^{-x}}{x}\right]u_\ell(x)=0, \\
\hat k=\frac{m_{\chi_1}v_{\rm rel}}{2m_s},\quad b=\frac{\alpha_{11}m_{\chi_1}}{m_s}.
\end{gathered}
\label{eq:radial}
\end{equation}
For the parameters of interest, $b$ ranges from $\mathcal O(0.1)$ to $\mathcal O(100)$, and $\hat k$ from $10^{-3}$ at dwarf-galaxy velocities to $\mathcal O(100)$ at cluster velocities. The scattering therefore spans the Born, resonant and semi-classical regimes. The Born regime corresponds to $b\lesssim1$; for $b\gtrsim1$ the scattering is resonant for $\hat k\lesssim1$ and semi-classical for $\hat k\gg1$, so that the same point in the $(m_{\chi_1},m_s)$ plane can be resonant at dwarf-galaxy velocities and semi-classical at cluster velocities. We solve Eq.~\eqref{eq:radial} exactly and integrate it outward for all partial waves, starting from the regular solution $u_\ell\propto x^{\ell+1}$, and obtain the phase shifts by matching at large $x$, where the potential is negligible, to
\begin{equation}
u_\ell(x)\propto\hat\jmath_\ell(\hat k x)\cos\delta_\ell-\hat n_\ell(\hat k x)\sin\delta_\ell ,
\label{eq:matching}
\end{equation}
where $\hat\jmath_\ell(z)=zj_\ell(z)$ and $\hat n_\ell(z)=zy_\ell(z)$ are
Riccati-Bessel functions. As an independent check we use the variable phase
method~\cite{Calogero:1967},
\begin{equation}
\begin{gathered}
\delta_\ell'(x)=\frac{b\,e^{-x}}{\hat k\,x}
\Big[\hat\jmath_\ell(\hat k x)\cos\delta_\ell(x)-\hat n_\ell(\hat k x)\sin\delta_\ell(x)\Big]^2, \\
\delta_\ell(0)=0,
\end{gathered}
\label{eq:vpm}
\end{equation}
which gives the same phase shifts wherever it can be integrated reliably. At large $b$ and low velocities, however, the phase function of Eq.~\eqref{eq:vpm} increases by $\pi$ over very short intervals in $x$, which makes the equation stiff and its numerical solution with a fixed step unreliable, whereas the direct integration of Eq.~\eqref{eq:radial} is not affected. For high partial waves, once the phase shifts are small, we use the analytic Born result $\delta_\ell=(b/2\hat k)\,Q_\ell\big(1+1/(2\hat k^2)\big)$, where $Q_\ell$ is the Legendre function of the second kind. The numerical implementation and its validation are described in Appendix~\ref{app:partialwave}. 

Because $\chi_1$ is a Majorana fermion, the two-body states are antisymmetric: even partial waves occur in the spin singlet and odd partial waves in the spin triplet. For unpolarized scattering, $d\sigma/d\Omega=|f_{\rm e}(\theta)|^2+3|f_{\rm o}(\theta)|^2$, where $f_{\rm e,o}$ are the even and odd parts of the scattering amplitude. As the figure of merit for halo phenomenology we use the viscosity cross section~\cite{Tulin:2013teo,Colquhoun:2020adl}, normalized such that it equals the total cross section for isotropic scattering,
\begin{align}
\sigma_V\equiv{}&\frac34\int d\Omega\,\sin^2\theta\,\frac{d\sigma}{d\Omega}
\nonumber\\
={}&\frac{3\pi}{k^2}\sum_{\ell}w_\ell\,\frac{(\ell+1)(\ell+2)}{2\ell+3}\,
\sin^2\!\big(\delta_{\ell+2}-\delta_\ell\big),
\label{eq:sigmaV}
\end{align}
where $w_\ell=1$ for even and $w_\ell=3$ for odd $\ell$, and $k=m_{\chi_1}v_{\rm rel}/2$. The factor $1/2$ that avoids double counting of identical final states is included in the prefactor.

The exact treatment matters quantitatively. Widely used analytic approximations~\cite{Tulin:2013teo} describe the transfer cross section of distinguishable particles in the Born, resonant and semi-classical regimes. Two effects separate them from Eq.~\eqref{eq:sigmaV}. First, the approximations themselves deviate from the exact transfer cross section by $20$-$40\%$ in the non-perturbative regime; for example, for $m_{\chi_1}=17.2$~GeV, $m_s=13.7$~MeV and $\alpha_{11}=2.5\times10^{-3}$ they underestimate it by about $20\%$ at dwarf-galaxy velocities and overestimate it by about $20\%$ at cluster velocities, after velocity averaging. Second, identical-particle statistics reduce the cross section in the $s$-wave regime by a factor of two, since only the spin singlet contributes, while at cluster velocities the viscosity weighting and the many contributing partial waves raise the ratio to the distinguishable transfer cross section above unity. At dwarf-galaxy velocities the two effects combine to reduce the cross section by about $40\%$ relative to the approximate formulae. Because the coupling is fixed by the relic abundance rather than fitted, such differences translate directly into shifts of the preferred masses.

\subsection{Fit to astrophysical data}
\label{sec:halo_data}

We compare the model with the self-interaction cross sections inferred in Ref.~\cite{Kaplinghat:2015aga} from the rotation curves of five dwarf galaxies of the THINGS survey~\cite{Oh:2010ea} and seven low-surface-brightness galaxies~\cite{KuziodeNaray:2007qi}, and from the density profiles of six relaxed galaxy clusters~\cite{Newman:2012nw,Newman:2012nv}. All of these were obtained by fitting SIDM halo models to the observed rotation curves and density profiles. 
These determinations are expressed as the velocity-weighted average
$\langle\sigma v\rangle/\langle v\rangle$ over a Maxwell-Boltzmann
distribution of relative velocities, $f(v)\propto v^2e^{-v^2/v_0^2}$ with
$\langle v\rangle=2v_0/\sqrt\pi$, which we compute from Eq.~\eqref{eq:sigmaV}.
We define
\begin{equation}
\chi^2=\sum_{i=1}^{N}\frac{\big[\log_{10}(\sigma/m)_{i}^{\rm th}-\log_{10}(\sigma/m)_{i}^{\rm inf}\big]^2}{\Delta_i^2},
\label{eq:chi2}
\end{equation}
where $\Delta_i$ is the upper or lower logarithmic uncertainty of data point $i$, depending on the sign of the deviation, and $N=18$. The coupling is not a free parameter: at every point we set $\alpha_{11}=\alpha_{11}^{\rm relic}(m_{\chi_1})$ from Sec.~\ref{sec:relic}, using the decoupled dark sector as the default and the common-temperature result to assess the dependence on the thermal history. The fit therefore involves only the two masses $m_{\chi_1}$ and $m_s$, and we quote the goodness of fit as the reduced chi-square $\chi^2/{\rm dof}$ with ${\rm dof}=N-2=16$. Confidence regions in the $(m_{\chi_1},m_s)$ plane are defined by $\Delta\chi^2=\chi^2-\chi^2_{\rm min}<2.30$, $6.18$ and $11.8$ for $68\%$, $95\%$ and $99.7\%$ confidence level (CL). 

The data scatter by more than their quoted uncertainties: neighboring systems at similar velocities differ in $\sigma/m$ by up to an order of magnitude, reflecting halo-to-halo diversity and systematic uncertainties of the inferred cross sections. This sets a floor to the attainable fit quality that is independent of any particle-physics model. Fitting purely phenomenological curves to the same data, we find $\chi^2/{\rm dof}=1.92$ for a power law $\sigma/m\propto v^{-n}$ (two parameters, best-fit $n\simeq1$) and $\chi^2/{\rm dof}=1.94$ for the more flexible form $\sigma_0/[1+(v/w)^2]^{n}$ (three parameters). A velocity-independent cross section gives $\chi^2/{\rm dof}=5.8$ and is excluded by $\Delta\chi^2\simeq68$ relative to a power law, which quantifies the preference of the data for a velocity-dependent cross section. Values of $\chi^2/{\rm dof}$ close to $1.9$ therefore represent fits as good as the data allow.

\subsection{Allowed parameter space}
\label{sec:selfint_results}

\begin{figure}[h]
\centering
\includegraphics[width=\columnwidth]{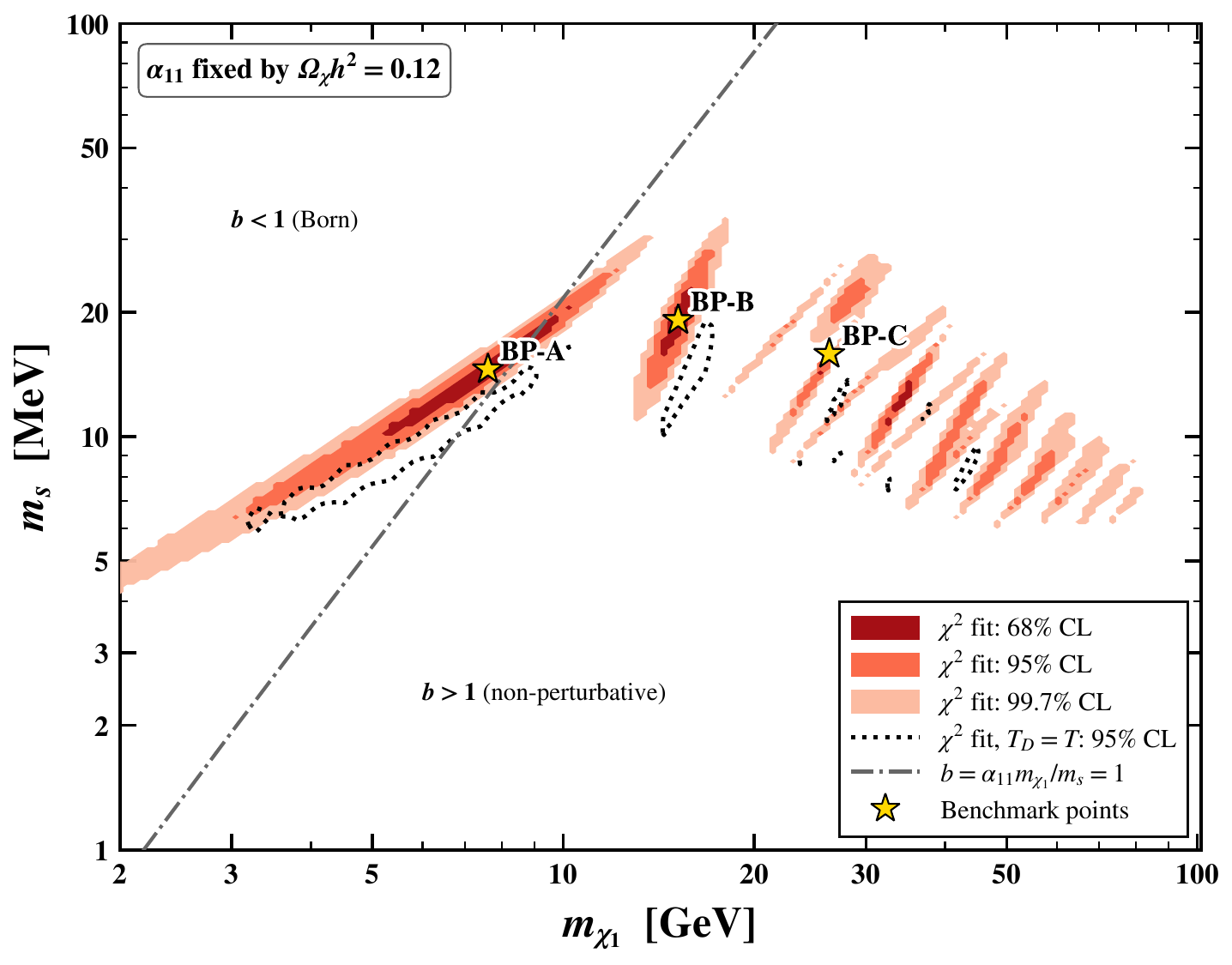}
\caption{Relic-constrained fit to the self-interaction data in the
$(m_{\chi_1},m_s)$ plane for a decoupled dark sector, with $\alpha_{11}$ fixed by $\Omega_\chi h^2=0.12$ at every point. Dotted: $95\%$ CL region for a common temperature, $T_D=T$. Dash-dotted: $b=1$, which separates the Born ($b<1$) and non-perturbative ($b>1$) regimes. For drawing, $\chi^2$ is interpolated between
the computed mass rows at fixed $b$.}
\label{fig:selfint_map}
\end{figure}

\paragraph{Narrow band at lower mass:}
Figure~\ref{fig:selfint_map} shows the result of the relic-constrained fit in the $(m_{\chi_1},m_s)$ plane, and Table~\ref{tab:selfint_profile} lists the best mediator mass for representative DM masses. 
The best fit,
\begin{equation}
\begin{gathered}
m_{\chi_1}\simeq7.6~{\rm GeV},\quad
m_s\simeq14.6~{\rm MeV},\quad
\alpha_{11}\simeq1.7\times10^{-3}, \\
{\rm with}  \quad
\chi^2/{\rm dof}=2.04
\end{gathered}
\label{eq:bestfit}
\end{equation}
lies within $\Delta\chi^2\simeq1.9$ of the power-law fit and within $\Delta\chi^2\simeq3.5$ of the flexible three-parameter curve, although its coupling is fixed by cosmology rather than fitted. At low masses the preferred region is a narrow band, $m_s\simeq1.9~{\rm MeV}\times(m_{\chi_1}/{\rm GeV})$, with $b\simeq0.7$-$1.1$ for $m_{\chi_1}=6$-$10$~GeV. Along the band the fit quality is nearly flat, while a change of $m_s$ by $20\%$ at fixed DM mass raises $\chi^2$ by $15$-$30$ units. Toward lighter DM, in the Born regime ($b<1$), the fit degrades: with the relic coupling, the cross section can no longer reproduce both the normalization at dwarf velocities and the suppression at cluster velocities.

\paragraph{Resonant islands:}
Above $m_{\chi_1}\simeq12$~GeV the relic coupling places the system in the non-perturbative regime, where the scattering is dominated by quasi-bound states of the Yukawa potential; an $s$-wave bound state appears at each zero-energy resonance, $b=1.68$, $6.45$, $14.3$, and so on~\cite{Tulin:2013teo}. The main band crosses the first resonance between $m_{\chi_1}\simeq12$ and $14$~GeV, where the fit degrades, and at higher masses the preferred region breaks up into narrow islands along lines of nearly constant $b$, since the positions of the resonances depend on $b$ alone. We classify the islands by the number $n$ of bound states in Table~\ref{tab:selfint_branches}; the individual minima are listed in Table~\ref{tab:selfint_branches_all} of Appendix~\ref{app:partialwave}.  They reproduce the data as well as the main band: the best point of each family lies within $\Delta\chi^2\simeq0.3$-$0.6$ of the best fit. 
\begin{table}[h]
\centering
\renewcommand{\arraystretch}{1.2}
\begin{tabular}{|c|c|c|c|c|c|}
\hline
$m_{\chi_1}$ [GeV] & $m_s$ [MeV] & $\alpha_{11}$ & $b$ & $\chi^2/{\rm dof}$ & $\Delta\chi^2$ \\
\hline
3.0  & 6.2  & $6.45\times10^{-4}$ & 0.31 & 2.40 & 5.8 \\
3.8  & 7.7  & $8.32\times10^{-4}$ & 0.41 & 2.27 & 3.7 \\
5.8  & 10.9 & $1.26\times10^{-3}$ & 0.67 & 2.10 & 1.0 \\
8.0  & 15.5 & $1.75\times10^{-3}$ & 0.91 & 2.04 & 0.0 \\
10.3 & 21.0 & $2.23\times10^{-3}$ & 1.09 & 2.18 & 2.2 \\
12.1 & 24.9 & $2.61\times10^{-3}$ & 1.27 & 2.39 & 5.7 \\
15.4 & 20.1 & $3.33\times10^{-3}$ & 2.56 & 2.07 & 0.4 \\
19.7 & 14.8 & $4.22\times10^{-3}$ & 5.63 & 2.37 & 5.4 \\
25.2 & 14.2 & $5.34\times10^{-3}$ & 9.50 & 2.11 & 1.1 \\
35.0 & 13.0 & $7.21\times10^{-3}$ & 19.4 & 2.10 & 0.9 \\
42.0 & 10.5 & $8.49\times10^{-3}$ & 34.1 & 2.11 & 1.2 \\
50.0 & 9.6  & $9.86\times10^{-3}$ & 51.4 & 2.28 & 3.8 \\
62.0 & 7.4  & $1.18\times10^{-2}$ & 98.9 & 2.35 & 4.9 \\
73.1 & 7.7  & $1.34\times10^{-2}$ & 127 & 2.51 & 7.5 \\
79.3 & 7.1  & $1.42\times10^{-2}$ & 160 & 2.67 & 10.1 \\
86.0 & 6.8  & $1.52\times10^{-2}$ & 193 & 2.82 & 12.5 \\
93.4 & 6.5  & $1.62\times10^{-2}$ & 232 & 2.94 & 14.4 \\
\hline
\end{tabular}
\caption{Best-fit mediator mass at fixed DM mass, at the grid points of Fig.~\ref{fig:selfint_map}. $\Delta\chi^2$ is measured from the best fit, Eq.~\eqref{eq:bestfit}.}
\label{tab:selfint_profile}
\end{table}

\begin{table*}[t]
\begin{center}
\renewcommand{\arraystretch}{1.2}
\begin{tabular}{|l|c|c|c|c|c|c|c|c|}
\hline
family & $m_{\chi_1}$ range [GeV] & $m_s$ range [MeV] & $m_{\chi_1}$ [GeV] & $m_s$ [MeV] & $\alpha_{11}$ & $b$ & $\chi^2/{\rm dof}$ & $\Delta\chi^2$ \\
\hline
main band & 3.01-12.1 & 6.21-26.0 & 7.61 & 14.6 & $1.65\times10^{-3}$ & 0.86 & 2.04 & 0.0 \\
resonant islands ($n=1$) & 14.2-23.3 & 13.6-28.4 & 15.2 & 19.2 & $3.27\times10^{-3}$ & 2.6 & 2.06 & 0.3 \\
resonant islands ($n=2$) & 23.3-32.3 & 10.0-23.9 & 26.3 & 15.9 & $5.54\times10^{-3}$ & 9.2 & 2.07 & 0.4 \\
resonant islands ($n\geq3$) & 29.7-70.0 & 6.49-16.1 & 33.6 & 11.7 & $6.95\times10^{-3}$ & 20 & 2.07 & 0.6 \\
\hline
\end{tabular}
\caption{Families of solutions within the $95\%$ CL region labeled by the number $n$ of $s$-wave bound states.}
\label{tab:selfint_branches}
\end{center}
\end{table*}

\paragraph{Upper limit on the dark matter mass:}
Above $m_{\chi_1}\simeq60$~GeV the best fit at fixed DM mass degrades (Table~\ref{tab:selfint_profile}), reaching $\Delta\chi^2=12.5$ at
$86$~GeV and $14.4$ at $93$~GeV. The data therefore bound the DM mass from above. Since the islands are narrower than the grid spacing, we determine the limit from the profile refined by minimizing $\chi^2$ in $m_s$ at fixed DM mass, which lowers $\Delta\chi^2$ by up to about $1.4$ units relative to the grid, for example to $6.5$ at $73$~GeV and $11.5$ at $86$~GeV. Projecting the two-dimensional regions of Sec.~\ref{sec:halo_data} onto $m_{\chi_1}$, we find
\begin{equation}
m_{\chi_1}\lesssim70~{\rm GeV}\ \ (95\%~{\rm CL}),\quad
m_{\chi_1}\lesssim87~{\rm GeV}\ \ (99.7\%~{\rm CL}).
\label{eq:mass_limit}
\end{equation}
For a dark sector at the SM temperature, whose relic coupling at $m_{\chi_1}=60$~GeV is $17\%$ smaller, the $95\%$ CL region (dotted in Fig.~\ref{fig:selfint_map}) ends at lower masses, near $m_{\chi_1}\simeq60$~GeV. The limit therefore depends on the thermal history of the dark sector: varying $T_{\rm dec}$ between $10$ and $40$~GeV changes the relic coupling at this mass by $-13\%$ to $+7\%$ (Sec.~\ref{sec:darktemp}).

The limit arises because the relic abundance fixes $\alpha_{11}/m_{\chi_1}$ to be nearly constant (Sec.~\ref{sec:relic}). In the semi-classical regime relevant for heavy DM, $\sigma_V\simeq(\pi/m_s^2)\,F(\beta)$ with $\beta=2\alpha_{11}m_s/(m_{\chi_1}v_{\rm rel}^2)$ and $F$ a universal function. The velocity at which the cross section turns over, fixed by the transition between galaxies and clusters, therefore determines $m_s$ nearly independently of the DM mass, while the normalization, $\sigma_V/m_{\chi_1}\propto1/(m_{\chi_1}m_s^2)$, falls with increasing DM mass. Heavier DM can recover $\sigma/m\sim1~{\rm cm^2/g}$ at dwarf velocities only by lowering $m_s$, which shifts the turnover away from the data; the best fits accordingly move to $m_s\simeq6$-$10$~MeV above $50$~GeV, close to the $N_{\rm eff}$ bound (Sec.~\ref{sec:cosmo}). The limit thus relies on the full velocity dependence of the data.

\begin{figure}[h]
\centering
\includegraphics[width=\columnwidth]{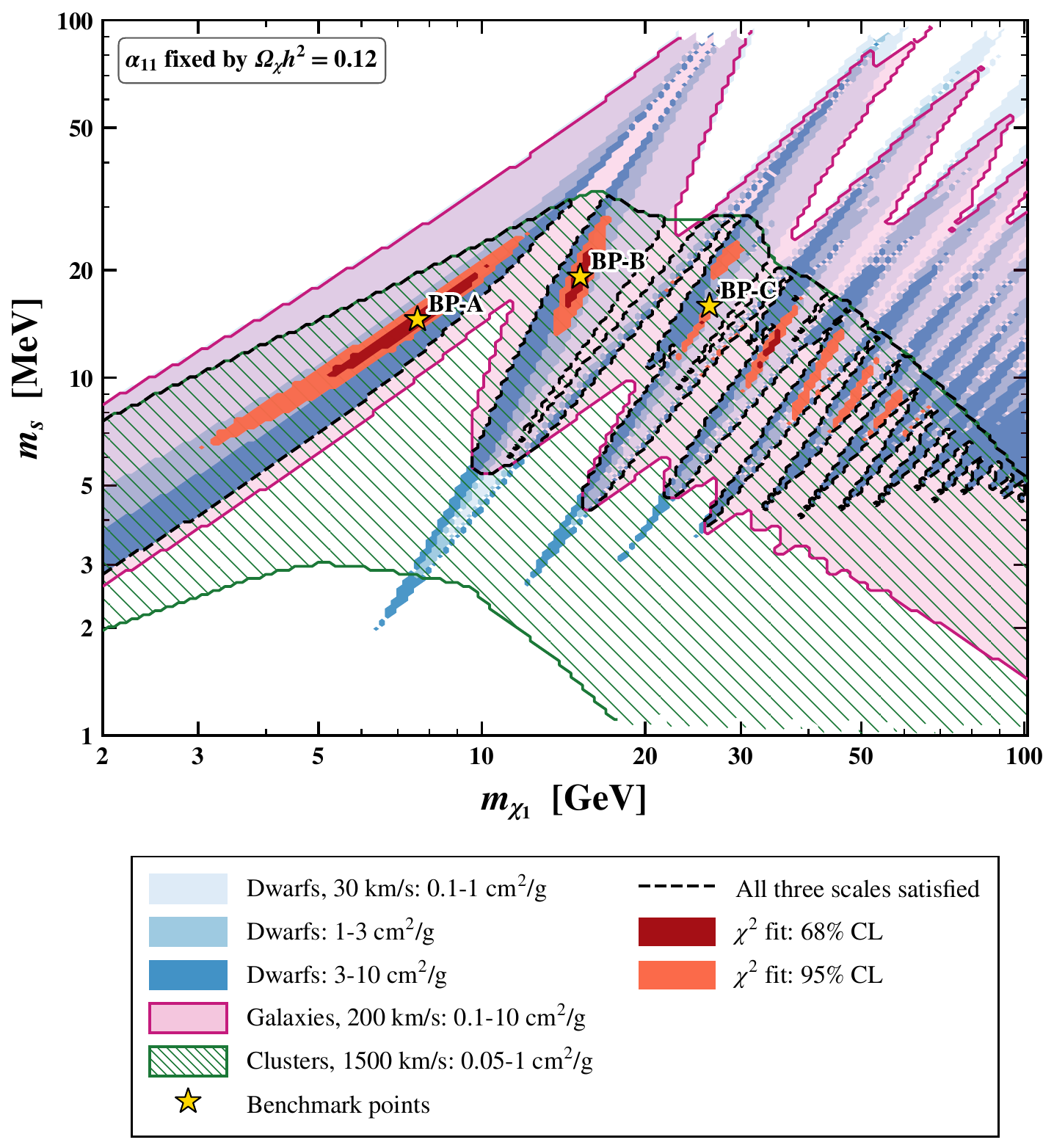}
\caption{Regions in which the self-interaction cross section lies in the ranges of Eq.~\eqref{eq:ballpark}, with $\alpha_{11}$ fixed by $\Omega_\chi h^2=0.12$, compared with the $\chi^2$ regions of Fig.~\ref{fig:selfint_map}.
}
\label{fig:ballpark}
\end{figure}

\paragraph{Comparison with ballpark constraints:}
Since the data scatter by more than their quoted uncertainties, we also apply the cruder test commonly used in the literature, which only requires the velocity-averaged cross section to lie in a broad range at a characteristic velocity of each class of systems~\cite{Tulin:2017ara,Kaplinghat:2015aga}:
\begin{equation}
\begin{aligned}
0.1&\leq\sigma/m_{\chi_1}\leq10~{\rm cm^2/g}\qquad({\rm dwarfs},\ 30~{\rm km/s}),\\
0.1&\leq\sigma/m_{\chi_1}\leq10~{\rm cm^2/g}\qquad({\rm galaxies},\ 200~{\rm km/s}),\\
0.05&\leq\sigma/m_{\chi_1}\leq1~{\rm cm^2/g}\ \qquad({\rm clusters},\ 1500~{\rm km/s}),
\end{aligned}
\label{eq:ballpark}
\end{equation}
where $\sigma/m_{\chi_1}=\langle\sigma_Vv\rangle/\langle v\rangle$ for a Maxwellian distribution with the stated mean velocity. We cap the dwarf range at $10~{\rm cm^2/g}$, since larger cross sections can trigger gravothermal collapse of dwarf halos within a Hubble time~\cite{Balberg:2002ue,Essig:2018pzq}. The upper edge of $1~{\rm cm^2/g}$ in clusters is of the order of the bounds from merging clusters such as 1E~0657-56~\cite{Markevitch:2003at,Randall:2007ph}. For clusters we lower the usual bound of $0.1~{\rm cm^2/g}$ to $0.05~{\rm cm^2/g}$, since two of the six cluster determinations of Ref.~\cite{Kaplinghat:2015aga} lie below $0.1~{\rm cm^2/g}$ with uncertainties extending to about $0.04~{\rm cm^2/g}$, and take $1500$~km/s as the middle of the range $\langle v\rangle\simeq1200$-$1750$~km/s covered by cluster data.  The resulting regions are shown in Fig.~\ref{fig:ballpark}.

The $\chi^2$ region and the three benchmark points of Table~\ref{tab:selfint_benchmarks} lie inside the region in which all three conditions are satisfied; for BP-A, BP-B and BP-C we find $\sigma/m_{\chi_1}\simeq2.2$, $2.3$ and $1.9$~cm$^2$/g at dwarf velocities, $1.5$, $1.5$ and $1.6$~cm$^2$/g at galaxy velocities, and $0.10$, $0.096$ and $0.097$~cm$^2$/g at cluster velocities, so their viability does not rest on the precise uncertainties of the data. The ballpark test cannot, however, bound the DM mass: on the resonant islands heavy DM gives $\sigma/m_{\chi_1}\simeq0.05$-$0.1~{\rm cm^2/g}$ in clusters, and with $m_s>6$~MeV the largest allowed mass moves from $50$ to $86$~GeV as the cluster edge is lowered from $0.10$ to $0.05~{\rm cm^2/g}$. The $\chi^2$ fit avoids this arbitrariness by comparing with the six cluster cross sections, $0.06$-$0.20~{\rm cm^2/g}$, which heavier DM falls below, leading to Eq.~\eqref{eq:mass_limit}.

\paragraph{Summary and benchmarks:}
In summary, the relic-constrained model reproduces the inferred velocity dependence of DM self-interactions at $95\%$ CL for
\begin{equation}
\begin{gathered}
3.0~{\rm GeV}\lesssim m_{\chi_1}\lesssim70~{\rm GeV}, \quad
6~{\rm MeV}\lesssim m_s\lesssim28~{\rm MeV},
\end{gathered}
\label{eq:allowed_range}
\end{equation}
on the main band at $m_{\chi_1}\simeq3.0$-$12$~GeV and on the resonant islands
at higher masses. 

\begin{table}[h]
\centering
\renewcommand{\arraystretch}{1}
\begin{tabular}{|l|c|c|c|}
\hline
 & BP-A & BP-B & BP-C \\
\hline
$m_{\chi_1}$ [GeV] & 7.61 & 15.2 & 26.3 \\
$m_s$ [MeV] & 14.6 & 19.2 & 15.9 \\
$\alpha_{11}$ & $1.65\times10^{-3}$ & $3.27\times10^{-3}$ & $5.54\times10^{-3}$ \\
$g_{11}$ & 0.144 & 0.203 & 0.264 \\
$b=\alpha_{11}m_{\chi_1}/m_s$ & 0.86 & 2.59 & 9.18 \\
$s$-wave bound states & 0 & 1 & 2 \\
$\chi^2/{\rm dof}$ (dof$=16$) & 2.04 & 2.06 & 2.07 \\
\hline
\end{tabular}
\caption{Benchmark points of the self-interaction fit: the best points of the main band (BP-A) and of the resonant islands with one (BP-B) and two (BP-C) $s$-wave bound states.}
\label{tab:selfint_benchmarks}
\end{table}

\begin{figure}[h]
\centering
\includegraphics[width=\columnwidth]{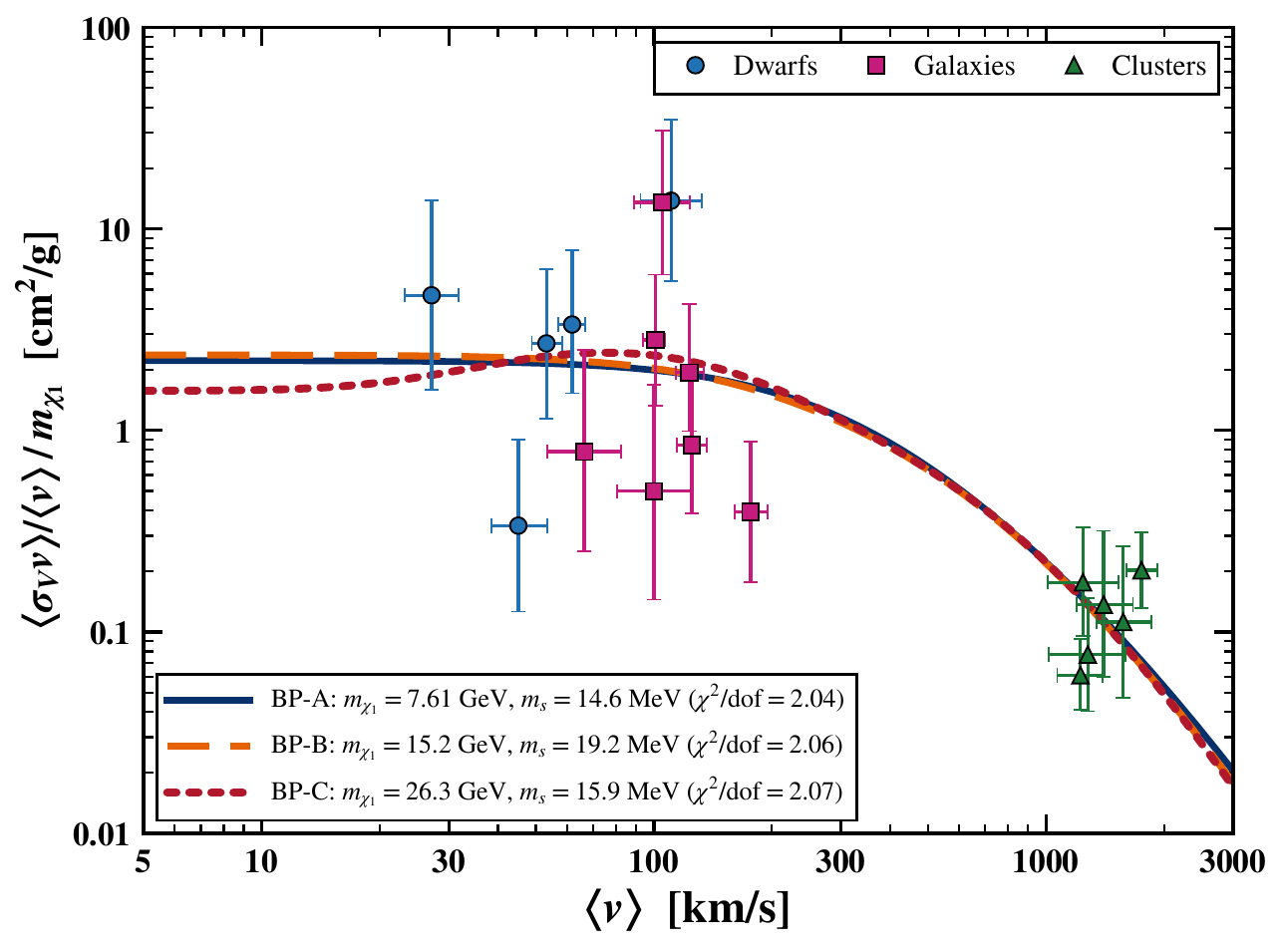}
\caption{Viscosity cross section of the benchmark points, averaged over a Maxwellian velocity distribution, compared with the inferred cross sections~\cite{Kaplinghat:2015aga}.}
\label{fig:selfint_bestfit}
\end{figure}

For the following sections we select three benchmark points (Table~\ref{tab:selfint_benchmarks}), one for each of the first three families of Table~\ref{tab:selfint_branches}: BP-A, the best fit, on the main band without a bound state, and BP-B and BP-C, the best points on the islands with one and two bound states. Since the minima of the different islands lie within $\Delta\chi^2\simeq0.6$ of each other, we label the benchmarks by the number of bound states rather than by their rank in $\chi^2$. Their viscosity cross sections, averaged over a Maxwellian distribution of relative velocities as in Sec.~\ref{sec:halo_data}, are compared with the cross sections inferred from dwarf galaxies, low-surface-brightness galaxies and clusters~\cite{Kaplinghat:2015aga} in Fig.~\ref{fig:selfint_bestfit}. All three benchmarks reproduce the data with similar quality, but their velocity dependence differs at low velocities. BP-B, with $b\simeq2.6$, retains a plateau at low velocities, like the main band, whereas BP-C, with $b\simeq9.2$, has a broad maximum near $\langle v\rangle\simeq80$~km/s, and its cross section decreases toward dwarf-galaxy velocities, a consequence of the nearby resonance. They illustrate rather than exhaust the allowed region: any point inside the contours of Fig.~\ref{fig:selfint_map} is equally consistent with the data.

\section{Neutrino masses and the lepton-number-breaking scale}
\label{sec:neutrino_pheno}

Sections~\ref{sec:relic} and~\ref{sec:selfint} fix the dark-sector parameters $m_{\chi_1}$, $m_s$ and $\alpha_{11}$. In this section we turn to
the remaining parameter, the lepton-number-breaking scale $v_\phi$. We use the oscillation data to fix the lepton-number-violating parameter $\mu_S$ of the inverse seesaw of Sec.~\ref{sec:neutrino}, derive the couplings of the singlet excitations to neutrinos, which fix the lifetime of the mediator, and summarize the constraints on the heavy neutral fermions. The resulting bounds on $v_\phi$ are derived in Sec.~\ref{sec:cosmo}.

\subsection{Fixing $\mu_S$ from oscillation data}
\label{sec:oscillations}

We work in the basis in which the charged-lepton mass matrix and the Dirac mass $M_N$ are diagonal and real. The light-neutrino mass matrix of
Eq.~\eqref{eq:inverse_seesaw} is diagonalized by the leptonic mixing matrix,
\begin{equation}
\begin{gathered}
m_\nu=U_{\rm PMNS}^*\,\hat m_\nu\,U_{\rm PMNS}^\dagger, \\
\hat m_\nu={\rm diag}(m_1,m_2,m_3),
\end{gathered}
\label{eq:mnu_diag}
\end{equation}
where $U_{\rm PMNS}$ contains the three mixing angles, the Dirac phase and two Majorana phases. Throughout we use the best-fit oscillation parameters of NuFIT~6.0~\cite{Esteban:2024eli}, $\sin^2\theta_{12}=0.308$,
$\sin^2\theta_{13}=0.0222$ $(0.0223)$, $\sin^2\theta_{23}=0.470$ $(0.550)$,
$\delta_{\rm CP}=212^\circ$ $(274^\circ)$,
$\Delta m^2_{21}=7.49\times10^{-5}~{\rm eV}^2$ and
$\Delta m^2_{3\ell}=+2.513\times10^{-3}~{\rm eV}^2$
$(-2.484\times10^{-3}~{\rm eV}^2)$ for normal (inverted) ordering, and treat
the Majorana phases as free. 

For given $m_D$ and $M_N$, Eq.~\eqref{eq:inverse_seesaw} can be inverted to express the lepton-number-violating matrix in terms of the oscillation data,
\begin{equation}
\mu_S=M_N\,m_D^{-1}\,U_{\rm PMNS}^*\,\hat m_\nu\,U_{\rm PMNS}^\dagger\,
\big(m_D^T\big)^{-1}M_N ,
\label{eq:mu_param}
\end{equation}
so that any choice of $(m_D,M_N)$ compatible with the constraints discussed below reproduces the measured mass splittings and mixing
angles exactly. Since $\mu_S=y_Sv_\phi/\sqrt2$, the Yukawa matrix of the singlet to $S_L$ is then determined by the lepton-number-breaking scale,
\begin{equation}
y_S=\frac{\sqrt2}{v_\phi}\,M_N\,m_D^{-1}\,U_{\rm PMNS}^*\,\hat m_\nu\,
U_{\rm PMNS}^\dagger\,\big(m_D^T\big)^{-1}M_N .
\label{eq:yS}
\end{equation}

The typical size of these quantities follows from the one-generation
estimate $\mu_S\simeq m_\nu M_N^2/m_D^2$:
\begin{equation}
\begin{gathered}
\mu_S\simeq16.5~{\rm keV}\,
\left(\frac{m_\nu}{0.05~{\rm eV}}\right)
\left(\frac{M_N}{1~{\rm TeV}}\right)^{2}
\left(\frac{0.01}{y_\nu}\right)^{2}, \\
y_S\simeq1.2\times10^{-4}
\left(\frac{\mu_S}{16.5~{\rm keV}}\right)
\left(\frac{200~{\rm MeV}}{v_\phi}\right).
\end{gathered}
\label{eq:muS_estimate}
\end{equation}
An MeV-scale lepton-number-breaking vev is therefore fully compatible with TeV-scale heavy neutrinos and perturbative couplings. The inverse-seesaw hierarchy $\|\mu_S\|\ll\|m_D\|\ll\|M_N\|$ is satisfied by a wide margin, and $y_S$ remains perturbative.

The dark-sector relation Eq.~\eqref{eq:prediction} links the two sectors through the ratio of Yukawa couplings, $\Delta m_\chi/\mu_S\simeq2\sqrt2\,|g_{11}|/|y_S|\sim10^{3}$-$10^{4}$ for the benchmark points. Although both originate from the same vev, the MeV-scale DM splitting and the keV-scale inverse-seesaw parameter therefore require no separation of scales, only a hierarchy $y_S\sim10^{-4}\ll g_{11}\sim0.2$ between the Yukawa couplings of $S_L$ and $\chi$ to $\phi$. 

The heavy states mix with the active neutrinos through $\Theta\simeq m_D\,M_N^{-1}$, which renders the leptonic mixing matrix non-unitary, $N=(1-\eta)U_{\rm PMNS}$ with $\eta=\tfrac12\Theta\Theta^\dagger$. Electroweak precision data and lepton-universality tests bound the diagonal elements of $\eta$ at the level of $10^{-3}$~\cite{Fernandez-Martinez:2016lgt}, and $\mu\to e\gamma$ bounds $\eta_{e\mu}$ more strongly~\cite{MEGII:2023ltw,MEGII:2025new}. For the reference values of Eq.~\eqref{eq:muS_estimate}, $|\Theta|\simeq1.7\times10^{-3}$ and $|\eta|\sim10^{-6}$, well within all bounds. These constraints restrict only $m_D/M_N$ and do not affect the dark sector, which depends on the neutrino sector only through $v_\phi$ and $\sum_im_{\nu_i}^2$ (Sec.~\ref{sec:nu_couplings}).

\subsection{Couplings of the singlet excitations to neutrinos}
\label{sec:nu_couplings}

\begin{figure}[!htbp]
\centering
\includegraphics[width=0.45\columnwidth]{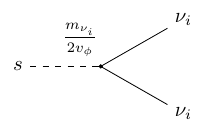}
\caption{Decay of the mediator into neutrinos.}
\label{fig:diag_sdecay}
\end{figure}

Because the light-neutrino masses are linear in $v_\phi$, the couplings of $s$ and $a$ to the light mass eigenstates follow directly from
Eq.~\eqref{eq:inverse_seesaw}. Expanding $\phi$ around its vev in $\mathcal L_\nu$ and rotating to the light mass basis, we obtain
\begin{equation}
\mathcal L\supset-\sum_i\frac{m_{\nu_i}}{2v_\phi}\,
\Big(s\,\overline{\nu_i}\nu_i+a\,\overline{\nu_i}\,i\gamma^5\nu_i\Big),
\label{eq:s_nu_coupling}
\end{equation}
up to corrections of order $\mu_S/M_N$ and $m_D^2/M_N^2$, where $\nu_i$ are Majorana fields. The coupling of $s$ equals $\partial m_{\nu_i}/\partial v_\phi$, as for the dark states, and the pseudoscalar coupling of $a$ is the one required by the Goldstone theorem. In the basis of
Eq.~\eqref{eq:mnu_diag} both couplings are flavor-diagonal. The resulting decay widths into neutrinos (Fig.~\ref{fig:diag_sdecay}) are
\begin{equation}
\begin{gathered}
\Gamma(s\to\nu\nu)\simeq\frac{m_s}{16\pi v_\phi^2}\sum_im_{\nu_i}^2, \\
\Gamma(a\to\nu\nu)\simeq\frac{m_a}{16\pi v_\phi^2}\sum_im_{\nu_i}^2 .
\end{gathered}
\label{eq:nu_decays}
\end{equation}
For the mediator this gives
\begin{align}
\tau_s\simeq{}&3.4\times10^{-2}~{\rm s}\,
\left(\frac{v_\phi}{200~{\rm MeV}}\right)^{2}
\left(\frac{15~{\rm MeV}}{m_s}\right)
\nonumber\\
&\times\left(\frac{2.6\times10^{-3}~{\rm eV^2}}{\sum_im_{\nu_i}^2}\right),
\label{eq:tau_s}
\end{align}
where the reference value of $\sum_im_{\nu_i}^2$ is the minimum allowed by oscillation data for normal ordering, $m_1\to0$. A heavier lightest neutrino only shortens the lifetime, so Eq.~\eqref{eq:tau_s} is an upper bound for given $v_\phi$ and $m_s$. The competing decay through Higgs mixing,
\begin{equation}
\Gamma(s\to e^+e^-)=\frac{\theta_{hs}^2m_e^2m_s}{8\pi v_h^2}
\left(1-\frac{4m_e^2}{m_s^2}\right)^{3/2},
\label{eq:s_ee}
\end{equation}
is negligible for the mixing angles allowed by direct detection (Sec.~\ref{sec:dd}), for which the branching ratio into $e^+e^-$ is below
$10^{-7}$. The mediator therefore decays almost exclusively into neutrinos, with a lifetime set by the neutrino masses and $v_\phi$.

The couplings in Eq.~\eqref{eq:s_nu_coupling} are of order $m_\nu/v_\phi\sim10^{-10}$ for $v_\phi\sim200$~MeV, far below laboratory and
astrophysical limits on neutrino couplings to light scalars~\cite{Berryman:2022hds}. Neutrinoless double-beta decay with emission of $s$ is kinematically forbidden for $m_s>Q_{\beta\beta}$, which holds throughout the parameter space of interest, since $Q_{\beta\beta}\lesssim4.3$~MeV for all double-beta emitters. The standard light-neutrino contribution to neutrinoless double-beta decay is unchanged, and the contribution of the heavy pseudo-Dirac pairs is suppressed by their small lepton-number-violating splitting.

\section{Early-Universe and supernova constraints}
\label{sec:cosmo}

In this section we derive the cosmological and astrophysical constraints on the dark sector and on the mediator. Together with the decay of $\chi_2$, the lifetime of the mediator and its emission in SN~1987A bound the lepton-number-breaking scale from above and from below, and we combine them into the allowed window of $v_\phi$ at the end of the section.

\subsection{Decays of the dark sector after freeze-out}
\label{sec:thermal_history}

As discussed in Sec.~\ref{sec:darktemp}, the dark sector decouples from the SM once the Higgs boson becomes non-relativistic and freezes out with its own temperature, $T_D/T\simeq1.3$-$1.5$. Its subsequent evolution is governed by the decays of $\chi_2$ and $s$.

After freeze-out the heavier state decays through $\chi_2\to\chi_1s$ via the pseudoscalar coupling $\eta_\chi=\delta y/(2\sqrt2)$ of Eq.~\eqref{eq:couplings}.
For $\Delta m_\chi>m_s$ the width is
\begin{equation}
\Gamma(\chi_2\to\chi_1s)\simeq\frac{\eta_\chi^2}{8\pi}\,
\frac{\big(\Delta m_\chi^2-m_s^2\big)^{3/2}}{m_{\chi_1}^2},
\label{eq:chi2_width}
\end{equation}
which gives a lifetime below one second for $|\delta y|\gtrsim10^{-8}$ at all three benchmark points. This is compatible with the symmetric regime, $|\delta y|\ll|y_L|\simeq0.2$-$0.4$, over many orders of magnitude, so that no $\chi_2$ population survives to the epoch of BBN.

The light mediator reconnects the dark sector with the SM through its decays into neutrinos. The decays and inverse decays $s\leftrightarrow\nu\nu$ come into equilibrium once $\Gamma(s\to\nu\nu)\,m_s/T$ exceeds the Hubble rate. This must happen before the neutrinos decouple from the electromagnetic plasma at $T\simeq2$~MeV, so that the entropy of the dark sector is shared with the SM plasma while all species are still in equilibrium, consistent with our treatment of the relic abundance in terms of the total comoving entropy. We impose this as $\tau_s\lesssim0.1$~s. Since $\tau_s\propto v_\phi^2$, Eq.~\eqref{eq:tau_s}, it gives an upper bound on the lepton-number-breaking scale, which we evaluate conservatively for the longest lifetime (normal ordering, $m_1=0$):
\begin{equation}
\begin{gathered}
v_\phi\lesssim338~{\rm MeV}\ \ ({\rm BP\text{-}A}), \\
v_\phi\lesssim387~{\rm MeV}\ \ ({\rm BP\text{-}B}), \\
v_\phi\lesssim352~{\rm MeV}\ \ ({\rm BP\text{-}C}).
\end{gathered}
\label{eq:vphi_tau}
\end{equation}
A heavier lightest neutrino or inverted ordering, for which $\sum_im_{\nu_i}^2$ is about twice its minimal normal-ordering value, only shortens the lifetime. Within the allowed window of $v_\phi$ (Sec.~\ref{sec:vphi_window}), equilibrium is reached at $T\simeq6$-$11$~MeV for the benchmark points.

\subsection{Cosmic microwave background and Big Bang nucleosynthesis}
\label{sec:bbn}
Energy injection from DM annihilation during recombination is strongly constrained for $s$-wave annihilation into visible final states. In our model the only $s$-wave channel, $\chi_1\chi_2\to sa$, is kinematically closed for a heavy pseudo-Majoron, and in any case no $\chi_2$ remains after its decay. The late-time annihilation $\chi_1\chi_1\to ss$ is $p$-wave and suppressed by the small DM velocities at recombination. Moreover, the annihilation products decay into neutrinos and do not deposit energy in the plasma. The model is therefore unconstrained by CMB limits on DM annihilation. 

The mediator decays almost exclusively into neutrinos: for the mixing angles allowed by direct detection, the branching ratio into $e^+e^-$ is below $10^{-7}$ (Sec.~\ref{sec:nu_couplings}). Its lifetime of at most $0.1$~s, together with the prompt decays of $\chi_2$ and of the heavy pseudo-Majoron, ensures that no unstable dark-sector state is present during BBN, and that no electromagnetic energy is injected that could photodissociate light nuclei. The only remaining effect of the dark sector on BBN and the CMB is its contribution to the energy density in neutrinos, which we discuss next. 
\paragraph*{Effective number of neutrino species:}
\label{sec:neff}

\begin{figure}[h]
\centering
\includegraphics[width=\columnwidth]{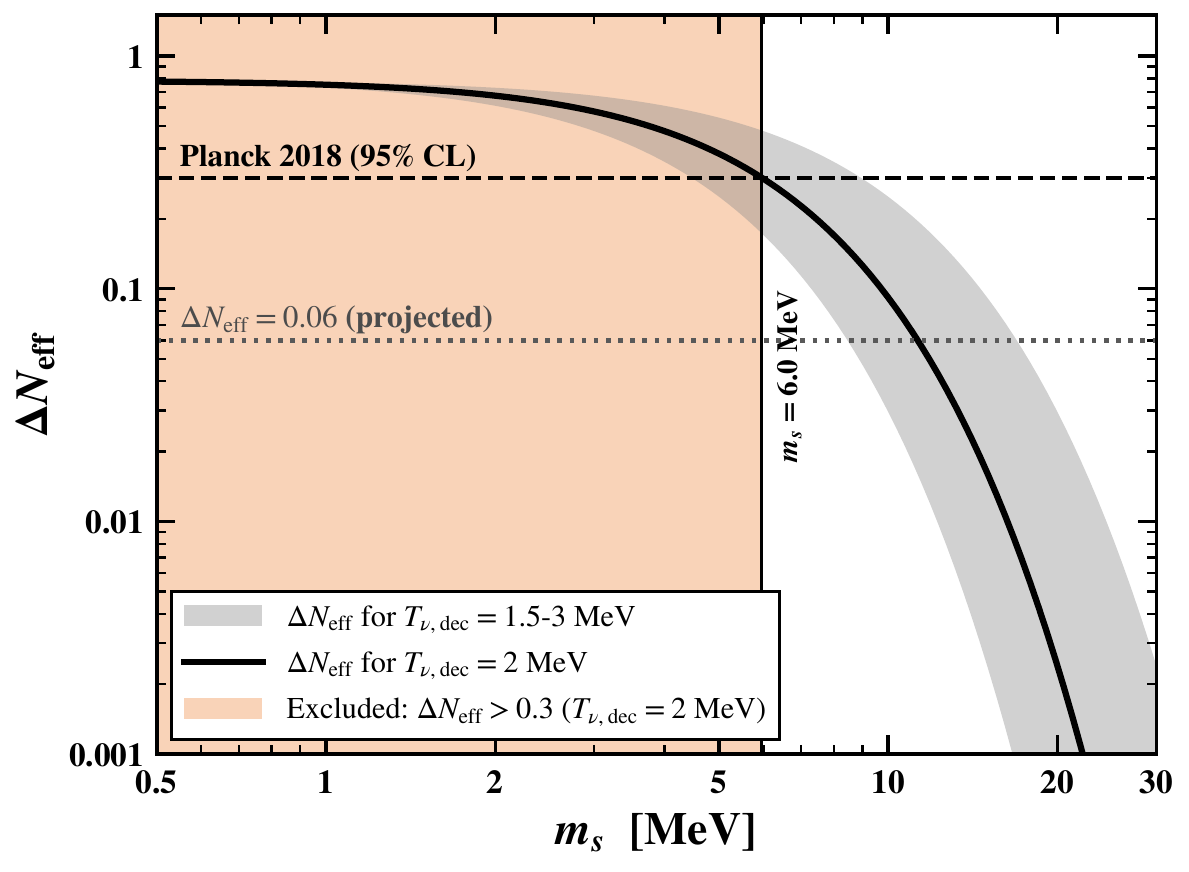}
\caption{Contribution of the mediator to \(\Delta N_{\rm eff}\) from Eq.~\eqref{eq:neff}, for $T_{\nu,{\rm dec}}=2$~MeV (solid) and in the range $1.5$-$3$~MeV (gray band). The dashed line shows the Planck bound, $\Delta N_{\rm eff}<0.3$, and the dotted line a projected next-generation sensitivity, $\Delta N_{\rm eff}=0.06$. Shaded: mediator masses excluded by the Planck bound for $T_{\nu,{\rm dec}}=2$~MeV.}
\label{fig:neff}
\end{figure}

Because $s$ is in equilibrium with the neutrinos when they decouple from the plasma, the entropy it carries at that time ends up in the neutrino sector when $s$ subsequently decays, and increases the neutrino temperature relative to the photon temperature. With instantaneous neutrino decoupling at $T_{\nu,{\rm dec}}$ and separate entropy conservation in the neutrino and photon-electron sectors afterwards, the entropy of $s$ is shared with the three neutrino species, with entropy weight $21/4$, and we obtain
\begin{equation}
\Delta N_{\rm eff}=3\left[\left(1+\frac{g_{*s}^{(s)}(m_s/T_{\nu,{\rm dec}})}{21/4}\right)^{4/3}-1\right],
\label{eq:neff}
\end{equation}
where $g_{*s}^{(s)}(x)$ is the entropy weight of a boson of mass $m_s$ at temperature $T_{\nu,{\rm dec}}$, which approaches one for $m_s\ll T_{\nu,{\rm dec}}$ and vanishes for $m_s\gg T_{\nu,{\rm dec}}$. Figure~\ref{fig:neff} shows the result for $T_{\nu,{\rm dec}}=1.5$-$3$~MeV, which brackets the range of the gradual decoupling in the SM. Requiring $\Delta N_{\rm eff}<0.3$, in accordance with Planck~\cite{Planck:2018vyg}, gives
\begin{equation}
\begin{gathered}
m_s\gtrsim6~{\rm MeV} \quad
(4.5\text{-}9~{\rm MeV\ for}\ T_{\nu,{\rm dec}}=1.5\text{-}3~{\rm MeV}),
\end{gathered}
\label{eq:neff_bound}
\end{equation}
consistent with dedicated analyses of neutrinophilic scalars in thermal equilibrium~\cite{Escudero:2018mvt,Escudero:2020dfa}. The benchmark points BP-A, BP-B and BP-C, with $m_s\simeq15$, $19$ and $16$~MeV, contribute $\Delta N_{\rm eff}\simeq0.019$, $0.003$ and $0.012$, respectively. The bound of Eq.~\eqref{eq:neff_bound} lies just below the $95\%$ CL region of the self-interaction fit, which begins at $m_s\simeq6.2$~MeV, and removes the lightest part of the band at the $99.7\%$ CL, $m_{\chi_1}\lesssim3$~GeV (Figs.~\ref{fig:selfint_map} and~\ref{fig:all_constraints}). A next-generation sensitivity of $\Delta N_{\rm eff}\simeq0.06$, the target of the CMB-S4 design~\cite{CMB-S4:2016ple}, would extend the bound to $m_s\gtrsim11$~MeV and thus probe the main band up to $m_{\chi_1}\simeq6$~GeV, as well as most of the resonant islands above $m_{\chi_1}\simeq35$~GeV, where the preferred mediator masses are $m_s\simeq7$-$11$~MeV (dashed line in Fig.~\ref{fig:all_constraints}). A $\Delta N_{\rm eff}$ within reach of next-generation CMB experiments arises also in other SIDM scenarios with light mediators; see, e.g., Ref.~\cite{Borah:2025cqj}.

\subsection{Domain walls}
\label{sec:walls}

The soft term $\mu'^2(\phi^2+\phi^{*2})$ preserves a $\mathbb Z_4$ subgroup of $U(1)_L$ under which $\phi\to-\phi$, so that the vacua $\langle\phi\rangle=\pm v_\phi/\sqrt2$ are degenerate in the absence of the tadpole, and domain walls with tension $\sigma\simeq\frac{2\sqrt2}{3}\sqrt{\lambda_\phi}\,v_\phi^3$ form when lepton number is broken~\cite{Saikawa:2017hiv}. The $\Delta L=2$ tadpole of Eq.~\eqref{eq:potential} lifts the degeneracy by $\Delta V\simeq2\kappa^3v_\phi$, and the resulting pressure makes the walls annihilate after a time $t\sim\sigma/\Delta V$. Requiring the walls never to dominate the energy density, $\Delta V\gtrsim\sigma^2/M_{\rm Pl}^2$, and to disappear before BBN, $\sigma/\Delta V\lesssim1$~s, gives for the benchmark points $\kappa\gtrsim1$~eV, the latter condition being the stronger. The tadpole leaves the spectrum unaffected as long as $\kappa^3\ll\lambda_\phi v_\phi^3$, i.e.\ $\kappa\ll10$~MeV. A tadpole anywhere in the wide range $1~{\rm eV}\lesssim\kappa\ll10~{\rm MeV}$ therefore removes the domain-wall problem without affecting the phenomenology.

\subsection{Supernova bound}
\label{sec:sn_bound}

The couplings of the mediator to neutrinos, Eq.~\eqref{eq:s_nu_coupling}, which make it decay before neutrino decoupling, also lead to its emission in supernovae and bound $v_\phi$ from below. Bosons coupled to neutrinos are produced in a supernova core by neutrino coalescence, $\nu\nu\to s$ and $\bar\nu\bar\nu\to s$, with energies of order $100$~MeV. If they decay outside the neutrinosphere, their decay products form a high-energy component of the neutrino flux. The absence of such events in the Kamiokande-II~\cite{Kamiokande-II:1987idp} and IMB~\cite{Bionta:1987qt} signals of SN~1987A limits the energy emitted in this way to about $1\%$ of the total, which for a flavor-universal coupling $g_{\alpha\beta}=g\,\delta_{\alpha\beta}$ implies~\cite{Fiorillo:2022cdq}
\begin{equation}
\begin{gathered}
g\lesssim10^{-9}~{\rm MeV}/m_s \\
(100~{\rm eV}\lesssim m_s\lesssim100~{\rm MeV}).
\end{gathered}
\label{eq:sn_universal}
\end{equation}
In our model the mediator couples to the light neutrinos through their mass matrix,
\begin{equation}
\begin{gathered}
\mathcal L\supset-\frac{s}{2v_\phi}\,(m_\nu)_{\alpha\beta}\,\nu_\alpha\nu_\beta+{\rm h.c.}, \\
g_i=\frac{m_i}{v_\phi}\ \ \text{(mass basis)},
\end{gathered}
\label{eq:s_nu_coupling_flavor}
\end{equation}
and the same couplings determine its lifetime, $\Gamma_s=m_s\sum_ig_i^2/(16\pi)$. Taken at face value, Eq.~\eqref{eq:sn_universal} conflicts with the requirement $\tau_s\lesssim0.1$~s of Sec.~\ref{sec:thermal_history}. The latter requires $g_3\gtrsim1.5\times10^{-10}$ for $m_s=15$~MeV, whereas Eq.~\eqref{eq:sn_universal} allows only $g\lesssim6.7\times10^{-11}$; since both bounds constrain the same couplings, the conflict cannot be removed by a choice of $v_\phi$, and the two would be compatible only for $m_s\lesssim3$~MeV. 

The bound of Eq.~\eqref{eq:sn_universal}, however, does not apply directly, because the couplings of Eq.~\eqref{eq:s_nu_coupling_flavor} are far from universal. For normal ordering with $m_1=0$ they are dominated by $\nu_3$, which consists mostly of $\nu_\mu$ and $\nu_\tau$. This affects both stages of the argument. First, the signal in Kamiokande-II and IMB arises from inverse beta decay of $\bar\nu_e$, and only $3\%$ of the decay products of $s$ are of electron flavor, compared with $1/3$ for universal couplings. Second, in the supernova core the production is dominated by the coalescence of the strongly degenerate $\nu_e$, whose coupling to $s$, $(m_\nu)_{ee}/v_\phi$, is suppressed by more than an order of magnitude relative to $g_3$. 

To quantify these effects, we compute the signal, i.e.\ the energy emitted by coalescence of flavor states in the core times the fraction of $\bar\nu_e$ among the decay products in the mass basis, for the couplings of Eq.~\eqref{eq:s_nu_coupling_flavor} and for universal couplings under identical core conditions. Since the signal scales as $g^2$, the bound on the largest coupling becomes
\begin{equation}
\begin{gathered}
g_3\,m_s\lesssim F\times10^{-9}~{\rm MeV}, \\
F=\left(\frac{S_{\rm universal}}{S_{\rm model}}\right)^{1/2},
\end{gathered}
\label{eq:sn_rescaled}
\end{equation}
where the astrophysical uncertainties largely cancel in the ratio. We scan the core temperature over $T=20$-$40$~MeV, the $\nu_e$ chemical potential over $0\leq\mu_{\nu_e}\leq4T$, and the unknown Majorana phase. For a non-degenerate core we find $F=5.7$, independently of $T$; $F^2$ follows analytically from the reduced production, by a factor $2.9$, and the reduced fraction of $\bar\nu_e$, by a factor $11$. Degeneracy of $\nu_e$ increases the rescaling to $F\simeq9$-$18$, since it enhances the production only in the universal case. The Majorana phase changes $F$ by less than $5\%$. We use the conservative value $F=5.7$ in the following; the calculation is described in Appendix~\ref{app:sn}. 

Combining Eq.~\eqref{eq:sn_rescaled} with $\tau_s<0.1$~s, the two requirements are compatible for $m_s\lesssim100$~MeV, and for $m_s\lesssim270$-$980$~MeV in a degenerate core. This includes the entire parameter space of Fig.~\ref{fig:all_constraints}. The bound nevertheless restricts the vev of lepton-number breaking from below,
\begin{equation}
v_\phi\gtrsim\frac{m_3\,m_s}{F\times10^{-9}~{\rm MeV}}
\simeq130~{\rm MeV}\left(\frac{m_s}{15~{\rm MeV}}\right)\left(\frac{5.7}{F}\right),
\label{eq:vphi_sn}
\end{equation}
which is stronger than the condition $\Delta m_\chi>m_s$ of Sec.~\ref{sec:model} for the benchmark points; the resulting window of $v_\phi$ is given in Sec.~\ref{sec:vphi_window}.

For inverted ordering the conclusion is reversed. The light-neutrino masses are then dominated by $\nu_1$ and $\nu_2$, which have a large electron-flavor component, so that about half of the decay products of $s$ are of electron flavor. We find $F\simeq0.9$-$1.4$, with $F\simeq1.0$ for a non-degenerate core, and the supernova bound combined with $\tau_s<0.1$~s requires $m_s\lesssim4$-$12$~MeV, or $m_s\lesssim6$~MeV for $F=1.0$. This is in conflict with the $N_{\rm eff}$ bound (Sec.~\ref{sec:bbn}) and with the self-interaction fit, both of which require $m_s\gtrsim6$~MeV; only a narrow window $m_s\simeq6$-$12$~MeV survives, and only for a strongly degenerate core. \emph{The model therefore strongly favors normal ordering of the neutrino masses}, which will be tested by JUNO~\cite{JUNO:2015zny} and DUNE~\cite{DUNE:2020ypp}. 

Our estimate treats the core in a one-zone approximation and the decay products in the mass basis, neglecting flavor conversion in the stellar envelope. These approximations affect the ratio $F$ much less than the absolute signal, and the conservative value $F=5.7$ does not rely on the uncertain $\nu_e$ degeneracy. A dedicated analysis with flavor-dependent couplings in supernova simulations would nevertheless be worthwhile, in particular for the heavier mediators, for which the lower bound on $v_\phi$, Eq.~\eqref{eq:vphi_sn}, is strongest.

\subsection{Window of the lepton-number-breaking scale}
\label{sec:vphi_window}

\begin{figure}[h]
\centering
\includegraphics[width=\columnwidth]{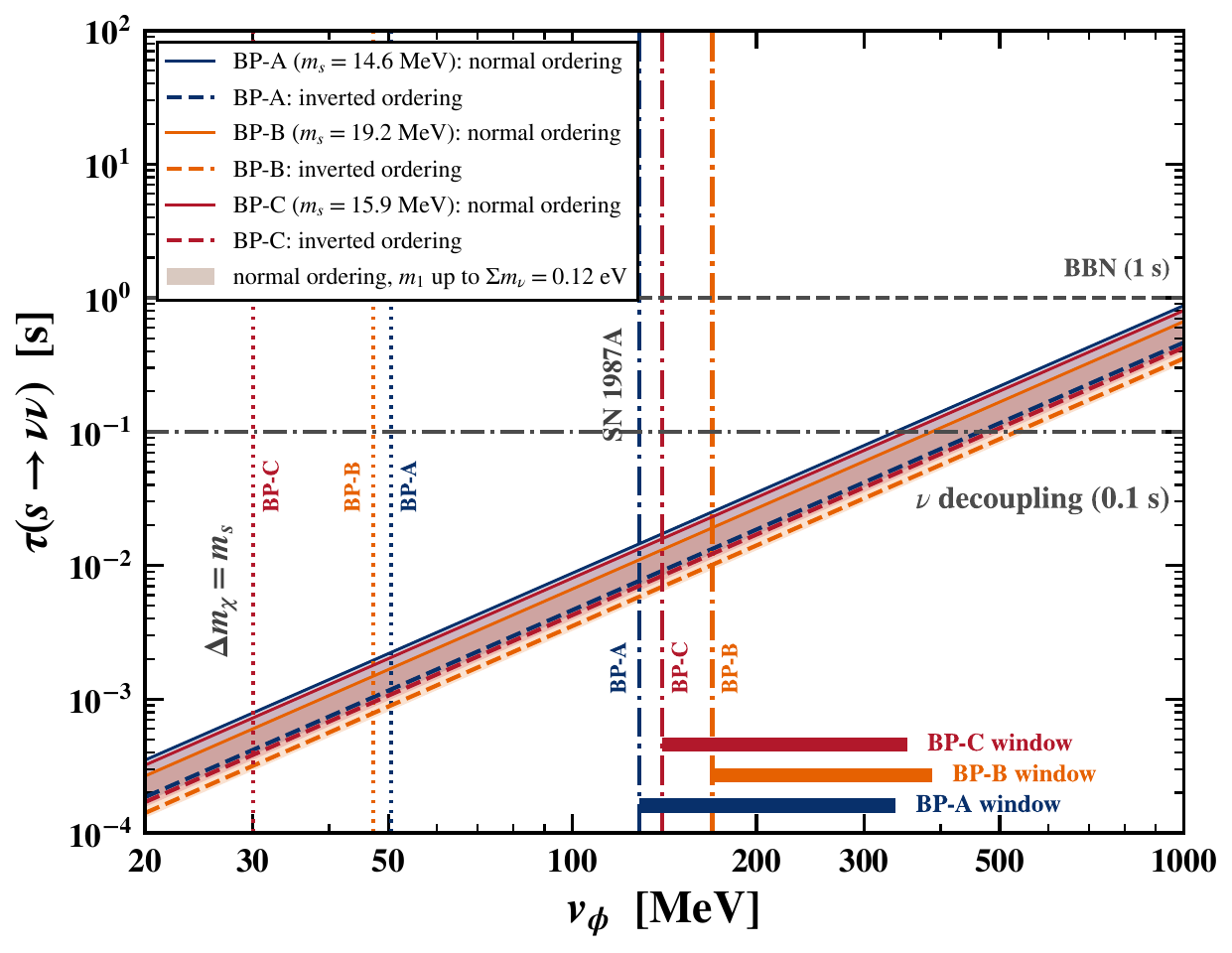}
\caption{Lifetime of the mediator as a function of $v_\phi$ for the benchmark points, and the resulting allowed windows of $v_\phi$ (horizontal
bars). Bands: lightest neutrino mass up to $\sum_im_{\nu_i}=0.12$~eV for normal ordering; inverted ordering is shown for $m_3=0$. Vertical lines: SN~1987A bound (dash-dotted) and $\Delta m_\chi=m_s$ (dotted); horizontal lines: $\tau_s=0.1$~s and $1$~s.}
\label{fig:tau_s}
\end{figure}

Figure~\ref{fig:tau_s} combines the bounds on $v_\phi$ for the three benchmark points: the upper bound from the lifetime of the mediator,
Eq.~\eqref{eq:vphi_tau}, and the lower bounds from the decay of $\chi_2$ and from SN~1987A, Eq.~\eqref{eq:vphi_sn}. It shows the lifetime of the mediator, Eq.~\eqref{eq:tau_s}; the bands span the range of the lightest neutrino mass allowed by oscillation data and by the cosmological bound
$\sum_im_{\nu_i}<0.12$~eV~\cite{Planck:2018vyg}.

\begin{table}[h]
\centering
\renewcommand{\arraystretch}{1.2}
\begin{tabular}{|l|c|c|c|}
\hline
 & BP-A & BP-B & BP-C \\
\hline
$m_{\chi_1}$ [GeV], $m_s$ [MeV] & 7.61, 14.6 & 15.2, 19.2 & 26.3, 15.9 \\
$g_{11}$ & 0.144 & 0.203 & 0.264 \\
allowed $v_\phi$ [MeV] & 129-338 & 169-387 & 140-352 \\
reference $v_\phi$ [MeV] & 208 & 256 & 222 \\
$\Delta m_\chi=2g_{11}v_\phi$ [MeV] & 60.1 & 103.7 & 117.4 \\
$\lambda_\phi$ & $2.4\times10^{-3}$ & $2.8\times10^{-3}$ & $2.5\times10^{-3}$ \\
$\mu_S$ [keV] & \multicolumn{3}{c|}{16.5} \\
$|\Theta|\simeq m_D/M_N$, $|\eta|$ & \multicolumn{3}{c|}{$1.7\times10^{-3}$, $1.5\times10^{-6}$} \\
$y_S=\sqrt2\mu_S/v_\phi$ & $1.1\times10^{-4}$ & $9.1\times10^{-5}$ & $1.1\times10^{-4}$ \\
$\Delta m_\chi/\mu_S$ & 3636 & 6278 & 7102 \\
$\tau_s$ [ms] & 38.1 & 43.7 & 39.8 \\
\hline
\end{tabular}
\caption{Neutrino-sector quantities for the benchmark points, for the reference values of Eq.~\eqref{eq:muS_estimate} ($M_N=1$~TeV, $y_\nu=0.01$, $m_\nu=0.05$~eV). The reference $v_\phi$ is the geometric mean of the allowed window.}
\label{tab:neutrino_benchmarks}
\end{table}

A lower bound follows from the dark sector itself. The heavier dark state can decay through $\chi_2\to\chi_1s$, and thus disappear before BBN (Sec.~\ref{sec:thermal_history}), only if $\Delta m_\chi=2g_{11}v_\phi>m_s$, Eq.~\eqref{eq:chi2_decay}, i.e.\ for $v_\phi>m_s/(2|g_{11}|)$. This gives $v_\phi>51$, $47$ and $30$~MeV for BP-A, BP-B and BP-C, shown by the vertical dotted lines in Fig.~\ref{fig:tau_s}. The supernova bound, Eq.~\eqref{eq:vphi_sn}, is about $2.5$-$5$ times stronger. Together with Eq.~\eqref{eq:vphi_tau}, it confines $v_\phi$ to a window of a few hundred MeV: $129$-$338$~MeV for BP-A, $169$-$387$~MeV for BP-B and $140$-$352$~MeV for BP-C. The corresponding neutrino-sector parameters are listed in Table~\ref{tab:neutrino_benchmarks}.  

The same couplings that make $s$ decay quickly enough also determine its emission in supernovae, so that the window is bounded from both sides by the coupling of the mediator to neutrinos. Within this window, direct detection constrains the Higgs-portal coupling (Sec.~\ref{sec:dd}).

\section{Direct detection}
\label{sec:dd}

\subsection{Scattering on nuclei}
\label{sec:dd_rate}

DM scatters on nuclei only through the CP-even singlet, via its mixing with the Higgs boson. The pseudo-Majoron is CP-odd and does not mix with $h$, so it has no tree-level coupling to quarks. The transition $\chi_1N\to\chi_2N$ is kinematically closed: it requires a splitting below the kinetic energy available in the DM-nucleus system, at most a few hundred keV, whereas $\Delta m_\chi=2|g_{11}|v_\phi\simeq60$-$120$~MeV for the benchmark points (Table~\ref{tab:neutrino_benchmarks}); moreover, it would proceed through the pseudoscalar coupling $\eta_\chi$ and be velocity suppressed. Direct detection therefore probes the elastic process $\chi_1N\to\chi_1N$, mediated by the scalar coupling $g_{11}$.

\begin{figure}[!htbp]
\centering
\includegraphics[width=0.45\columnwidth]{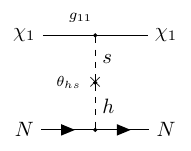}
\caption{DM-nucleon scattering through $h$-$s$ mixing.}
\label{fig:diag_dd}
\end{figure}

Through $h$-$s$ mixing (Fig.~\ref{fig:diag_dd}), the singlet couples to nucleons with $y_{sNN}\simeq\theta_{hs}f_Nm_n/v_h$, where $f_N\simeq0.3$~\cite{Cline:2013gha} encodes the Higgs coupling to the nucleon and $\theta_{hs}\simeq\lambda_{H\phi}v_hv_\phi/m_h^2$
(Eq.~\eqref{eq:mixing}). The resulting DM-nucleon cross section is spin-independent and isospin-conserving. Because the mediator mass is
comparable to or smaller than the momentum transfer $q=\sqrt{2m_NE_R}\simeq10$-$60$~MeV in xenon, the differential rate keeps the
full propagator,
\begin{equation}
\begin{gathered}
\frac{dR}{dE_R}\propto\sigma_{\rm ref}\,
\left(\frac{m_s^2}{q^2+m_s^2}\right)^{2}F^2(q)\,\eta(v_{\min}), \\
\sigma_{\rm ref}=\frac{\mu_{\chi n}^2}{\pi}\,
\frac{g_{11}^2\,\theta_{hs}^2\,f_N^2m_n^2}{v_h^2\,m_s^4},
\end{gathered}
\label{eq:dd_rate}
\end{equation}
where $F(q)$ is the Helm form factor~\cite{Lewin:1995rx}, $\eta(v_{\min})$ the mean inverse speed of the standard halo model with $v_0=238$~km/s and $v_{\rm esc}=544$~km/s~\cite{Baxter:2021pqo}, and $\sigma_{\rm ref}$ the DM-nucleon cross-section at zero momentum transfer. For Majorana DM with the coupling normalization of Eq.~\eqref{eq:couplings}, Eq.~\eqref{eq:dd_rate} coincides with the result for a Dirac fermion with coupling $g_{11}$.

\subsection{Recast of experimental limits}
\label{sec:dd_recast}

\begin{table*}[t]
\centering
\renewcommand{\arraystretch}{1.3}
\begin{tabular}{|l|l|c|c|c|c|c|}
\hline
& analysis & $R$ & $S$ [cm$^2$] & $\sigma_{\rm lim}$ [cm$^2$] & $\theta_{\max}$ & $\lambda_{H\phi}^{\max}$ \\
\hline
BP-A & LZ 4.2 t\,yr & 0.038 & $2.0\times10^{-30}$ & - & - & - \\
BP-A & XENONnT 3.1 t\,yr & 0.029 & $1.5\times10^{-30}$ & - & - & - \\
BP-A & DarkSide-50 & 0.34 & $1.8\times10^{-29}$ & - & - & - \\
BP-A & XENONnT light DM & 0.13 & $7.0\times10^{-30}$ & $4.5\times10^{-46}$ & $8.0\times10^{-9}$ & $2.5\times10^{-6}$ \\
BP-A & XENONnT light DM (LM)$^\dagger$ & 0.31 & $7.3\times10^{-25}$ & $1.6\times10^{-40}$ & $1.5\times10^{-8}$ & $4.5\times10^{-6}$ \\
BP-A & LZ light DM 5.7 t\,yr & 0.10 & $5.2\times10^{-30}$ & $2.3\times10^{-46}$ & $6.6\times10^{-9}$ & $2.0\times10^{-6}$ \\
BP-B & LZ 4.2 t\,yr & 0.045 & $1.7\times10^{-30}$ & $7.1\times10^{-48}$ & $2.0\times10^{-9}$ & $5.0\times10^{-7}$ \\
BP-B & XENONnT 3.1 t\,yr & 0.040 & $1.5\times10^{-30}$ & $3.4\times10^{-47}$ & $4.7\times10^{-9}$ & $1.2\times10^{-6}$ \\
BP-B & DarkSide-50 & 0.30 & $1.2\times10^{-29}$ & - & - & - \\
BP-B & XENONnT light DM & 0.15 & $5.9\times10^{-30}$ & - & - & - \\
BP-B & XENONnT light DM (LM)$^\dagger$ & 0.27 & $1.4\times10^{-24}$ & - & - & - \\
BP-B & LZ light DM 5.7 t\,yr & 0.14 & $5.2\times10^{-30}$ & - & - & - \\
BP-C & LZ 4.2 t\,yr & 0.015 & $2.2\times10^{-30}$ & $2.6\times10^{-48}$ & $1.1\times10^{-9}$ & $3.1\times10^{-7}$ \\
BP-C & XENONnT 3.1 t\,yr & 0.014 & $2.0\times10^{-30}$ & $1.8\times10^{-47}$ & $2.9\times10^{-9}$ & $8.4\times10^{-7}$ \\
BP-C & DarkSide-50 & 0.17 & $2.5\times10^{-29}$ & - & - & - \\
BP-C & XENONnT light DM & 0.085 & $1.2\times10^{-29}$ & - & - & - \\
BP-C & XENONnT light DM (LM)$^\dagger$ & 0.39 & $3.6\times10^{-24}$ & - & - & - \\
BP-C & LZ light DM 5.7 t\,yr & 0.076 & $1.1\times10^{-29}$ & - & - & - \\
\hline
\end{tabular}
\caption{Direct-detection recast for the benchmark points: $R$ and $S$ of Eqs.~\eqref{eq:dd_R} and~\eqref{eq:dd_S}, published limit $\sigma_{\rm lim}$, and the resulting $\theta_{\max}$ and $\lambda_{H\phi}^{\max}$ (for the reference $v_\phi$ of Table~\ref{tab:neutrino_benchmarks}). Dashes: benchmark mass outside the published range; for DarkSide-50 at BP-A the limit is much weaker than the xenon limits and is not recast. $^\dagger$Light-mediator reference; $S$ and $\sigma_{\rm lim}$ in cm$^2$\,MeV$^4$. Analyses: LZ 4.2 t\,yr~\cite{LZ:2024zvo,LZ:2024hepdata}, XENONnT 3.1 t\,yr~\cite{XENON:2025vwd}, DarkSide-50~\cite{DarkSide-50:2022qzh}, XENONnT light DM~\cite{XENON:2024hup,XENON:2024lightdata}, and LZ light DM 5.7 t\,yr~\cite{LZ:2025light}, with limit and efficiency from Ref.~\cite{LZ:2025hepdata}.}
\label{tab:dd}
\end{table*}
Experimental limits are reported for a contact interaction. To compare with Eq.~\eqref{eq:dd_rate}, we compute for each experiment the ratio $R$ of the event rate with the propagator factor to that of a contact interaction, both integrated over the recoil-energy window of the analysis,
\begin{equation}
R=\frac{\int dE_R\,\epsilon(E_R)\,\big[m_s^2/(q^2+m_s^2)\big]^2F^2(q)\,\eta(v_{\min})}
{\int dE_R\,\epsilon(E_R)\,F^2(q)\,\eta(v_{\min})},
\label{eq:dd_R}
\end{equation}
with the published efficiency where available and otherwise a smooth efficiency $\epsilon(E_R)=\{1+\exp[-(E_R-E_{50})/(0.25E_{50})]\}^{-1}$, where $E_{50}$ is the energy at which the efficiency of the analysis reaches $50\%$. For the light-mediator reference spectrum, the denominator contains $(m_s/q)^4$ instead of one. The model then corresponds to an effective contact cross section
\begin{equation}
\sigma_{\rm eff}=R\,\sigma_{\rm ref}\equiv S\,\theta_{hs}^2 ,
\label{eq:dd_S}
\end{equation}
which is compared with the published limit $\sigma_{\rm lim}(m_{\chi_1})$, giving $\theta_{hs}<\theta_{\max}=\sqrt{\sigma_{\rm lim}/S}$. This procedure accounts for the change of the total rate but not for the change in the shape of the recoil spectrum. For analyses with keV thresholds the momentum transfer is $q\simeq20$-$60$~MeV, and for the light-DM search of XENONnT~\cite{XENON:2024hup} with a threshold of $0.5$~keV it is $q\simeq11$-$35$~MeV. In both cases $q\sim m_s$, so that the benchmarks lie in the intermediate regime in which neither the contact nor the light-mediator ($m_s\ll q$) limits of the experiments apply directly; the propagator changes the spectrum by an order of magnitude across the analysis window. Where both are available, we therefore recast the contact and the light-mediator limits of the same analysis, each relative to its own reference spectrum. The two results coincide for a pure rate comparison, so that their difference measures the sensitivity to the spectral shape. The light mediator reduces the rate substantially: for xenon with a threshold of a few keV we find $R\simeq0.03$-$0.05$ for BP-A and BP-B and $R\simeq0.015$ for BP-C, while low-threshold searches are affected less, with $R\simeq0.08$-$0.4$ for xenon and $R\simeq0.2$-$0.3$ for argon. The coefficients $S$ are
\begin{equation}
\begin{gathered}
S\simeq(1.5\text{-}18)\times10^{-30}~{\rm cm^2}\ ({\rm BP\text{-}A}), \\
S\simeq(1.5\text{-}12)\times10^{-30}~{\rm cm^2}\ ({\rm BP\text{-}B}), \\
S\simeq(2.0\text{-}25)\times10^{-30}~{\rm cm^2}\ ({\rm BP\text{-}C}),
\end{gathered}
\label{eq:dd_S_values}
\end{equation}
where the ranges span xenon and argon targets and thresholds. For limits of $\sigma_{\rm lim}\sim10^{-48}$-$10^{-46}~{\rm cm^2}$, typical of current xenon experiments at these masses, the mixing angle must therefore satisfy $\theta_{hs}\lesssim10^{-9}$-$10^{-8}$.

\begin{figure*}[t]
\centering
\includegraphics[width=\textwidth]{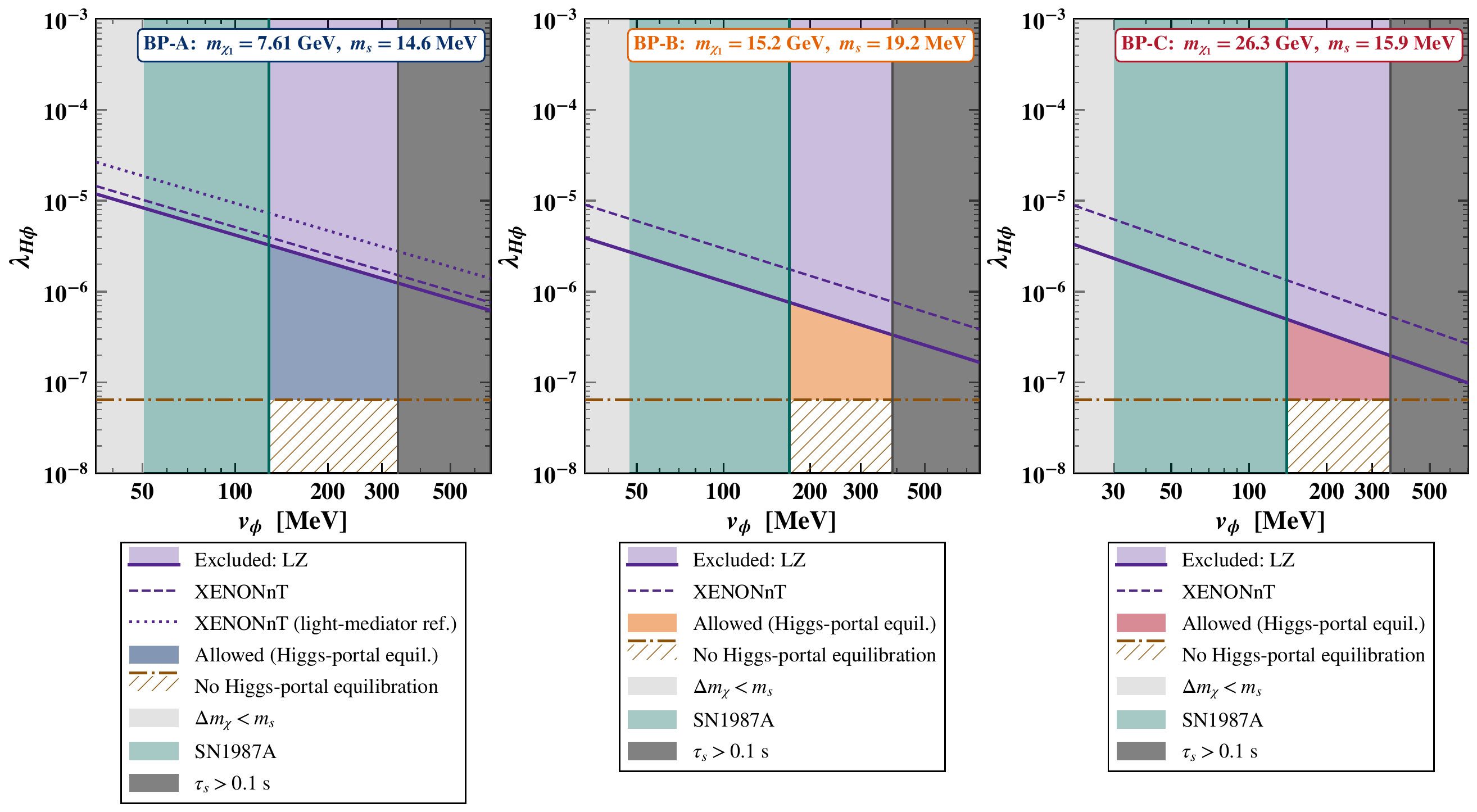}
\caption{Allowed range of the Higgs-portal coupling as a function of $v_\phi$ for the benchmark points. Shaded: excluded by LZ, using the light-DM search ($5.7$~t\,yr) for BP-A and the $4.2$~t\,yr search for BP-B and BP-C. Dashed: the weaker XENONnT bound, from the light-DM search (contact reference) for BP-A and the $3.1$~t\,yr search for BP-B and BP-C; dotted (BP-A): the light-mediator recast of the XENONnT light-DM search. Hatched: no equilibration through the Higgs portal, Eq.~\eqref{eq:lambda_th}. Vertical bands: values of $v_\phi$ excluded by the decay of $\chi_2$, SN~1987A and $\tau_s>0.1$~s. The colored wedge is allowed for a thermally populated dark sector.}
\label{fig:dd}
\end{figure*}

The benchmarks probe different experimental regimes. BP-B and BP-C, with $m_{\chi_1}\simeq15.2$ and $26.3$~GeV, lie in the mass range of the LZ search with a $4.2$~tonne-year exposure~\cite{LZ:2024zvo}, which sets the limits $\sigma_{\rm lim}=7.1\times10^{-48}$ and $2.6\times10^{-48}~{\rm cm^2}$ at these masses~\cite{LZ:2024hepdata}. With $S=1.7\times10^{-30}$ and $2.2\times10^{-30}~{\rm cm^2}$, this gives
\begin{equation}
\begin{gathered}
\theta_{hs}<2.0\times10^{-9}\ \ ({\rm BP\text{-}B}), \\
\theta_{hs}<1.1\times10^{-9}\ \ ({\rm BP\text{-}C}).
\end{gathered}
\label{eq:theta_BPB}
\end{equation}
The XENONnT search with a $3.1$~tonne-year exposure~\cite{XENON:2025vwd}, which has a slightly higher energy threshold, gives the independent but weaker bounds $\theta_{hs}<4.7\times10^{-9}$ and $2.9\times10^{-9}$, respectively. BP-A, with $m_{\chi_1}\simeq7.6$~GeV, lies below the lowest masses of $9$ and $10$~GeV reported in the high-mass searches of LZ and XENONnT, and is constrained by two low-threshold searches. The LZ search for light DM with a $5.7$~tonne-year exposure~\cite{LZ:2025light} covers $3$-$9$~GeV and sets the limit $\sigma_{\rm lim}=2.3\times10^{-46}~{\rm cm^2}$ at $7.6$~GeV; we use the published limit and nuclear-recoil efficiency~\cite{LZ:2025hepdata}, with the tabulated efficiency in Eq.~\eqref{eq:dd_R}. The XENONnT search for light DM with a threshold of $0.5$~keV~\cite{XENON:2024hup} covers $3$-$12$~GeV, with published contact and light-mediator limits~\cite{XENON:2024lightdata}, $\sigma_{\rm lim}=4.5\times10^{-46}~{\rm cm^2}$ and $\sigma\,m_{\rm med}^4<1.6\times10^{-40}~{\rm cm^2\,MeV^4}$ at $7.6$~GeV. We find
\begin{equation}
\begin{gathered}
\theta_{hs}<6.6\times10^{-9}\ \ ({\rm LZ\ light\ DM}), \\
\theta_{hs}<8.0\times10^{-9}\ \ ({\rm XENONnT,\ contact}), \\
\theta_{hs}<1.5\times10^{-8}\ \ ({\rm XENONnT,\ light\ mediator}) \\
({\rm BP\text{-}A}).
\end{gathered}
\label{eq:theta_BPA}
\end{equation}

Although the LZ limit is $2.0$ times stronger than the XENONnT contact limit, its efficiency extends to higher recoil energies, where the propagator suppression is larger ($R=0.10$ instead of $0.13$), so that the bound improves only by $20\%$. The light-mediator recast is weaker by a factor of two; this difference between the two XENONnT results reflects the sensitivity of the analysis to the shape of the recoil spectrum in the intermediate regime and provides an estimate of the uncertainty of the recast. BP-A is thus constrained about three to seven times more weakly than BP-B and six to fourteen times more weakly than BP-C. The results are summarized in Table~\ref{tab:dd}.

\subsection{Implications for the Higgs portal}
\label{sec:dd_portal}

In terms of the portal coupling, the bound reads
\begin{equation}
\lambda_{H\phi}<\frac{\theta_{\max}\,m_h^2}{v_h\,v_\phi}
\simeq5\times10^{-6}\left(\frac{\theta_{\max}}{10^{-8}}\right)
\left(\frac{130~{\rm MeV}}{v_\phi}\right).
\label{eq:lambda_max}
\end{equation}
Because $\theta_{hs}$ is proportional to $v_\phi$, an MeV-scale lepton-number-breaking vev automatically suppresses the Higgs mixing for a given portal coupling. The small mixing required by direct detection is consistent with cosmology, since the mediator decays into neutrinos and does not need the Higgs mixing to decay before BBN (Sec.~\ref{sec:bbn}). 

For a dark sector brought into equilibrium through the portal (Sec.~\ref{sec:darktemp}), the portal coupling is bounded from below by the equilibration threshold of Eq.~\eqref{eq:lambda_th} and from above by direct detection, Eq.~\eqref{eq:lambda_max}, and the DM-nucleon cross section is bounded from below. In terms of the mixing angle, Eq.~\eqref{eq:mixing}, the threshold corresponds to $\theta_{hs}\simeq(1.2$-$4.9)\times10^{-10}$ across the windows of $v_\phi$ of the benchmark points.  

The allowed range in the $(v_\phi,\lambda_{H\phi})$ plane is shown in Fig.~\ref{fig:dd} together with the window of $v_\phi$ derived in Sec.~\ref{sec:vphi_window}. For BP-C, the LZ bound of Eq.~\eqref{eq:theta_BPB} corresponds to $\lambda_{H\phi}<5.0\times10^{-7}$ at the lower end of the window, $v_\phi\simeq140$~MeV, and to $\lambda_{H\phi}<2.0\times10^{-7}$ at its upper end, $v_\phi\simeq352$~MeV; for BP-B it corresponds to $\lambda_{H\phi}<7.6\times10^{-7}$ at $v_\phi\simeq169$~MeV and $\lambda_{H\phi}<3.3\times10^{-7}$ at $v_\phi\simeq387$~MeV. For BP-A, the LZ bound of Eq.~\eqref{eq:theta_BPA} corresponds to $\lambda_{H\phi}<3.2\times10^{-6}$ at $v_\phi\simeq129$~MeV and $\lambda_{H\phi}<1.2\times10^{-6}$ at $v_\phi\simeq338$~MeV. In all cases the bound lies above the equilibration threshold of Eq.~\eqref{eq:lambda_th} across the entire window, so that a thermally populated dark sector is compatible with current limits and direct detection does not yet restrict $v_\phi$. The portal coupling is thereby confined to $(6$-$8)\times10^{-8}\lesssim\lambda_{H\phi}\lesssim(2.0$-$5.0)\times10^{-7}$ for BP-C, less than one decade, to $(3.3$-$7.6)\times10^{-7}$ for BP-B, up to about one decade, and to $(1.2$-$3.2)\times10^{-6}$ for BP-A, almost two decades. At the equilibration threshold, the effective cross section of Eq.~\eqref{eq:dd_S} ranges from about $4\times10^{-50}~{\rm cm^2}$ to $10^{-48}~{\rm cm^2}$ across the $v_\phi$ windows of the three benchmarks, inside the neutrino fog~\cite{OHare:2021utq} at these masses. A thermal origin of the dark sector therefore does not guarantee a signal in direct detection; the range between the threshold and current limits will be probed progressively by next-generation experiments, from the top down. If instead the dark sector were populated directly, for instance by inflaton decay, the portal coupling could be arbitrarily small, and the model would evade direct detection entirely.

The same reasoning can be applied across the whole parameter space preferred by the relic abundance and the self-interactions. At each point of the $(m_{\chi_1},m_s)$ plane, with $\alpha_{11}$ fixed by the relic abundance, the window of $v_\phi$ extends from the larger of $m_s/(2g_{11})$, required by the decay of $\chi_2$, and the supernova bound of Eq.~\eqref{eq:vphi_sn}, to the upper bound set by $\tau_s<0.1$~s. For $\lambda_{H\phi}\geq\lambda_{\rm th}$ the smallest scattering rate is obtained at the lower end of this window. Using at each DM mass the strongest applicable limit (and, for the XENONnT light-DM search, the more conservative of its two recasts), we find that current limits do not restrict the window of $v_\phi$ at any point of the plane shown in Fig.~\ref{fig:selfint_map}: for a thermally populated dark sector, direct detection does not yet constrain the masses of DM and of the mediator. 

\begin{figure}[h]
\centering
\includegraphics[width=\columnwidth]{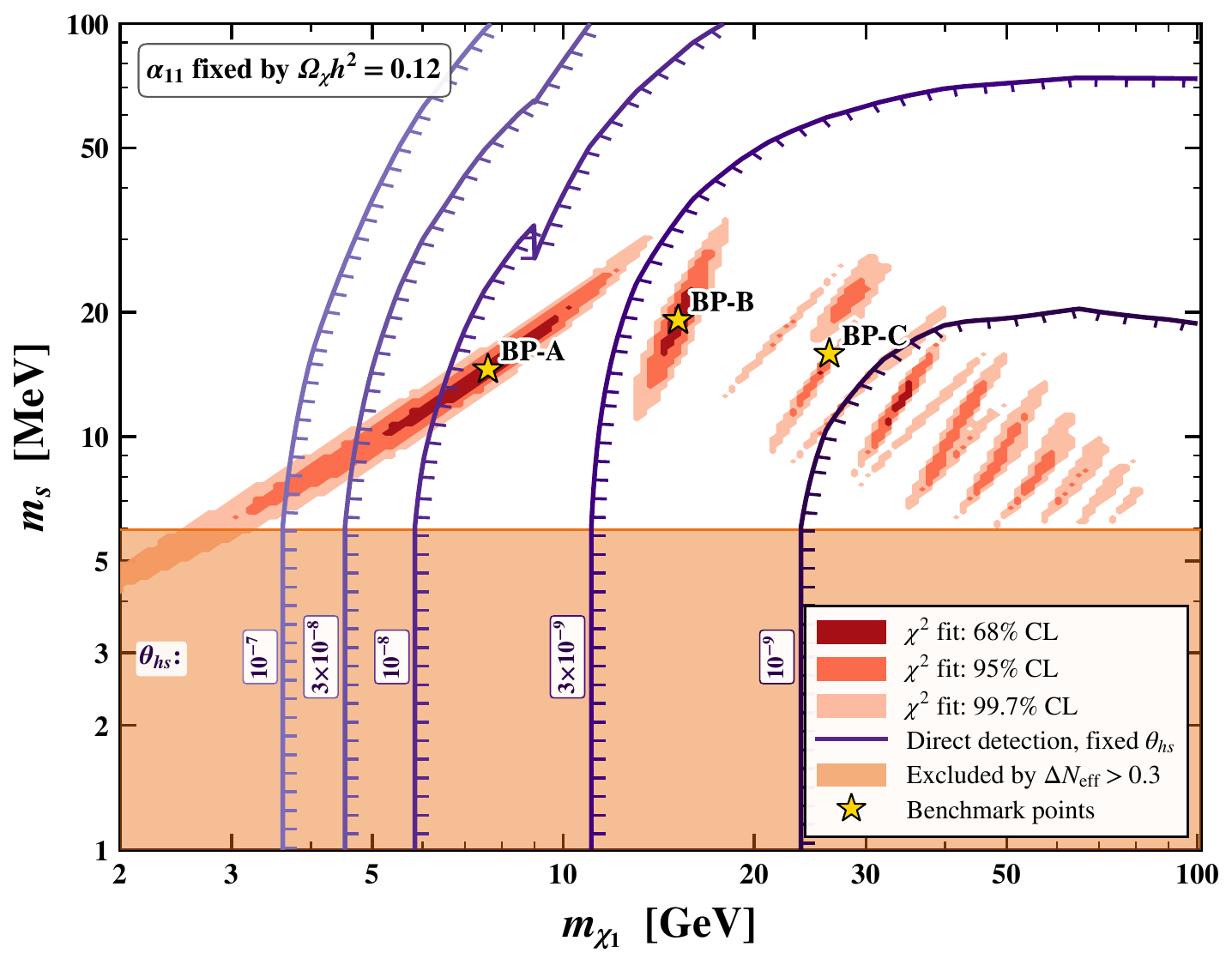}
\caption{Direct-detection limits for fixed $h$-$s$ mixing in the plane of Fig.~\ref{fig:selfint_map}, from the strongest applicable LZ or XENONnT limit (the region to the right of each line is excluded). The steps at $9$ and $12$~GeV occur where the strongest limit changes; inside the band excluded by $N_{\rm eff}$ the lines are continued vertically.}
\label{fig:dd_theta}
\end{figure}

Independently of the thermal history, the bounds can be displayed for fixed values of the mixing angle. Figure~\ref{fig:dd_theta} shows the lines $\theta_{\max}(m_{\chi_1},m_s)=\theta_{hs}$ for $\theta_{hs}=10^{-9}$-$10^{-7}$ in the plane of Fig.~\ref{fig:selfint_map}; for each value, the region to the right of the line is excluded. For a thermally populated dark sector the mixing is at least $\theta_{hs}\simeq(1$-$5)\times10^{-10}$, below all lines shown. 

Two features of the full propagator shape these lines. For $m_s\lesssim20$~MeV, below the typical momentum transfer, the propagator $1/(q^2+m_s^2)$ is dominated by $q^2$, so that the scattering rate and the bound become independent of $m_s$ and the lines run vertically. For heavier mediators the propagator approaches $1/m_s^2$, the cross section falls as $1/m_s^4$, and the lines bend toward larger DM masses. In the contact approximation the rate would instead grow as $1/m_s^4$ without limit for light mediators, and the predicted rates would be overestimated by factors of about $7$-$10$ for BP-A, depending on the analysis, $22$ for BP-B and $68$ for BP-C (the inverse of the suppression factors $R$ in Table~\ref{tab:dd}), and by more for lighter mediators. No point of the plane is excluded for $\theta_{hs}\lesssim8\times10^{-10}$. The main band is probed for $\theta_{hs}\gtrsim7\times10^{-9}$ near BP-A, $m_{\chi_1}\simeq8$~GeV, for $\theta_{hs}\gtrsim2\times10^{-8}$ at $m_{\chi_1}\simeq5$~GeV and for $\theta_{hs}\gtrsim4\times10^{-9}$ at its upper end, while the resonant islands are probed for $\theta_{hs}\gtrsim(1$-$2)\times10^{-9}$ at $m_{\chi_1}\simeq14$-$23$~GeV and for $\theta_{hs}\gtrsim(0.8$-$1.2)\times10^{-9}$ at higher masses.

\section{Further constraints}
\label{sec:other}

In this section we discuss the remaining theoretical, laboratory and astrophysical requirements, all of which are satisfied by large margins once the bounds of the previous sections are imposed.

\paragraph*{Theoretical consistency:} 
All couplings are perturbative: $g_{11}\simeq0.14$-$0.26$ for the benchmark points and $g_{11}\lesssim0.40$ in the whole allowed region, and $\lambda_\phi=m_s^2/(2v_\phi^2)\sim10^{-3}$-$10^{-2}$ for the benchmark points. The soft term $\mu'^2$ gives the pseudo-Majoron a mass far above $m_s$ without affecting the boundedness of the potential, which requires $\lambda_\phi>0$ and, for a negative portal coupling, $\lambda_{H\phi}>-2\sqrt{\lambda_H\lambda_\phi}$; both are satisfied by a wide margin for the values of $\lambda_{H\phi}$ allowed by direct detection. Since $\mu'^2$ and the tadpole $\kappa^3$ break lepton number softly, the hierarchy $m_a\gg m_s$ is stable against radiative corrections. 

\paragraph*{Higgs decays:} 
The portal coupling induces the decays $h\to ss$ and, for $m_a<m_h/2$, $h\to aa$ (Sec.~\ref{sec:darktemp}). Both final states decay into neutrinos and contribute to the invisible width. The bound ${\rm BR}(h\to{\rm inv})<0.107$~\cite{ATLAS:2023tkt} requires $\lambda_{H\phi}<7\times10^{-3}$, more than three orders of magnitude weaker than the bound from direct detection (Sec.~\ref{sec:dd_portal}). Decays $h\to\chi_1\chi_1$ through the mixing angle are suppressed by $\theta_{hs}^2\lesssim2\times10^{-16}$. The pseudo-Majoron, with $m_a>m_{\chi_1}+m_{\chi_2}$, decays promptly into $\chi_1\chi_2$, with a subdominant branching ratio into neutrinos, and is otherwise produced only in these Higgs decays.

 
\paragraph*{Meson decays and beam-dump experiments:} 
Light scalars that mix with the Higgs boson~\cite{Winkler:2018qyg} are probed by rare meson decays such as $K\to\pi s$ and $B\to Ks$, and by fixed-target and beam-dump experiments in which they are produced through the mixing and detected through their decays into $e^+e^-$ or photons. For an invisibly decaying scalar, the searches for $K^+\to\pi^+X$~\cite{NA62:2021zjw} and $B^+\to K^+\nu\bar\nu$~\cite{Belle-II:2023esi} are sensitive only to mixing angles far above the bound $\theta_{hs}\lesssim1.5\times10^{-8}$ from direct detection.

 Beam-dump experiments are insensitive to the mediator for two reasons. Its production rate is proportional to $\theta_{hs}^2\lesssim2\times10^{-16}$, and it decays almost exclusively into neutrinos: the branching ratio into $e^+e^-$ is $\Gamma(s\to e^+e^-)/\Gamma(s\to\nu\nu)\simeq\theta_{hs}^2m_e^2v_\phi^2/(\tfrac12v_h^2\sum m_\nu^2)$, which for the bound on $\theta_{hs}$ from the LZ searches and the reference $v_\phi$ of Table~\ref{tab:neutrino_benchmarks} is $6\times10^{-9}$ for BP-A, $9\times10^{-10}$ for BP-B and $2\times10^{-10}$ for BP-C. Moreover, with lifetimes of about $40$~ms, mediators with energies of order GeV travel about $10^9$~m before decaying. Electron beam-dump and missing-energy experiments are even less sensitive, since the coupling to electrons is further suppressed by $m_e/v_h$. The DM itself, with $m_{\chi_1}\gtrsim3$~GeV, is too heavy to be produced in kaon or $B$-meson decays. The same suppression by $\theta_{hs}^2$ renders the emission of $s$ through nucleon bremsstrahlung in supernovae negligible~\cite{Dev:2020eam}. 
 
\paragraph*{Small-scale structure, bound states and indirect detection:} 
The DM remains in kinetic equilibrium with the mediators until these disappear at temperatures of a few MeV. The subsequent scattering on neutrinos through $s$ exchange is suppressed by $g_\nu^2\lesssim10^{-19}$, where $g_\nu=m_\nu/v_\phi$ is the coupling of Eq.~\eqref{eq:s_nu_coupling_flavor}, so that kinetic decoupling~\cite{Bringmann:2016din} occurs at $T\sim{\rm MeV}$, many orders of magnitude above the keV temperatures to which the Lyman-$\alpha$ forest~\cite{Irsic:2017ixq} is sensitive. For the same reason, neutrino self-interactions mediated by $s$ have a strength $G_{\rm eff}\simeq g_\nu^2/m_s^2\sim10^{-11}G_F$ and are unobservable. 

On the resonant islands the Yukawa potential supports bound states, for example one $s$-wave bound state for BP-B with $b\simeq2.6$ and two for BP-C with $b\simeq9.2$. Their binding energies, at most of order $\alpha_{11}^2m_{\chi_1}/4\sim0.2$~MeV for BP-C, and the kinetic energies in halos, below $1$~MeV even in clusters, are far below $m_s$, so that bound-state formation through the emission of $s$ is kinematically forbidden~\cite{Cirelli:2016rnw}. Present-day annihilation, $\chi_1\chi_1\to ss$ followed by $s\to\nu\nu$, is $p$-wave suppressed by $v^2\sim10^{-6}$ in galactic halos. Solving the Schr\"odinger equation with the Yukawa potential, we find a $p$-wave Sommerfeld factor $S_1\lesssim10$ for BP-A and BP-B. BP-C lies just above the threshold $b\simeq9.1$ for a $p$-wave bound state, so that the enhancement is resonant, $S_1\simeq700$ at $v_{\rm rel}=300$~km/s, saturating at $10^6$ for $v_{\rm rel}\lesssim10$~km/s. The resulting cross section, $\langle\sigma v\rangle\lesssim5\times10^{-28}~{\rm cm^3/s}$ in the Milky Way and $\lesssim3\times10^{-27}~{\rm cm^3/s}$ at dwarf-galaxy velocities, lies more than three orders of magnitude below the limits from neutrino telescopes~\cite{Super-Kamiokande:2020sgt}. Since the annihilation products are neutrinos, CMB limits on energy injection do not apply.

\section{Combined constraints and surviving parameter space}
\label{sec:window}
\begin{figure}[h]
\centering
\includegraphics[width=\columnwidth]{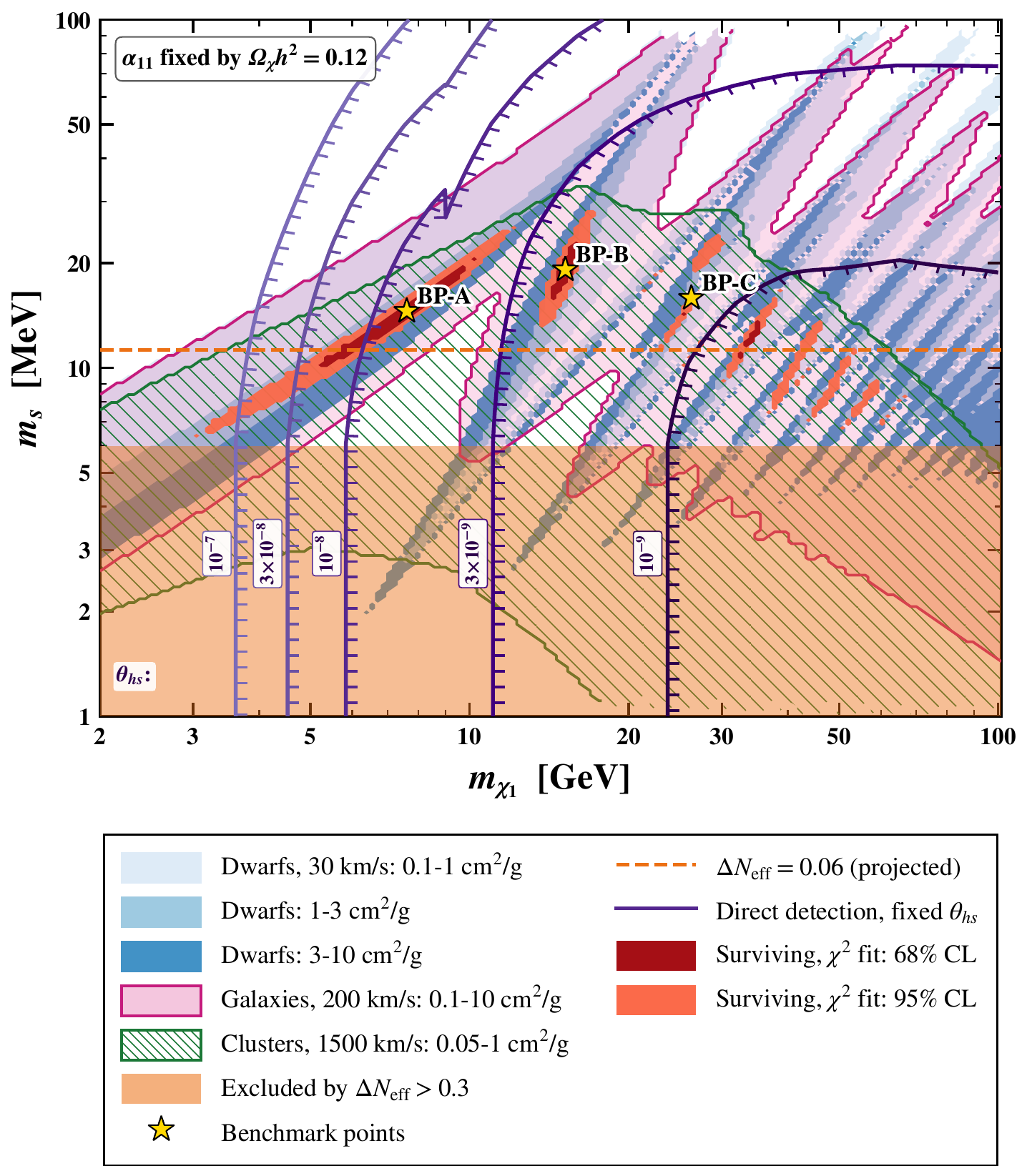}
\caption{Combined constraints in the $(m_{\chi_1},m_s)$ plane: self-interaction fit, ballpark ranges of Eq.~\eqref{eq:ballpark}, $N_{\rm eff}$ bound and projected sensitivity $\Delta N_{\rm eff}=0.06$, and direct-detection lines for fixed $\theta_{hs}$ as in Fig.~\ref{fig:dd_theta}.}
\label{fig:all_constraints}
\end{figure}

\begin{table*}[t]
\centering
\renewcommand{\arraystretch}{1.2}
\begin{tabular}{|l|c|c|c|}
\hline
 & BP-A & BP-B & BP-C \\
\hline
$m_{\chi_1}$, $m_s$ & $7.6$~GeV, $14.6$~MeV & $15.2$~GeV, $19.2$~MeV & $26.3$~GeV, $15.9$~MeV \\
$s$-wave bound states & 0 & 1 & 2 \\
$\alpha_{11}$ (relic abundance) & $1.7\times10^{-3}$ & $3.3\times10^{-3}$ & $5.5\times10^{-3}$ \\
$\chi^2/{\rm dof}$ (self-interactions) & $2.04$ & $2.06$ & $2.07$ \\
$\Delta N_{\rm eff}$ & $0.019$ & $0.003$ & $0.012$ \\
$v_\phi$: $\chi_2$ decay, SN~1987A, $\tau_s$ & $129$-$338$~MeV & $169$-$387$~MeV & $140$-$352$~MeV \\
$\theta_{hs}$ bound (direct detection) & $<(0.66$-$1.5)\times10^{-8}$ & $<2.0\times10^{-9}$ & $<1.1\times10^{-9}$ \\
$\lambda_{H\phi}$ for Higgs-portal equilibration & $(6$-$8)\times10^{-8}$-$(1.2$-$3.2)\times10^{-6}$ & $(6$-$8)\times10^{-8}$-$(3.3$-$7.6)\times10^{-7}$ & $(6$-$8)\times10^{-8}$-$(2.0$-$5.0)\times10^{-7}$ \\
\hline
\end{tabular}
\caption{Summary of the benchmark points and the constraints on them.}
\label{tab:summary}
\end{table*}
Because the relic abundance fixes $\alpha_{11}$ as a function of the DM mass, the model is described at low energies by $m_{\chi_1}$ and $m_s$, together with $v_\phi$ and $\lambda_{H\phi}$. Figure~\ref{fig:all_constraints} combines the constraints in the $(m_{\chi_1},m_s)$ plane: the self-interaction fit (Sec.~\ref{sec:selfint_results}), the ballpark ranges of Eq.~\eqref{eq:ballpark}, the $N_{\rm eff}$ bound (Sec.~\ref{sec:bbn}), and the direct-detection limits for fixed mixing (Sec.~\ref{sec:dd_portal}). The remaining constraints act on $v_\phi$ rather than on this plane: a window satisfying $\Delta m_\chi>m_s$ and $\tau_s<0.1$~s exists for $m_s<\alpha_{11}\tau_{\max}\sum m_\nu^2/\hbar$, i.e.\ up to $170$~MeV at $m_{\chi_1}=2$~GeV and $6$~GeV at $80$~GeV, and the supernova bound, which raises its lower end, leaves it open for $m_s\lesssim100$~MeV (Sec.~\ref{sec:sn_bound}).

The surviving parameter space consists of the main band,
\begin{equation}
\begin{gathered}
3.0~{\rm GeV}\lesssim m_{\chi_1}\lesssim12~{\rm GeV}, \quad
6~{\rm MeV}\lesssim m_s\lesssim26~{\rm MeV},
\end{gathered}
\label{eq:surviving_main}
\end{equation}
and of the resonant islands between $m_{\chi_1}\simeq14$ and $70$~GeV, with $m_s\simeq6.5$-$28$~MeV ($95\%$ CL). All of it satisfies the ballpark ranges on all three scales. The $N_{\rm eff}$ bound removes only the lightest part of the band, $m_{\chi_1}\lesssim3$~GeV at $99.7\%$ CL, and the upper end of the mass range is set by the cluster data. Direct detection removes no part of this region, since the mixing angle can be chosen small enough and the equilibration threshold of Eq.~\eqref{eq:lambda_th} lies below current limits everywhere in the plane, so that a thermal origin of the dark sector is compatible with all constraints. 

Table~\ref{tab:summary} summarize the different types of solutions. BP-A, on the main band, is constrained only by the self-interaction fit and, through $v_\phi$, by SN~1987A. BP-B and BP-C, on resonant islands in the mass range of the high-mass xenon searches, leave about one decade or less in the portal coupling for a thermally populated dark sector.

\section{Conclusions}
\label{sec:conclusions}

We have studied a minimal framework in which a single scalar, whose vev breaks lepton number at the MeV scale, generates both the light-neutrino masses through the inverse seesaw and the pseudo-Dirac structure of fermionic dark matter, and whose CP-even excitation mediates the self-interactions of DM. 

Our first result concerns the structure of the couplings. We have shown, for arbitrary Yukawa couplings and phases, that every coupling able to produce a central force is fixed by the splitting, $|g^{\rm central}|\simeq\Delta m_\chi/(2v_\phi)$. A scalar that generates the pseudo-Dirac splitting therefore cannot mediate a strong inelastic self-interaction; the viable regime is an elastic Yukawa interaction of the lighter state, with the prediction $\Delta m_\chi\simeq2g_{11}v_\phi$. 

Second, the neutrino sector removes the usual tension between BBN and direct detection for light scalar mediators. The mediator decays into neutrinos within tens of milliseconds, independently of its mixing with the Higgs boson, and all annihilation products end up as neutrinos, so that the CMB and indirect searches impose no constraint. 

Third, with the self-coupling fixed by the relic abundance, the fit to the self-interaction cross sections inferred from dwarf galaxies to clusters involves only the masses of DM and of the mediator. The data are reproduced along a narrow band, $m_s\simeq1.9~{\rm MeV}\times(m_{\chi_1}/{\rm GeV})$, with a best fit at $m_{\chi_1}\simeq7.6$~GeV and $m_s\simeq15$~MeV and a quality comparable to that of purely phenomenological fits, and on resonant islands at higher masses. The cluster data bound the DM mass, $m_{\chi_1}\lesssim70$~GeV at $95\%$ CL, a limit not captured by the broad ranges commonly used for the different scales. The separate temperature of the dark sector and the exact partial-wave treatment change the relic coupling and the cross section by up to $33\%$ and $40\%$, respectively. 

Fourth, $N_{\rm eff}$ requires $m_s\gtrsim6$~MeV. Because the coupling of the mediator to neutrinos is dominated by the heaviest neutrino, the SN~1987A bound is weakened by a factor of about six for normal ordering and, together with the lifetime of the mediator, confines $v_\phi$ to a window of a few hundred MeV, for example $129$-$338$~MeV for the best-fit benchmark; for inverted ordering it excludes the model except in a strongly degenerate core. Direct detection limits the mixing to $\theta_{hs}\lesssim10^{-9}$-$10^{-8}$, and a thermal origin of the dark sector through the Higgs portal is compatible with current limits throughout the parameter space.

The surviving parameter space, Fig.~\ref{fig:all_constraints}, can be tested in several independent ways:
\begin{itemize}
\item the neutrino mass ordering at JUNO~\cite{JUNO:2015zny} and DUNE~\cite{DUNE:2020ypp}, since the model strongly favors normal ordering;
\item a measurement of $N_{\rm eff}$ with a precision of $0.06$, which would probe mediator masses up to $11$~MeV;
\item low-threshold direct detection, which will reach the main band and probe the portal coupling down to the thermal threshold, inside the neutrino fog;
\item improved self-interaction cross sections on cluster scales, which control the upper limit on the DM mass;
\item the neutrino signal of a future galactic supernova.
\end{itemize}

Several aspects would merit further study: the effect of the bound states, including the $p$-wave resonance near BP-C, on the inferred cross sections and a finer resolution of the resonant islands; a supernova analysis with flavor-dependent couplings, which would sharpen the lower bound on $v_\phi$; and a treatment of the dark-sector decoupling beyond the instantaneous approximation. None of these refinements affects the central conclusion: the scale at which lepton number is broken can simultaneously account for neutrino masses and provide the light mediator of self-interacting dark matter, in a framework that is predictive and testable with near-future experiments.

\begin{acknowledgments}
The author thanks the LZ and XENON collaborations for making their limits publicly available through HEPData and the XENONnT data release. All
computations were performed in Python with NumPy~\cite{Harris:2020xlr} and SciPy~\cite{Virtanen:2019joe}, and the figures were produced with
Matplotlib~\cite{Hunter:2007ouj}.
\end{acknowledgments}

\appendix

\section{Couplings of the singlets for arbitrary phases}
\label{app:takagi}

In this appendix we derive Eq.~\eqref{eq:central_bound} for arbitrary complex Yukawa couplings.

\subsection{Mass matrix and couplings}
\label{app:setup}

For complex $y_{L,R}$, inserting Eq.~\eqref{eq:vevs} into Eq.~\eqref{eq:Lchi_weyl} gives Eq.~\eqref{eq:Lchi_matrix} with
\begin{equation}
\begin{gathered}
\mathcal M_\chi=\begin{pmatrix}m_L&M_\chi\\ M_\chi&m_R\end{pmatrix},\\
m_L=\frac{y_L\,v_\phi}{\sqrt2},\qquad m_R=\frac{y_R^*\,v_\phi}{\sqrt2},\\
\mathcal S=\frac{1}{\sqrt2}\begin{pmatrix}y_L&0\\0&y_R^*\end{pmatrix},\\
\mathcal A=\frac{i}{\sqrt2}\begin{pmatrix}y_L&0\\0&-y_R^*\end{pmatrix}.
\end{gathered}
\label{eq:app_MSA}
\end{equation}
The complex conjugates arise because $\xi_2\xi_2$ couples to $\phi^*=(v_\phi+s-ia)/\sqrt2$. A phase of $M_\chi$ can be removed by rephasing
$\chi$, so we take $M_\chi>0$. The remaining rephasing $\xi_1\to e^{i\alpha}\xi_1$, $\xi_2\to e^{-i\alpha}\xi_2$ leaves $M_\chi$ invariant and multiplies both $y_L$ and $y_R$ by $e^{2i\alpha}$. Physical quantities therefore depend only on $|y_L|$, $|y_R|$ and the CP-violating phase $\arg(y_Ly_R^*)$. For real couplings, Eq.~\eqref{eq:app_MSA} reduces to
Eq.~\eqref{eq:MSA}.

Two exact relations follow from the structure of the Lagrangian. First, the Majorana entries of $\mathcal M_\chi$ are linear in $v_\phi$, while the Dirac entry does not depend on it, so that
\begin{equation}
\mathcal S=\frac{\partial\mathcal M_\chi}{\partial v_\phi}
=\frac{\mathcal M_\chi-M_\chi\sigma_1}{v_\phi}.
\label{eq:app_Srel}
\end{equation}
Second, the coupling of $a$ is generated by an infinitesimal $U(1)_L$ rotation of the Majorana entries. With $Q={\rm diag}(1,-1)$ the lepton numbers of $\xi_{1,2}$, and since the Dirac entry is invariant,
\begin{equation}
\mathcal A=\frac{i}{2v_\phi}\big(Q\mathcal M_\chi+\mathcal M_\chi Q\big).
\label{eq:app_Arel}
\end{equation}
Both relations hold at tree level for any values of $\mu'$ and $\kappa$, which enter only the scalar potential.

\subsection{Scalar and pseudoscalar couplings}
\label{app:bilinears}

The mass eigenstates $\xi=U\eta$ are defined by the Takagi factorization~\cite{Dreiner:2008tw}
\begin{equation}
U^T\mathcal M_\chi U=\Sigma\equiv{\rm diag}(m_{\chi_1},m_{\chi_2}),\qquad
m_{\chi_i}\ge0,
\label{eq:app_takagi}
\end{equation}
with $U$ unitary. In the mass basis,
\begin{equation}
\begin{gathered}
\mathcal L\supset-\frac12\big(s\,C^s_{ij}+a\,C^a_{ij}\big)\eta_i\eta_j
+{\rm h.c.},\\
C^s=U^T\mathcal S\,U,\qquad C^a=U^T\mathcal A\,U .
\end{gathered}
\label{eq:app_C}
\end{equation}
For the Majorana fields $\chi_i=(\eta_i,\bar\eta_i)^T$, in the chiral basis with $\gamma^5={\rm diag}(-1,1)$,
\begin{equation}
\begin{gathered}
\overline{\chi_i}\chi_j=\eta_i\eta_j+\bar\eta_i\bar\eta_j,\\
\overline{\chi_i}\,i\gamma^5\chi_j=-i\big(\eta_i\eta_j-\bar\eta_i\bar\eta_j\big),
\end{gathered}
\end{equation}
so that $C\,\eta_i\eta_j+C^*\bar\eta_i\bar\eta_j={\rm Re}\,C\;
\overline{\chi_i}\chi_j-{\rm Im}\,C\;\overline{\chi_i}\,i\gamma^5\chi_j$ and
\begin{equation}
\mathcal L\supset-\frac{s}{2}\sum_{i,j}\Big[{\rm Re}\,C^s_{ij}\,
\overline{\chi_i}\chi_j-{\rm Im}\,C^s_{ij}\,\overline{\chi_i}\,i\gamma^5\chi_j\Big],
\label{eq:app_bilinears}
\end{equation}
and likewise for $a$. The real parts are scalar couplings, which give a central Yukawa potential in the non-relativistic limit. The imaginary parts are pseudoscalar couplings; their non-relativistic vertex, $\bar u_j\,i\gamma^5u_i\simeq\boldsymbol\sigma\cdot\mathbf q/(2m_\chi)$, gives
a spin-dependent potential suppressed by $q^2/m_\chi^2$.

This decomposition is physical. Once the masses are real and positive, the only remaining field redefinitions are $\eta_i\to\pm\eta_i$. These change the signs of the $C_{ij}$ but cannot move a coupling between its real and imaginary parts. In particular, the redefinition $\eta_1\to i\eta_1$, which would turn a pseudoscalar transition coupling into a scalar one, reverses the sign of
$m_{\chi_1}$ and is therefore not available.

Since $U$ is unitary, the Frobenius norms of the coupling matrices do not depend on the basis:
\begin{equation}
\begin{aligned}
\sum_{i,j}|C^s_{ij}|^2=\sum_{i,j}|C^a_{ij}|^2
&=\frac{|y_L|^2+|y_R|^2}{2}\\
&=2\big(\kappa_\chi^2+\eta_\chi^2\big),
\end{aligned}
\label{eq:app_norm}
\end{equation}
where, up to signs, we generalize Eq.~\eqref{eq:kappa_eta} to
\begin{equation}
\kappa_\chi\equiv\frac{|y_L+y_R|}{2\sqrt2},\qquad
\eta_\chi\equiv\frac{|y_L-y_R|}{2\sqrt2}.
\label{eq:app_kappa_eta}
\end{equation}

\subsection{Mass splitting}
\label{app:splitting}

The masses follow from the invariants
${\rm tr}(\mathcal M_\chi^\dagger\mathcal M_\chi)=m_{\chi_1}^2+m_{\chi_2}^2$ and
$|\det\mathcal M_\chi|=m_{\chi_1}m_{\chi_2}$, which give the exact result
\begin{equation}
\big(m_{\chi_2}\mp m_{\chi_1}\big)^2=2M_\chi^2+|m_L|^2+|m_R|^2\mp2\,|M_\chi^2-m_Lm_R|.
\label{eq:app_dm_exact}
\end{equation}
Expanding
$|M_\chi^2-m_Lm_R|=M_\chi^2-{\rm Re}(m_Lm_R)+[{\rm Im}(m_Lm_R)]^2/(2M_\chi^2)
+\dots$, and using $|m_L+m_R^*|^2=|m_L|^2+|m_R|^2+2\,{\rm Re}(m_Lm_R)$, we
obtain
\begin{equation}
\begin{gathered}
\Delta m_\chi=|m_L+m_R^*|\left[1+\mathcal O\!\left(\frac{|m|^2}{M_\chi^2}\right)\right],\\
|m_L+m_R^*|=2\kappa_\chi v_\phi,\\
\bar M\equiv\frac{m_{\chi_1}+m_{\chi_2}}{2}
=M_\chi+\mathcal O\!\left(\frac{|m|^2}{M_\chi}\right),
\end{gathered}
\label{eq:app_dm}
\end{equation}
where $|m|$ stands for $|m_L|$ or $|m_R|$. The relative correction remains small as $\Delta m_\chi\to0$, because ${\rm Im}(m_Lm_R)$ vanishes together with $m_L+m_R^*$. For real couplings with $M_\chi^2>m_Lm_R$, ${\rm Im}(m_Lm_R)=0$ and $\Delta m_\chi=|m_L+m_R|$ exactly, in agreement with Eq.~\eqref{eq:masses}.

\subsection{Couplings of the pseudo-Majoron}
\label{app:majoron}

Inserting Eq.~\eqref{eq:app_Arel} and $\mathcal M_\chi=U^*\Sigma U^\dagger$
into Eq.~\eqref{eq:app_C} gives
\begin{equation}
C^a=\frac{i}{2v_\phi}\big(\tilde Q\,\Sigma+\Sigma\,\tilde Q^T\big),\qquad
\tilde Q\equiv U^TQ\,U^* .
\label{eq:app_Ca}
\end{equation}
The matrix $\tilde Q$ is Hermitian, unitary and traceless, so that $\tilde Q=\mathbf n\cdot\boldsymbol\sigma$ with $\mathbf n$ a real unit vector. Using $\tilde Q_{ji}=\tilde Q_{ij}^*$, we find
\begin{equation}
\begin{gathered}
{\rm Re}\,C^a_{ij}=\frac{{\rm Im}\,\tilde Q_{ij}}{2v_\phi}\,
\big(m_{\chi_i}-m_{\chi_j}\big),\\
{\rm Im}\,C^a_{ij}=\frac{{\rm Re}\,\tilde Q_{ij}}{2v_\phi}\,
\big(m_{\chi_i}+m_{\chi_j}\big),
\end{gathered}
\label{eq:app_goldstone}
\end{equation}
exactly. The scalar couplings of $a$ are therefore proportional to mass differences. They vanish on the diagonal, and the transition coupling satisfies
\begin{equation}
|{\rm Re}\,C^a_{12}|=|n_2|\,\frac{\Delta m_\chi}{2v_\phi}
\le\frac{\Delta m_\chi}{2v_\phi}.
\end{equation}
This is the relation implied by the Goldstone theorem~\cite{Goldstone:Majorana:2002}. Here it follows from the $U(1)_L$ invariance of the Yukawa interactions, Eq.~\eqref{eq:app_Arel}, and therefore holds for arbitrary soft breaking in the scalar potential.

The bound is saturated. The pseudoscalar couplings
${\rm Im}\,C^a_{11}=n_3m_{\chi_1}/v_\phi$,
${\rm Im}\,C^a_{22}=-n_3m_{\chi_2}/v_\phi$ and
${\rm Im}\,C^a_{12}=n_1(m_{\chi_1}+m_{\chi_2})/(2v_\phi)$ are bounded by
Eq.~\eqref{eq:app_norm}, so that $n_{1,3}=\mathcal O(|y|v_\phi/M_\chi)$ and
$|n_2|=1-\mathcal O(|y|^2v_\phi^2/M_\chi^2)$. Hence
\begin{equation}
|{\rm Re}\,C^a_{12}|=\frac{\Delta m_\chi}{2v_\phi}
\left[1+\mathcal O\!\left(\frac{|y|^2v_\phi^2}{M_\chi^2}\right)\right],
\label{eq:app_Ca_result}
\end{equation}
and, by Eq.~\eqref{eq:app_norm}, the pseudoscalar couplings of $a$ carry the
remaining strength, $\sum_{i,j}({\rm Im}\,C^a_{ij})^2\simeq2\eta_\chi^2$.

\subsection{Couplings of the radial mode}
\label{app:radial}

Inserting Eq.~\eqref{eq:app_Srel} into Eq.~\eqref{eq:app_C} gives, exactly,
\begin{equation}
C^s=\frac{\Sigma-M_\chi T}{v_\phi},\qquad T\equiv U^T\sigma_1U .
\label{eq:app_Cs}
\end{equation}
The matrix $T$ is symmetric and unitary. From Eq.~\eqref{eq:app_takagi},
$M_\chi T=\Sigma-U^T{\rm diag}(m_L,m_R)\,U$, so that $T=\mathbf 1+X$ with
$X=\mathcal O(|m|/M_\chi)$. Unitarity, $T^\dagger T=\mathbf 1$, together with
$X^T=X$, gives $2\,{\rm Re}\,X=-X^\dagger X$, and hence
${\rm Re}\,T=\mathbf 1+\mathcal O(|m|^2/M_\chi^2)$. With
$m_{\chi_{1,2}}=\bar M\mp\Delta m_\chi/2$ and Eq.~\eqref{eq:app_dm},
\begin{equation}
{\rm Re}\,C^s_{ij}=\mp\frac{\Delta m_\chi}{2v_\phi}\,\delta_{ij}
+\mathcal O\!\left(\frac{|y|^2v_\phi}{M_\chi}\right),
\label{eq:app_Cs_result}
\end{equation}
with the upper (lower) sign for $i=1$ ($i=2$). The diagonal entries agree with
the Hellmann-Feynman relation
${\rm Re}\,C^s_{ii}=\partial m_{\chi_i}/\partial v_\phi$, which follows from $\mathcal S=\partial\mathcal M_\chi/\partial v_\phi$. For real couplings they reduce to $g_{11}$ and $g_{22}$ of Eq.~\eqref{eq:g11}, in which $\eta_\chi c$ is the $\mathcal O(y^2v_\phi/M_\chi)$ correction. The transition coupling of $s$ has no scalar part at this order, and by Eq.~\eqref{eq:app_norm} the pseudoscalar couplings of $s$ carry
$\sum_{i,j}({\rm Im}\,C^s_{ij})^2\simeq2\eta_\chi^2$.

Equations~\eqref{eq:app_Ca_result} and~\eqref{eq:app_Cs_result} show that every scalar coupling of $s$ and $a$ satisfies
\begin{equation}
|g^{\rm central}|\le\frac{\Delta m_\chi}{2v_\phi}
+\mathcal O\!\left(\frac{|y|^2v_\phi}{M_\chi}\right),
\end{equation}
with equality for the diagonal couplings of $s$ and the transition coupling of $a$. This proves Eq.~\eqref{eq:central_bound} for arbitrary phases. The couplings of order $\eta_\chi$, which do not vanish as $\Delta m_\chi\to0$, are all pseudoscalar. In the antisymmetric regime $y_R\simeq-y_L$, where $\Delta m_\chi\to0$ and $\eta_\chi\simeq|y_L|/\sqrt2$, neither $s$ nor $a$ can therefore mediate a strong central interaction, for any value of the CP-violating phase.

For real couplings, the leading-order mass eigenstates reduce to Eq.~\eqref{eq:Takagi} at $\vartheta=\pi/4$ and reproduce the couplings of
Eq.~\eqref{eq:couplings}.

\subsection{Numerical check}
\label{app:numerics}

We have verified these results by performing the Takagi factorization numerically for $2\times10^4$ randomly chosen complex $y_L$ and $y_R$, with $|y_{L,R}|\in[0.1,1]$, uniformly distributed phases and $M_\chi/v_\phi=10^3$. Equations~\eqref{eq:app_dm_exact} and~\eqref{eq:app_goldstone} hold to a relative accuracy of \(10^{-11}\), and the diagonal scalar couplings of $a$ vanish identically. The
largest scalar coupling of either $s$ or $a$ exceeds $\Delta m_\chi/(2v_\phi)$
by at most $0.5\,|y|_{\rm max}^2v_\phi/M_\chi$, with a median excess of $7\times10^{-5}$, in agreement with Eq.~\eqref{eq:central_bound}.

\section{Partial-wave computation of the self-interaction cross section}
\label{app:partialwave}

\paragraph{Direct integration.}
For each relative velocity we integrate Eq.~\eqref{eq:radial} for all partial waves $\ell<L_{\rm num}$ simultaneously, as one linear system for $u_\ell$ and $u_\ell'$, with an adaptive eighth-order Runge-Kutta scheme (DOP853) with relative tolerance $10^{-10}$ and step size limited to $\min(0.5,0.3/\hat k)$. Partial wave $\ell$ starts at $x_0=10^{-3}(\ell+1)/\max(b,1)$, or at $\hat k x_0=\ell/5$, $\ell/2$ and $0.8\,\ell$ for $10\leq\ell<300$, $300\leq\ell<1000$ and $\ell\geq1000$, respectively, well inside the centrifugal barrier, with the regular behavior $u_\ell\propto x^{\ell+1}[1-bx/(2\ell+2)]$ and an arbitrary small normalization that prevents overflow in the classically forbidden region; the $s$ wave starts at $x_0=10^{-4}/\max(b,1)$. The integration extends to $x_{\max}=30+\ln\max(b,1)$, where the potential is negligible, and the phase shifts follow from Eq.~\eqref{eq:matching},
\begin{equation}
\tan\delta_\ell=\frac{\hat\jmath_\ell(z)\,u_\ell'-\hat k\,\hat\jmath_\ell'(z)\,u_\ell}
{\hat n_\ell(z)\,u_\ell'-\hat k\,\hat n_\ell'(z)\,u_\ell}\bigg|_{z=\hat k x_{\max}} ,
\end{equation}
which determines $\delta_\ell$ modulo $\pi$, sufficient for Eq.~\eqref{eq:sigmaV}. Partial waves for which the Born phase shift $\delta_\ell^{\rm B}=(b/2\hat k)\,Q_\ell\big(1+1/(2\hat k^2)\big)$ is below $10^{-2}$ are taken from the Born expression; $Q_\ell$ is evaluated with Miller's downward recurrence normalized to $Q_0(z)=\frac12\ln\frac{z+1}{z-1}$. The velocity average is performed by interpolating $\ln\sigma_V$ between 28 velocity nodes from $3$ to $5500$~km/s.

\paragraph{Validation.}
For weak coupling, $b=10^{-3}$, the numerical phase shifts agree with the Born result to within $10^{-4}$-$10^{-3}$, consistent with higher-order corrections of relative size $b/\hat k$, and the partial-wave expression Eq.~\eqref{eq:sigmaV} agrees with a direct angular integration of $d\sigma/d\Omega$ to $10^{-6}$. For $b=0.9$-$250$ and $\hat k=0.03$-$20$, the phase shifts agree with those of the variable phase method, Eq.~\eqref{eq:vpm}, integrated separately for each partial wave with the same adaptive scheme (relative tolerance $10^{-9}$), to better than $10^{-7}$ wherever the latter converges, and up to $\hat k=100$ the numerical phase shifts of the highest partial waves reproduce the Born values. Tightening the tolerances of the variable phase method by two orders of magnitude changes $\sigma_V$ by less than $10^{-4}$. At large $b$ and low velocities the variable phase method becomes stiff, as discussed in Sec.~\ref{sec:scattering}, and the comparison is restricted to the region where it converges.

\paragraph{Map of the parameter space.}
The $\chi^2$ map of Fig.~\ref{fig:selfint_map} is computed on a grid of $53$ values of $m_{\chi_1}$ between $2$ and $101$~GeV, logarithmically spaced up to $35$~GeV with additional rows above, and $70$ logarithmically spaced values of $m_s$ between $5$ and $100$~MeV, extended with the same spacing down to $1$~MeV, for the decoupled dark sector, and on every second point in each direction for a common temperature. At large $b$ the resonant islands are narrower than the grid spacing, so that the grid underestimates their extent. 

\begin{table}[!htbp]
\centering
\renewcommand{\arraystretch}{1.15}
\begin{tabular}{|c|c|c|c|c|c|}
\hline
& $m_{\chi_1}$ [GeV] & $m_s$ [MeV] & $\alpha_{11}$ & $\chi^2/{\rm dof}$ & $\Delta\chi^2$ \\
\hline
1 & 7.605 & 14.56 & $1.65\times10^{-3}$ & 2.039 & 0.00 \\
2 & 15.17 & 19.16 & $3.27\times10^{-3}$ & 2.058 & 0.30 \\
3 & 26.27 & 15.87 & $5.54\times10^{-3}$ & 2.065 & 0.42 \\
4 & 27.18 & 17.04 & $5.72\times10^{-3}$ & 2.067 & 0.44 \\
5 & 25.83 & 15.25 & $5.46\times10^{-3}$ & 2.068 & 0.45 \\
6 & 8.729 & 16.68 & $1.90\times10^{-3}$ & 2.069 & 0.47 \\
7 & 33.63 & 11.66 & $6.95\times10^{-3}$ & 2.074 & 0.55 \\
8 & 33.81 & 11.76 & $6.99\times10^{-3}$ & 2.075 & 0.57 \\
9 & 33.88 & 11.78 & $7.00\times10^{-3}$ & 2.077 & 0.60 \\
10 & 41.86 & 10.47 & $8.47\times10^{-3}$ & 2.100 & 0.96 \\
11 & 40.29 & 9.665 & $8.18\times10^{-3}$ & 2.105 & 1.05 \\
12 & 55.29 & 7.810 & $1.07\times10^{-2}$ & 2.127 & 1.40 \\
13 & 55.24 & 7.794 & $1.07\times10^{-2}$ & 2.127 & 1.40 \\
14 & 54.87 & 7.699 & $1.07\times10^{-2}$ & 2.128 & 1.42 \\
15 & 40.77 & 10.00 & $8.27\times10^{-3}$ & 2.132 & 1.49 \\
16 & 40.19 & 9.503 & $8.16\times10^{-3}$ & 2.141 & 1.63 \\
17 & 27.89 & 12.12 & $5.86\times10^{-3}$ & 2.165 & 2.01 \\
18 & 27.77 & 12.05 & $5.84\times10^{-3}$ & 2.166 & 2.03 \\
19 & 45.90 & 7.596 & $9.18\times10^{-3}$ & 2.187 & 2.36 \\
20 & 29.52 & 13.36 & $6.18\times10^{-3}$ & 2.199 & 2.55 \\
21 & 47.28 & 8.014 & $9.41\times10^{-3}$ & 2.200 & 2.57 \\
22 & 43.94 & 11.97 & $8.84\times10^{-3}$ & 2.213 & 2.77 \\
23 & 25.85 & 10.62 & $5.46\times10^{-3}$ & 2.225 & 2.98 \\
24 & 27.97 & 21.04 & $5.88\times10^{-3}$ & 2.267 & 3.65 \\
25 & 22.03 & 17.70 & $4.69\times10^{-3}$ & 2.279 & 3.84 \\
26 & 21.82 & 17.46 & $4.65\times10^{-3}$ & 2.281 & 3.86 \\
27 & 52.97 & 11.05 & $1.04\times10^{-2}$ & 2.284 & 3.91 \\
28 & 62.68 & 7.392 & $1.19\times10^{-2}$ & 2.290 & 4.01 \\
29 & 63.41 & 7.547 & $1.20\times10^{-2}$ & 2.292 & 4.04 \\
30 & 63.39 & 7.546 & $1.20\times10^{-2}$ & 2.292 & 4.04 \\
31 & 23.10 & 19.17 & $4.91\times10^{-3}$ & 2.300 & 4.17 \\
32 & 36.50 & 11.11 & $7.48\times10^{-3}$ & 2.321 & 4.50 \\
33 & 58.29 & 9.067 & $1.12\times10^{-2}$ & 2.332 & 4.69 \\
34 & 30.03 & 8.987 & $6.28\times10^{-3}$ & 2.371 & 5.30 \\
35 & 70.47 & 7.140 & $1.30\times10^{-2}$ & 2.421 & 6.11 \\
\hline
\end{tabular}
\caption{Refined minima of all connected regions of the exact $\chi^2$ map inside the 95\% confidence region (decoupled dark sector), ordered by $\chi^2$. Minima of neighboring regions can coincide after refinement.}
\label{tab:selfint_branches_all}
\end{table}

The minima of all connected
regions inside the $95\%$ CL region are therefore refined by a two-dimensional Nelder-Mead minimization; these refined minima are listed in
Table~\ref{tab:selfint_branches_all} and define the families of Table~\ref{tab:selfint_branches}. The upper limit of Eq.~\eqref{eq:mass_limit} is obtained by minimizing $\chi^2$ in $m_s$ at fixed $m_{\chi_1}$, as described in Sec.~\ref{sec:selfint_results}.

\section{Thermalization through the Higgs portal}
\label{app:thermal}

The portal coupling produces dark-sector particles in the decays $h\to ss$
and, for $m_a<m_h/2$, $h\to aa$, each with width
\begin{equation}
\Gamma_h=\frac{\lambda_{H\phi}^2v_h^2}{32\pi m_h}.
\label{eq:app_gamma_h}
\end{equation}
In a plasma of temperature $T$ the Higgs bosons have the equilibrium density $n_h=m_h^2T K_2(x)/(2\pi^2)$, with $x=m_h/T$, and their decays are slowed by the time-dilation factor $\langle m_h/E_h\rangle=K_1(x)/K_2(x)$. Each decay produces two dark-sector particles, so that the production rate density is
\begin{equation}
\dot n_D=2N_{\rm ch}\,\Gamma_h\,n_h\,\frac{K_1(x)}{K_2(x)}
=\frac{N_{\rm ch}\,\Gamma_h\,m_h^2\,T\,K_1(x)}{\pi^2},
\end{equation}
where $N_{\rm ch}=2$ if both channels are open and $N_{\rm ch}=1$ otherwise. We regard the dark sector as thermalized if, at some temperature, this rate exceeds the Hubble rate, $H=1.66\sqrt{g_*}\,T^2/M_{\rm Pl}$, times the equilibrium density of the dark sector, $n_D^{\rm eq}=g_D\zeta(3)T^3/\pi^2$:
\begin{equation}
\frac{\dot n_D}{H\,n_D^{\rm eq}}
=\frac{N_{\rm ch}\,\Gamma_h\,M_{\rm Pl}}{1.66\sqrt{g_*}\,g_D\,\zeta(3)\,m_h^2}\;
x^4K_1(x)>1 .
\label{eq:app_therm_ratio}
\end{equation}
The function $x^4K_1(x)$ is maximal, with value $3.34$, at $x\simeq3.4$, i.e.\ at $T\simeq m_h/3$, where the production is most efficient before the Higgs bosons become Boltzmann suppressed. The condition is therefore first
satisfied for
\begin{equation}
\lambda_{\rm th}=\left[\frac{32\pi\times1.66\sqrt{g_*}\,g_D\,\zeta(3)\,m_h^3}
{3.34\,N_{\rm ch}\,v_h^2\,M_{\rm Pl}}\right]^{1/2}.
\label{eq:app_lambda_th}
\end{equation}
With $g_*=106.75$ and $g_D=5$ effective degrees of freedom ($s$, $a$ and the two Majorana fermions, the latter weighted by $3/4$), we obtain $\lambda_{\rm th}=6.4\times10^{-8}$ for $N_{\rm ch}=2$. If $h\to aa$ is closed and $a$ is absent from the plasma ($N_{\rm ch}=1$, $g_D=4$), $\lambda_{\rm th}=8.1\times10^{-8}$. Using the lattice value $g_*\simeq91$ at $T\simeq m_h/3$ instead lowers both values by $4\%$. Scattering processes $HH^\dagger\to\phi\phi^*$ above the electroweak scale add to the production, so that these values are conservative. 

For $\lambda_{H\phi}>\lambda_{\rm th}$, the ratio of Eq.~\eqref{eq:app_therm_ratio} exceeds one over a range of temperatures and falls below one again once the Higgs bosons are Boltzmann suppressed. The dark sector then decouples at the temperature at which
\begin{equation}
\left(\frac{\lambda_{H\phi}}{\lambda_{\rm th}}\right)^2\frac{x^4K_1(x)}{3.34}=1,
\qquad x>3.4 .
\label{eq:app_tdec}
\end{equation}
This gives $T_{\rm dec}\simeq32$, $19$, $11$ and $9$~GeV for $\lambda_{H\phi}=6.5\times10^{-8}$, $10^{-7}$, $5\times10^{-7}$ and $10^{-6}$, respectively, the range quoted in Sec.~\ref{sec:darktemp}; for the largest couplings allowed by direct detection for BP-A, $\lambda_{H\phi}\simeq3\times10^{-6}$, $T_{\rm dec}\simeq8$~GeV, still far above freeze-out. The default value $T_{\rm dec}=20$~GeV corresponds to $\lambda_{H\phi}\simeq10^{-7}$. 

For $\lambda_{H\phi}<\lambda_{\rm th}$, the same decays populate the dark sector only by freeze-in, without bringing it into equilibrium with the SM; the dark sector may then thermalize among itself at a temperature below that of the SM, and the relic coupling differs from the one derived in Sec.~\ref{sec:relic}. We do not consider this case; for self-interacting dark matter produced by freeze-in, see, e.g., Ref.~\cite{Ghosh:2021wrk}.

\section{Rescaling of the supernova bound}
\label{app:sn}

The bound of Eq.~\eqref{eq:sn_universal} applies to a coupling $g_{\alpha\beta}=g\,\delta_{\alpha\beta}$. In our model the couplings in the
flavor basis are $G_{\alpha\beta}=(m_\nu)_{\alpha\beta}/v_\phi$, Eq.~\eqref{eq:s_nu_coupling_flavor}, and in the mass basis $g_i=m_{\nu_i}/v_\phi$. We rescale the bound by comparing the signal in Kamiokande-II and IMB for these couplings with that for universal couplings
under identical core conditions.

In the core, oscillations are suppressed and the mediator is produced by coalescence of flavor states, $\nu_\alpha\nu_\beta\to s$ and
$\bar\nu_\alpha\bar\nu_\beta\to s$. For $m_s\ll T$ the emitted energy is, up to a factor common to both cases,
\begin{equation}
\begin{gathered}
Q\propto\sum_{\alpha,\beta}\frac{|G_{\alpha\beta}|^2}{2}
\Big[I(\mu_\alpha,\mu_\beta)+I(-\mu_\alpha,-\mu_\beta)\Big], \\
I(\mu_a,\mu_b)=\int_0^\infty\!dE\,E\!\int_0^E\!dE'\,
f(E';\mu_a)f(E-E';\mu_b),
\end{gathered}
\label{eq:app_sn_Q}
\end{equation}
where the factor $1/2$ accounts for identical neutrinos for $\alpha=\beta$ and for the double counting of $\alpha\neq\beta$ in the symmetric sum, and $f(E;\mu)=[e^{(E-\mu)/T}+1]^{-1}$ is the Fermi-Dirac distribution with chemical potentials $\mu_{\nu_e}$ for $\nu_e$ and zero for $\nu_{\mu,\tau}$. The mediators decay far outside the neutrinosphere, into mass eigenstates, so that the fraction of their decay products detected as $\bar\nu_e$ is
\begin{equation}
f_{\bar e}=\frac{\sum_ig_i^2|U_{ei}|^2}{\sum_ig_i^2},
\label{eq:app_sn_fe}
\end{equation}
compared with $1/3$ for universal couplings. For the couplings of interest the decay length is $\gamma c\tau_s\gtrsim3\times10^{9}$~cm, inside the envelope of the progenitor, so that flavor conversion of the decay products in the envelope, neglected here, could modify $f_{\bar e}$. The signal is $S\propto Q\,f_{\bar e}$ and scales as $g^2$, so that the bound on the largest coupling becomes $g_3m_s\lesssim F\times10^{-9}$~MeV with
\begin{equation}
F=\left(\frac{S_{\rm universal}(g=1)}{S_{\rm model}(g_3=1)}\right)^{1/2}.
\end{equation}

For a non-degenerate core, $\mu_{\nu_e}=0$, all integrals $I$ are equal, and $\sum_{\alpha\beta}|G_{\alpha\beta}|^2=\sum_ig_i^2$. The rescaling factor then takes the simple form
\begin{equation}
F=\frac{m_3}{\big(\sum_im_{\nu_i}^2|U_{ei}|^2\big)^{1/2}}
\qquad(\mu_{\nu_e}=0),
\label{eq:app_sn_F}
\end{equation}
independent of the temperature and of the Majorana phase. Its square combines the reduced production, $3m_3^2/\sum_im_{\nu_i}^2\simeq2.9$, with the reduced fraction of $\bar\nu_e$, $(1/3)/f_{\bar e}\simeq11$, so that $F\simeq(2.9\times11)^{1/2}$. For normal ordering with $m_1=0$ and the oscillation parameters of Sec.~\ref{sec:oscillations}, Eq.~\eqref{eq:app_sn_F} gives $F=5.7$; for inverted ordering, with $m_2$ in place of $m_3$, it gives $F\simeq1.0$. For $\mu_{\nu_e}>0$ we evaluate Eq.~\eqref{eq:app_sn_Q} numerically for $T=20$-$40$~MeV and $\mu_{\nu_e}\leq4T$, scanning the Majorana phase. At fixed $\mu_{\nu_e}/T$ the result is independent of $T$: for normal ordering, $F=5.7$, $9.2$ and $17.7$ for $\mu_{\nu_e}/T=0$, $2$ and $4$, and the Majorana phase changes these values by less than $5\%$; for inverted ordering, $F\simeq0.9$-$1.4$. These are the ranges quoted in Sec.~\ref{sec:sn_bound}. Combined with $\tau_s<0.1$~s, Eq.~\eqref{eq:tau_s}, the two requirements on $g_3$ are compatible for
\begin{equation}
m_s\lesssim\frac{F^2(10^{-9}~{\rm MeV})^2}{16\pi\hbar}\,
\tau_{\max}\sum_i\left(\frac{g_i}{g_3}\right)^2 ,
\end{equation}
which gives $m_s\lesssim100$~MeV for $F=5.7$ and normal ordering.

\bibliography{references}

\end{document}